\documentclass[twocolumn,aps,prb,10pt,nofootinbib]{revtex4-2}

\usepackage{amsmath}
\usepackage{amssymb}
\usepackage{wasysym}
\usepackage{graphicx}
\usepackage{hyperref}
\usepackage{dsfont}
\usepackage{newtxtext}
\usepackage[varvw]{newtxmath}
\usepackage{mathtools}
\usepackage[shortlabels]{enumitem}
\usepackage[percent]{overpic}
\usepackage{tabularx}
\usepackage{physics}

\hypersetup{
    colorlinks,
    linkcolor={blue},
    citecolor={blue},
    urlcolor={blue}
}

\graphicspath{{./}{./figs/}}

\newcommand{\rme}{\mathrm{e}}
\newcommand{\rmi}{\mathrm{i}}

\newcommand{\NCTO}{$\mathrm{Na_2Co_2TeO_6}$}
\newcommand{\NCSO}{$\mathrm{Na_3Co_2SbO_6}$}
\newcommand{\BCAO}{$\mathrm{BaCo_2(AsO_4)_2}$}

\newcommand{\llangle}{\langle\!\langle}
\newcommand{\rrangle}{\rangle\!\rangle}
\newcommand{\expvaltext}[1]{\langle #1 \rangle}
\newcommand{\unitvec}[1]{\hat{\mathbf{#1}}}
\newcommand{\ssop}[2]{\left[ #1 || #2 \right]}

\begin{document}

\title{%
Ferrimagnetism from quantum fluctuations in Kitaev materials
}

\author{Niccol\`o Francini}
\author{Pedro M. C\^{o}nsoli}
\author{Lukas Janssen}

\affiliation{Institut f\"ur Theoretische Physik and W\"urzburg-Dresden Cluster of Excellence ct.qmat, TU Dresden, 01062 Dresden, Germany}

\begin{abstract}
Ferrimagnetism appears in the temperature-field phase diagrams of several candidate Kitaev materials, such as the honeycomb cobaltates Na$_2$Co$_2$TeO$_6$ and Na$_3$Co$_2$SbO$_6$. In a number of instances, however, the exact nature of the corresponding ground states remains the subject of ongoing debate. We show that general symmetry considerations can rule out candidate states that are incompatible with the observed ferrimagnetic behavior. In particular, we demonstrate that a ferrimagnetic response cannot be reconciled with a collinear zigzag ground state, owing to the combined time-reversal and translational symmetry inherent to that configuration.
Instead, the observed behavior is fully compatible with the symmetries of noncollinear multi-$\mathbf q$ states, such as the triple-$\mathbf q$ discussed in the context of Na$_2$Co$_2$TeO$_6$.
We exemplify this general result by computing the ferrimagnetic response of an extended Heisenberg-Kitaev-Gamma model with explicit sublattice symmetry breaking within linear spin-wave theory. If the model realizes a triple-$\mathbf q$ ground state, the calculated magnetization curve is well consistent with the low-temperature behavior observed in Na$_2$Co$_2$TeO$_6$. 
In this case, a finite magnetization remains in the zero-temperature limit as a consequence of quantum fluctuations, even if the $g$-factors on the different sublattices are identical.
For a zigzag ground state, by contrast, the total magnetization vanishes both at zero and finite temperatures, independent of possible sublattice-dependent $g$-factors, as expected from the symmetry analysis. The implications of our general result for other Kitaev materials exhibiting ferrimagnetic behavior are also briefly discussed.
\end{abstract}

\date{July 14, 2025}

\maketitle

\section{Introduction}
\label{sec:intro}

Since its introduction~\cite{kitaev06}, the Kitaev honeycomb model has attracted significant interest in the study of frustrated magnetism.
This is motivated not only by the fact that the model hosts different quantum spin-liquid phases, characterized by fractionalized excitations and emergent gauge fields~\cite{savary16,zhou17,knolle19,broholm20}, but also by the prospect of realizing its bond-dependent, Ising-like interactions in transition-metal compounds with partially filled $d$ shells~\cite{jackeli09,chaloupka10,chaloupka13}.
Among the various candidates proposed to exhibit such behavior, the honeycomb cobaltates \NCTO~\cite{yao20, songvilay20, lin21, chen21, lee21, hong21, samarakoon21, kim22, mukherjee22, sanders22, yang22, yao22, krueger23, yao23, xiang23, zhang23, hong23, pilch23, bera23, gillig23, miao24, zhou24, lin24, arneth24, bischof25}, \NCSO~\cite{yan19, songvilay20, kim22, sanders22, li22, gu24, hu24, miao24}, and \BCAO~\cite{zhong20, shi21, zhang22, maksimov22b, maksimov25} have emerged as a promising class of Kitaev materials.
In the cobaltates, the combined effect of crystal-field environments, strong spin-orbit coupling, and Coulomb repulsion on the $d^7$ electronic configuration of $\mathrm{Co^{2+}}$ ions leads to the formation of a Mott-insulating phase with effective $j=1/2$ local magnetic moments~\cite{liu18,sano18,liu20,winter22,rousochatzakis24}. The geometry of the edge-sharing CoO$_6$ octahedra enhances the role of the Kitaev interactions~\cite{jackeli09}.
However, as in other Kitaev materials~\cite{trebst22}, magnetic interactions beyond the nearest-neighbor Kitaev exchange drive these honeycomb cobaltates away from the ideal Kitaev limit and stabilize long-range magnetic order at low temperatures and small external magnetic fields.

Nevertheless, the nature of the magnetic order in the honeycomb cobaltates remains a subject of ongoing debate.
In the case of \NCTO, early neutron scattering data obtained from powder samples were interpreted in favor of a collinear zigzag ground state~\cite{songvilay20, lin21, kim22, sanders22}. Later experiments on single crystals, however, demonstrated that the symmetry of the magnon spectrum cannot be explained within the zigzag scenario~\cite{krueger23}. This led to the proposal that \NCTO\ rather displays a noncollinear triple-$\mathbf{q}$ order~\cite{chen21, krueger23, gu25} and called for further experiments capable of distinguishing both scenarios. Electric polarization measurements employed to this end found conflicting results~\cite{zhang23,kocsis24}. Recent Faraday rotation measurements have revealed a sizable internal field, interpreted as a consequence of the spin vorticity associated with the noncollinear triple-$\mathbf{q}$ order~\cite{jin25}.
A similar ambiguity arises in \NCSO, where the observed Bragg-peak pattern is consistent with either a domain mixture of two single-$\mathbf{q}$ zigzag states or a noncollinear double-$\mathbf{q}$ order~\cite{li22, gu24, hu24}.

Amid ongoing uncertainty, \NCTO\ exhibits a remarkable low-temperature ferrimagnetic response when cooled in a weak out-of-plane magnetic field~\cite{yao20}. When the field is turned off at low temperatures, a small remanent magnetization of $-0.01\mu_\text{B}$/Co$^{2+}$ remains along the out-of-plane direction, oriented \emph{opposite} to the training field. Upon warming, the magnetization grows and eventually reverses sign at a \emph{compensation point} $T_\star \simeq 12.5$~K. The magnetization reaches its maximum of $+0.01\mu_\text{B}$/Co$^{2+}$ slightly before dropping to zero at the ordering temperature $T_{\mathrm{c}}\simeq 26.7$~K. Above this temperature, the system enters the paramagnetic phase, and the magnetization remains zero.
In \NCSO, the residual magnetization vanishes once the external field is removed~\cite{yan19}. However, applying a finite field along an in-plane direction perpendicular to a Co-Co bond induces a metamagnetic transition to an intermediate phase, referred to as ``AFM $\tfrac13$,'' which shows features of ferrimagnetic behavior and suggests the system may lie in proximity to a ferrimagnetic state~\cite{li22}.

Ferrimagnetism typically arises when magnetic moments on inequivalent sublattices of an antiferromagnet do not compensate each other~\cite{kim22}.
On the honeycomb lattice, a zigzag ground state can give rise to ferrimagnetism only if the zigzag order is canted due to a coexisting Néel order~\cite{yao20}.
However, no known model within the extended Heisenberg-Kitaev parameter space supports such a peculiar form of canted zigzag order.
Moreover, a coexisting Néel order would give rise to an additional Bragg peak, which has not been observed within the experimental resolution in neutron diffraction measurements on \NCTO~\cite{lefrancois16, chen21}.
This points to an alternative mechanism underlying the ferrimagnetic behavior observed in this compound.

Recently, an extended Heisenberg-Kitaev model featuring a noncollinear triple-$\mathbf{q}$ classical ground state was introduced~\cite{francini24ferri}.
By explicitly accounting for the sublattice symmetry breaking in \NCTO, it was shown that the triple-$\mathbf{q}$ order generically develops ferrimagnetism at finite temperatures below the ordering transition.
In the classical model, however, fluctuations vanish as temperature approaches zero, leaving the remanent magnetization determined solely by the difference in $g$-factors between the $A$ and $B$ sublattices.
As a result, the existence of a compensation point $T_\star$ at a finite temperature below the ordering transition depends sensitively on the sublattice $g$-factor ratio, introducing a fine-tuning problem.
Moreover, in the low-temperature limit, the classical magnetization varies linearly with temperature, which is inconsistent with the activated behavior expected for a gapped magnetic excitation spectrum and observed experimentally in \NCTO~\cite{yao20}.
These artifacts of the classical model highlight the importance of quantum fluctuations in understanding the low-temperature behavior of ferrimagnetic Kitaev materials, an aspect that has been neglected in previous work~\cite{francini24ferri}.

In this study, we seek to address this gap. We explore extensions of the Heisenberg-Kitaev model featuring either collinear single-$\mathbf{q}$ zigzag order or noncollinear triple-$\mathbf{q}$ order, and analyze the effects of quantum fluctuations using linear spin-wave theory.
On one hand, we explain how symmetries forbid the zigzag ground state from exhibiting ferrimagnetism, providing insight into why this long-range order cannot acquire a uniform magnetization without further symmetry breaking in Heisenberg-Kitaev systems.
On the other hand, we show that the triple-$\mathbf{q}$ ground state exhibits ferrimagnetism both at zero and at finite temperatures below the ordering transition, under a broad set of generic and physically realistic symmetry assumptions.
In particular, our results are independent of specific $g$-factor ratios and naturally give rise to a compensation point $T_\star$, thereby substantially broadening the range of scenarios in which ferrimagnetism can emerge from a triple-$\mathbf{q}$ ground state.
Our results are directly applicable to the low-temperature physics of \NCTO, but may in principle also be of relevance to other honeycomb cobaltates, such as \NCSO\ in in-plane magnetic fields.

The remainder of the paper is organized as follows: In Sec.~\ref{sec:symmetries}, we discuss general symmetries constraints on the magnetization corrections. Our approach for computing the magnetization corrections within spin-wave theory for a noncollinear ground state is summarized in Sec.~\ref{sec:lsw}. In Sec.~\ref{sec:model}, we present a comprehensive analysis of magnetization corrections in the original Heisenberg-Kitaev model, augmented with sublattice-symmetry-breaking terms, using linear spin-wave theory.
The effects of an off-diagonal $\Gamma$ interaction present in extended Heisenberg-Kitaev models are discussed for the cases of single-$\mathbf q$ zizag and noncollinear triple-$\mathbf q$ ground states in Sec.~\ref{sec:ext-HK}.
We close with a summary and outlook in Sec.~\ref{sec:conclusion}.
Two appendices offer additional technical details of the spin-wave calculations.

\section{Symmetry constraints}
\label{sec:symmetries}
We start by discussing how symmetries constrain the corrections to sublattice magnetizations that arise due to quantum and thermal fluctuations.
Consider a Hamiltonian $\mathcal{H}$ describing a magnetic system with density matrix $\rho$. In the canonical ensemble, where the system is coupled to a heat bath at inverse temperature $\beta = 1/(k_\mathrm{B} T)$, the density matrix is given by the standard expression $\rho=e^{-\beta \mathcal{H}}/\Tr(\rme^{-\beta \mathcal{H}})$. A symmetry operation is defined as a transformation $\mathcal{T}$ that commutes with the Hamiltonian, $\comm{\mathcal H}{\mathcal T} = 0$. The subset of symmetry operations $\mathcal{T}$ that remain unbroken in the low-temperature ordered phase also commute with the density matrix, $\comm{\rho}{\mathcal{T}} = 0$.
At zero temperature, the density matrix reduces to a projector onto the ground state, $\rho = |\psi_0\rangle \langle\psi_0|$, and the condition $\comm{\rho}{\mathcal{T}} = 0$ implies that $\mathcal{T}$ can only change $\ket{\psi_0}$ by a harmless global phase. 

A symmetry $\mathcal{T} = \ssop{\mathcal{A}}{\mathcal{B}}$ of a magnetic Hamiltonian can generally be specified in terms of two operations, $\mathcal{A}$ and $\mathcal{B}$, that act respectively on spin and real space.
However, the inevitable presence of spin-orbit coupling implies that these operations are not independent: any proper or improper rotation in $\mathcal{B}$ must also be performed in $\mathcal{A}$. For a given magnetic system, the set of symmetries $\mathcal{T}$ that satisfy this property constitute a magnetic space group \cite{bradley_book}.
Recent developments \cite{corticelli22,smejkal22} have nonetheless demonstrated the utility of considering a broader class of symmetry groups, dubbed spin-space groups \cite{brinkman66,litvin74,litvin77}, in which $\mathcal{A}$ and $\mathcal{B}$ are completely decoupled.
A key advantage of such a generalization is that it enables one to predict new magnetic phenomena from enhanced symmetries that appear within \emph{minimal} models for a system of interest, but are weakly broken once all symmetry-allowed interactions are included. 
This link has successfully been explored to identify, e.g., instances of nontrivial magnon band topology \cite{corticelli22} as well as the existence of spin-split electronic bands in metallic altermagnets \cite{smejkal22}.
Here, we will also consider the effects of general spin-space transformations $\mathcal{T} = \ssop{\mathcal{A}}{ \{ \mathcal{R} | \mathbf{t} \}}$, where the real-space operation is expressed in terms of a translation $\mathbf{t}$ and a rotation $\mathcal{R}$ which is in principle independent of $\mathcal{A}$. However, we will ultimately show in Sec.~\ref{subsec:HKG-triple-q} that it is the \emph{breaking} of a spin-space symmetry that allows ferrimagnetism to emerge from a triple-$\mathbf{q}$ ordered state in Heisenberg-Kitaev magnets.

\begin{figure}[t]
    \centering
    \includegraphics[width=0.7\linewidth]{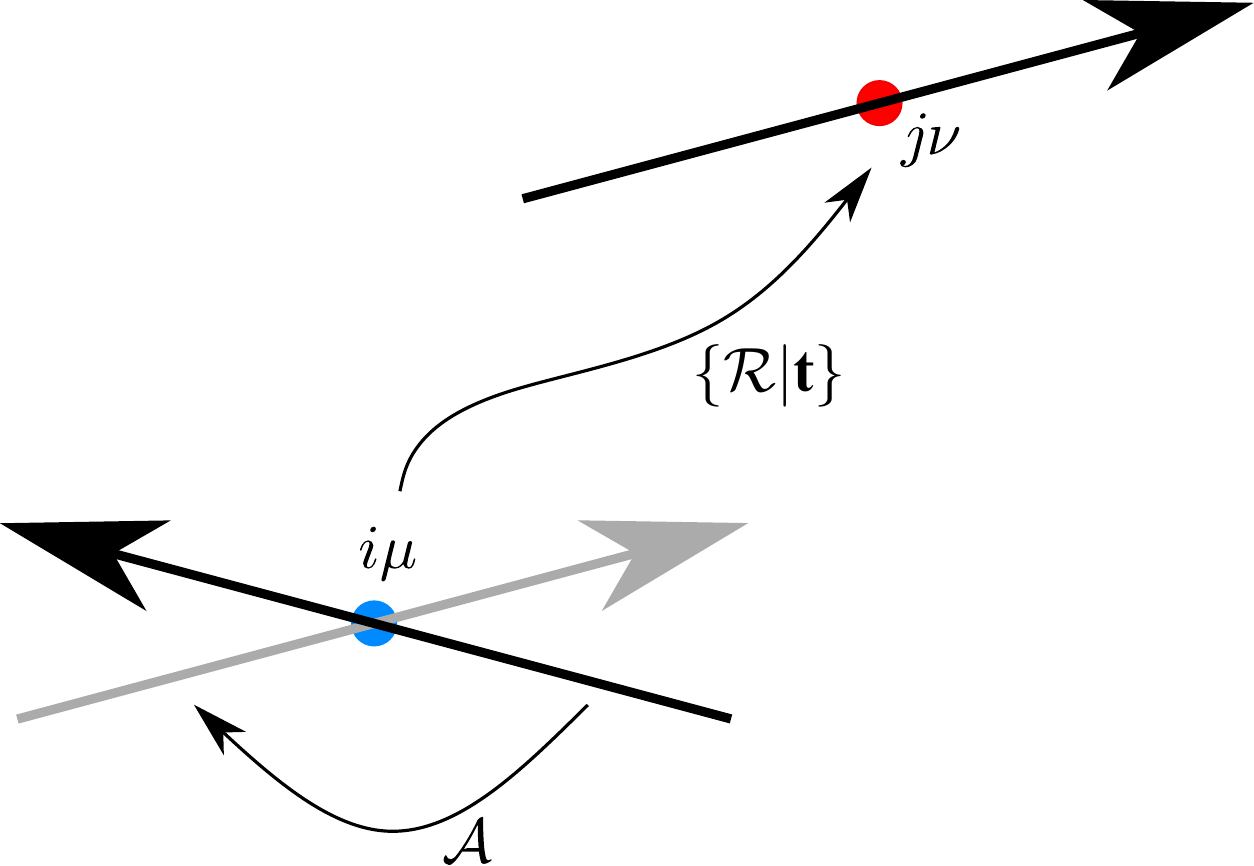}
\caption{Illustration of spin-space transformation mapping two inequivalent spins (large black arrows). A spin at site $i\mu$ (blue dot), where $i$ denotes the magnetic unit cell and $\mu$ the magnetic sublattice, is mapped to a spin at site $j\nu$ (red dot) in a two steps: First, the spin is rotated in spin space via the operation $\mathcal{A}$, with the resulting transformed spin shown as a gray arrow. Second, the spin transformation is accompanied by a lattice transformation $\{\mathcal R | \mathbf t\}$, consisting of a lattice rotation $\mathcal R$ and a translation $\mathbf t$, mapping site $i\mu$ to $j\nu$.
}
\label{fig:symmetry-sketch}
\end{figure}

To begin, consider the spins $\mathbf{S}_{i\mu}$ and $\mathbf{S}_{j\nu}$ located at lattice sites labeled by $i\mu$ and $j \nu$, where $i$ and $j$ specify the magnetic unit cells and $\mu$ and $\nu$ denote the magnetic sublattices.
If $\mathcal T$ represents an unbroken symmetry, the spin expectation values satisfy the relation
\begin{align}
        \langle \mathbf{S}_{i\mu} \rangle 
        &= \operatorname{Tr}\left[ \rho \mathbf{S}_{i\mu} \right] 
        = \operatorname{Tr}\left[\mathcal{T}^{-1}\mathcal{T} \rho \mathcal{T}^{-1}\mathcal{T}\mathbf{S}_{i\mu} \right]
        = \mathsf{A}\operatorname{Tr}\left[ \rho \mathbf{S}_{j\nu} \right]
        \nonumber\\& 
        = \mathsf{A}\langle \mathbf{S}_{j\nu}\rangle,
\label{eq:symm-thermal-avrg}
\end{align}
where the matrix $\mathsf{A}$ represents the action of the operation $\mathcal{A}$ in spin space, so that the transformation $\mathcal{T}$ acts on the spins as
\begin{equation}
    \label{eq:symmetry-on-spin}
    \mathcal{T}\mathbf{S}_{i\mu}\mathcal{T}^{-1} = \mathsf{A} \mathbf{S}_{j\nu},
\end{equation}
with sites $i\mu$ and $j\nu$ connected by $\mathbf{R}_{j\nu}=\mathsf{R} \mathbf{R}_{i\mu} + \mathbf{t}$, where $\mathsf{R}$ is the matrix representation of the proper or improper rotation operation $\mathcal{R}$, see Fig.~\ref{fig:symmetry-sketch}.
%
%
The result in Eq.~\eqref{eq:symm-thermal-avrg} holds exactly at all temperatures, and can therefore be used to significantly constrain possible magnetization corrections within the different candidate ground states.
In the following, we discuss two examples relevant for different phases within the extended Heisenberg-Kitaev parameter space.

\subsection{Translation and time reversal}
\label{subsub:translation+time-rev}
%
Assume that two collinear but oppositely aligned spins, $\mathbf{S}_{i\mu}$ and $\mathbf{S}_{j\nu}$, located at sites $i\mu$ and $j\nu$, are related by a symmetry $\mathcal{T}$ that combines a translation by a lattice vector $\mathbf t = \mathbf d$ with time reversal, $\mathcal A = \Theta$.
If $E$ denotes the identity operation, then the corresponding spin-space operation is $\mathcal{T}=[\Theta || \{E|\mathbf{d}\}]$, which acts as $\mathcal{T} \mathbf{S}_{i\mu} \mathcal{T}^{-1} = -\mathbf{S}_{j\nu}$. As a result, Eq.~\eqref{eq:symm-thermal-avrg} reduces to
\begin{equation}
    \label{eq:translation+TR-symmetry}
        \langle \mathbf{S}_{i\mu} \rangle = -\langle \mathbf{S}_{j\nu} \rangle.
\end{equation}
This nonperturbative result constrains the expectation value of the sum of the two spins to zero, thereby prohibiting any uncompensated moment from arising due to quantum or thermal fluctuations.

Below, we show that the collinear zigzag order found in both the original and extended Heisenberg-Kitaev models possesses precisely such kind of spin-space symmetry combining lattice translation and time reversal. As a result, the magnetization on each of the crystallographic sublattices $A$ and $B$ of the honeycomb lattice is individually constrained to vanish by symmetry. Importantly, this conclusion holds to all orders in the $1/S$ expansion, provided the spin-space symmetry remains unbroken in the ordered state.

\subsection{Translation and spin rotation}
%
As a second example, consider a state featuring a symmetry given by a combination of a lattice translation $\mathbf t = \mathbf{d}$ with a spin rotation $\mathcal{U}$. Rotations of by angle $\alpha$ around a unit vector $\hat{\mathbf{n}}$ are described by unitary operators $\mathcal{U}=\prod_{i\mu}e^{i\alpha\hat{\mathbf{n}}\cdot \mathbf{S}_{i\mu}}$. Under these transformations, the spin vector $\mathbf{S}_{j\nu}$ transforms as a vector, meaning that $\mathcal{U}\mathbf{S}_{j\nu} \mathcal{U}^{-1}=\mathsf{O}_{\hat{\mathbf{n}}}(\alpha)\mathbf{S}_{j\nu}$, where $\mathsf{O}_{\hat{\mathbf{n}}}(\alpha)$ is an orthogonal matrix describing the rotation of the spin vector by an angle $\alpha$ around the axis $\hat{\mathbf{n}}$. As a consequence, the spin-space transformation $\mathcal{T}=[\mathcal{U}||\{E|\mathbf{d}\}]$ acts as $\mathcal{T} \mathbf{S}_{i\mu} \mathcal{T}^{-1} = \mathsf{O}\mathbf{S}_{j\nu}$, where $\mathsf{O}$ is the rotation matrix corresponding to $\mathcal U$, and $\mathbf{R}_{j\nu}=\mathbf{R}_{i\mu}+\mathbf{d}$. We dropped the axis $\hat{\mathbf{n}}$ and the angle $\alpha$ in $\mathsf{O}$ for simplicity of notation. Hence, Eq.~\eqref{eq:symm-thermal-avrg} leads to the nonperturbative result
\begin{equation}
    \label{eq:translation+rotation-symmetry}
        \langle \mathbf{S}_{i\mu} \rangle = \mathsf{O}\langle \mathbf{S}_{j\nu} \rangle.
\end{equation}
Since the rotation matrix $\mathsf{O}$ is orthogonal, it preserves the spin length, implying that the fluctuation-induced spin corrections on the two inequivalent sites have equal magnitude. 

Below, we argue that the noncollinear triple-$\mathbf{q}$ order possesses this type of spin-space symmetry, provided that off-diagonal terms such as a $\Gamma$ interaction are absent from the spin interaction matrix. As a result, the magnetization is symmetry-constrained to vanish in these pure Heisenberg-Kitaev models. Once off-diagonal interactions are taken into account, such as in current models for \NCTO\ featuring a triple-$\mathbf{q}$ ground state~\cite{krueger23, francini24vestigial, francini24ferri}, the spin-space symmetry is broken. In this case, a sublattice imbalance, as observed in \NCTO~\cite{yao20}, generically gives rise to ferrimagnetism.
%

\section{Spin-wave theory}
\label{sec:lsw}
%
To illustrate the point within an explicit calculation, we compute the correction to the magnetization from quantum and thermal fluctuations within spin-wave theory.
Formally, spin-wave theory can be understood as an expansion in $1/S$, where $S$ is the spin magnitude and $S\rightarrow\infty$ represents the classical limit.

\subsection{$\boldsymbol{1/S}$ expansion}

Consider a bilinear spin Hamiltonian of the form
\begin{equation}
    \label{eq:bilinear-hamiltonian}
    \mathcal{H}= \sum_{i\mu,j\nu} \mathbf{S}_{i\mu} \mathsf{J}_{i\mu,j\nu}\mathbf{S}_{j\nu},
\end{equation}
with a $3 \times 3$ interaction matrix $\mathsf J_{i\mu,j\nu}$ for any given pair of sites $i\mu$ and $j\nu$.
Assume that the classical ground state contains $p$ sites per magnetic unit cell. Then, the magnetic-sublattice indices are $\mu, \nu = 1, \dots, p$, and the magnetic-unit-cell indices are $i, j = 1, \dots, N/p$, where $N$ is the total number of magnetic sites.
Due to translational invariance, the interaction matrix depends only on the relative position between the magnetic unit cells $i$ and $j$, $\mathsf{J}_{i\mu,j\nu}=\mathsf{J}_{\mu\nu}(\mathbf{R}_i - \mathbf{R}_j)$.
We express each spin $\mathbf{S}_{i\mu}$ in a sublattice-dependent reference frame $\{\mathbf{e}_{\mu}^{(1)},\mathbf{e}_{\mu}^{(2)},\mathbf{e}_{\mu}^{(3)}\}$, where the third axis is aligned with the classical spin direction, $\mathbf{e}_{\mu}^{(3)} = \langle \mathbf{S}_{i\mu}/S \rangle_{S \to \infty}$. In terms of spin ladder operators $S^\pm_{i\mu}$, the spin operators read
\begin{align}
\mathbf S_{i\mu} = \frac12 \left(S^{(+)}_{i\mu} + S^{(-)}_{i\mu}\right) \mathbf e^{(1)}_{\mu} + \frac{1}{2\rmi} \left(S^{(+)}_{i\mu} - S^{(-)}_{i\mu}\right) \mathbf e^{(2)}_{\mu} + S_{i\mu}^{(3)} \mathbf e^{(3)}_{\mu}.
\end{align}
In this basis, we employ a Holstein-Primakoff transformation~\cite{holstein40},
\begin{equation}
\label{eq:HP-bosons}
    \begin{split}
    S_{i\mu}^{(3)} &= S - a^{\dagger}_{i\mu} a_{i\mu}, \\
    S_{i\mu}^{(+)} &= \sqrt{2S} \left(1 -  \frac{a^{\dagger}_{i\mu} a_{i\mu}}{2S} \right)^{1/2} a_{i\mu}, \\
    S_{i\mu}^{(-)} &= a^{\dagger}_{i\mu} \sqrt{2S}\left(1 -  \frac{a^{\dagger}_{i\mu} a_{i\mu}}{2S} \right)^{1/2},
    \end{split}
\end{equation}
in which the spin operators are represented in terms of bosonic operators satisfying $[a_{i\mu},a^{\dagger}_{j\nu}]=\delta_{\mu\nu}\delta_{ij}$. 
The square roots in the spin ladder operators can be expanded in powers of $1/S$, resulting in the series
\begin{equation}
\label{eq:bosonic-expansion}
    \mathcal{H}=\sum_{n=0}^{\infty} S^{2-\frac{n}{2}} \mathscr{h}_n,
\end{equation}
where $\mathscr{h}_n$ contains $n$ bosonic operators.
Here, the zeroth-order term $\mathscr{h}_0$ describes the ground-state energy of the classical Hamiltonian.
The linear term $\mathscr{h}_1$ vanishes as long as we are expanding around a minimum of the energy.
As a consequence, the quadratic term $\mathscr{h}_2$ contains the leading quantum corrections within the $1/S$ expansion.
Higher-order terms $\mathscr{h}_{n \geq 3}$ describe boson-boson interactions. The cubic interactions $\mathscr{h}_3$ vanish if the spin interaction matrix $\mathsf{J}_{i\mu,i\nu}$ is diagonal for all $i\mu$ and $j\nu$, and if the classical ground is collinear. When off-diagonal spin interactions are present, such as the $\Gamma$ term in extended Heisenberg-Kitaev models~\cite{rau14a}, the cubic term generally does not vanish.

In practical calculations, the expansion is usually truncated after some order $n$, which may be expected to represent a reliable approximation as long as the boson density $\langle a_{i\mu}^\dagger a_{i\mu} \rangle$ is small compared to the spin length, $\expvaltext {a_{i\mu}^\dagger a_{i\mu}} \ll 2S$.
Since $\langle a_{i\mu}^\dagger a_{i\mu} \rangle$ grows with temperature, the spin-wave expansion remains valid only below a cutoff temperature scale $T_{\mathrm{max}}$, which depends on the spin magnitude $S$ and the system's level of frustration. $T_{\mathrm{max}}$ increases with $S$ but decreases with frustration.
Therefore, in strongly frustrated systems with significant quantum fluctuations, especially for $S=1/2$, the cutoff temperature scale can be low.

In linear spin-wave theory, Eq.~\eqref{eq:bosonic-expansion} is truncated after the quadratic order $n=2$, neglecting any boson-boson interactions $\mathscr{h}_{n \geq 3}$. In momentum space, Eq.~\eqref{eq:bilinear-hamiltonian} then acquires the compact expression
\begin{equation}
    \label{eq:momentum-hamiltonian}
    \mathcal{H}_\text{LSW} = S(S+1) \mathscr{h}_0 + 
    \frac{1}{2} \sum_{\mathbf{k}\in \mathrm{BZ}} \mathbf{x}^{\dagger}(\mathbf{k}) \mathsf{M}(\mathbf{k}) \mathbf{x}(\mathbf{k}) ,
\end{equation}
where the summation is over all momenta $\mathbf k$ within the magnetic Brillouin zone (BZ).
The $2p$-component vector 
\begin{equation*}
    \mathbf{x}^{\dagger}(\mathbf{k}) = \left(a^{\dagger}_1(\mathbf{k}), \dots, a^{\dagger}_p(\mathbf{k}), a_1(-\mathbf{k}), \dots , a_{p}(-\mathbf{k})) \right)
\end{equation*}
accounts for $p$ bosonic modes, one for each site in the magnetic unit cell. The Hamiltonian is diagonalized using a Bogoliubov transformation,
\begin{equation}
    \label{eq:diagonal-hamiltonian}
    \mathcal{H}_\text{LSW} = S(S+1)\mathscr{h}_0 + \sum_{\mathbf{k},\mu}\omega_{\mathbf{k}\mu}
    \left( \psi_{\mathbf{k}\mu}^\dagger\psi_{\mathbf{k}\mu} + \frac{1}{2} \right),
\end{equation}
where $\omega_{\mathbf{k}\mu}$ represents the magnon spectrum and $\psi_{\mathbf{k}\mu}$ are the Bogoliubov quasiparticle ladder operators; details are given in Appendix~\ref{appendix:LSW-details}.

\subsection{Magnetization corrections}
%
In the classical limit, the sublattice magnetization $\mathbf m_\mu \coloneqq \frac{p}{N} \sum_{i} \expval{\mathbf S_{i\mu}}$ is aligned with the $\mathbf e^{(3)}_\mu$ axis by definition, $\mathbf m_\mu \to m^{(3)}_\mu \mathbf{e}_{\mu}^{(3)}$, and fully saturated in the zero-temperature limit, so that $m^{(3)}_\mu/S \to 1$ for $S \to \infty$ and $T\to 0$.
Quantum fluctuations present at finite $S$ generally reduce the sublattice magnetization.

For a collinear ground state in the absence of any off-diagonal term in the spin interaction matrix $\mathsf{J}_{i\mu,j\nu}$, the sublattice magnetization at order $\mathcal O(1/S^0)$ of the semiclassical expansion remains parallel to the classical spin configuration, $\mathbf m_\mu = m^{(3)}_\mu \mathbf e^{(3)}_\mu + \mathcal O(1/S)$.
The total magnetization is then given by 
\begin{align}
\mathbf m & \coloneqq \frac{1}{N} \sum_{i,\mu} \expval{\mathbf S_{i\mu}} 
\nonumber\\&
=  \frac{S}{p} \sum_\mu \mathbf e^{(3)}_\mu - \frac{1}{N} \sum_{\mathbf{k},\mu}\langle 0 | a_{\mu}^\dagger (\mathbf{k}) a_{\mu}(\mathbf{k}) | 0 \rangle \mathbf e^{(3)}_\mu + \mathcal O(1/S).
\label{eq:magn-corrections}
\end{align}
For a collinear antiferromagnet, the first term vanishes. From symmetry, we expect that the second term yields a finite correction only if the spin-space symmetry connecting the two opposite spins is broken, such as in a N\'eel state on a honeycomb lattice with two inequivalent crystallographic sublattices.
For the zigzag state, the second term is constrained by symmetry to vanish, even if the two crystallographic sublattices on the honeycomb lattice are inequivalent. The same remains true for the higher-order terms $\mathcal O(1/S)$, as long as the spin-space symmetry is preserved. 
Below, we confirm this expectations within explicit calculations at the level of linear spin-wave theory.

If the classical ground state is noncollinear and/or the spin interaction matrix $\mathsf{J}_{i\mu,i\nu}$ contains off-diagonal terms, such as the $\Gamma$ term in extended Heisenberg-Kitaev models~\cite{rau14a}, then Eq.~\eqref{eq:magn-corrections} becomes incomplete. Three-boson terms $\mathscr{h}_3$ in Eq.~\eqref{eq:bosonic-expansion} correct the classical direction $\mathbf{e}_{i\mu}^{(3)}$ of the magnetic moments and produce a contribution at order $S^0$ that is absent in models with collinear ground states and diagonal spin interaction matrices~\cite{zhitomirsky98,coletta12,consoli20}.
To account for the missing contribution, we introduce a small external magnetic field $\mathbf{h} = h \mathbf n$ along a fixed but otherwise arbitrary direction $\mathbf{n}$ by adding a Zeeman term to the Hamiltonian,
\begin{align}
\mathcal{H} \mapsto \mathcal{H} - \mathbf{h} \cdot \sum_{i\mu} \mathbf{S}_{i\mu}.
\end{align}
This allows us to determine the total magnetization along the direction of $\mathbf n$ at zero temperature in the low-field limit as
\begin{equation}
    \label{eq:magn-external-field}
    m 
    \coloneqq \mathbf n \cdot \mathbf m \bigr|_{h \to 0} 
    = \mathbf n \cdot \frac{1}{N} \sum_{i\mu} \expval{\mathbf{S}_{i\mu}}\biggr|_{h\rightarrow0}
    = -\frac{1}{N} \frac{\partial E_{\mathrm{gs}}}{\partial h} \biggr|_{h\rightarrow0},
\end{equation}
where $E_\mathrm{gs}$ corresponds to the ground-state energy in linear spin-wave theory.
Importantly, the above equation includes all contributions required to obtain the magnetization correction at the leading order $S^0$~\cite{chernyshev09}, see Appendix~\ref{appendix:LSW-details} for more details.

No systematic method exists to treat the problem within the $1/S$ expansion at finite temperatures. However, given that Eq.~\eqref{eq:momentum-hamiltonian} is a free-boson Hamiltonian, one can readily write the Gibbs free energy
\begin{equation}
    \label{eq:free-energy}
    \mathcal{F} =
    S(S+1) \mathscr{h}_0 + \sum_{\mathbf{k}\mu} \left[ \frac{\omega_{\mathbf{k}\mu}}{2} + T \ln (1 - \rme^{-\omega_{\mathbf{k}\mu}/T})
    \right],
\end{equation}
with the temperature $T$ given in units in which $k_\mathrm{B} = 1$. From the above, we obtain the total magnetization at finite temperature as
\begin{equation}
    \label{eq:explicit-magn-correction}
    m(T) = - \frac{1}{N} \frac{\partial \mathcal F}{\partial h} \biggr|_{h \to 0}
     = m(0)-\frac{1}{N}\sum_{\mathbf{k},\mu}\left.\frac{\partial\omega_{\mathbf{k}\mu}/\partial h}{\rme^{\omega_{\mathbf{k}\mu}/T}-1}\right|_{h\rightarrow0},
\end{equation}
with $m(0)$ being the result from Eq.~\eqref{eq:magn-external-field}. This expression reveals that the magnetization at low temperatures varies if two conditions are met. First, the temperature has to be at least of the order of the energy gap, $T\gtrsim \min{\omega_{\mathbf{k}\mu}}$.
Second, the main contribution to the temperature variation of the magnetization comes from those $\omega_{\mathbf{k}\mu}$ which have the highest response to an external field, effectively introducing weights in the sum of Eq.~\eqref{eq:explicit-magn-correction}. This fact will play a role in analyzing the finite-temperature magnetization in the Sec.~\ref{sec:ext-HK}.
%

\section{Heisenberg-Kitaev model}
\label{sec:model}
In this section, we investigate how ferrimagnetism can emerge in a nearest-neighbor Heisenberg-Kitaev model supplemented with sublattice-dependent, next-nearest neighbor Heisenberg couplings, $J_{2A}$ and $J_{2B}$. By making $J_{2A} \ne J_{2B}$, we can emulate the asymmetry between the crystallographic sublattices, $A$ and $B$, of materials such as $\mathrm{Na}_2\mathrm{Co}_2\mathrm{TeO}_6$ \cite{yao20, francini24ferri}.
Our calculations were carried out for systems comprising $N = 2L^2$ spins of magnitude $S$, arranged on the sites of a honeycomb lattice with $L^2$ unit cells and primitive lattice vectors $\mathbf{t}_{\pm} = (\pm \sqrt{3}/2, 3/2)$. Unless otherwise specified, all numerical results presented below correspond to the case $S = 1/2$.

\subsection{Model}
We consider a $C_3^*$-symmetric Heisenberg-Kitaev model~\cite{chaloupka10,chaloupka13}
\begin{equation}
\label{eq:KH-term}
    \begin{split}
    \mathcal H_{\mathrm{HK}} = &\sum_{\gamma=x,y,z} \sum_{\langle ij \rangle_{\gamma}} 
    \Bigl[ J \mathbf{S}_i \cdot \mathbf{S}_j + K S_{i}^{\gamma}S_{j}^{\gamma} \Bigr]\, \\
     &+J_{2A}\sum_{\llangle ij \rrangle_A}\mathbf{S}_i \cdot \mathbf{S}_j + J_{2B}\sum_{\llangle ij \rrangle_B}\mathbf{S}_i \cdot \mathbf{S}_j,
    \end{split}
\end{equation}
where $J$ and $K$ parameterize the nearest-neighbor Heisenberg and Kitaev terms, respectively, while $J_{2A}$ and $J_{2B}$ characterize sublattice-dependent next-nearest-neighbor interactions. Note that $\langle ij \rangle_{\gamma}$ labels a pair of sites along a $\gamma$ bond, $\gamma = x,y,z$. In the following, we adopt the parametrization
\begin{align}
(J, K) = A (\cos\phi, \sin\phi)
\end{align}
with the overall energy scale set to $A = 1$ for convenience. At each site $i$, the spins $\mathbf{S}_i$ are represented  in a global cubic coordinate system, $\mathbf{S}_i=S_{i}^x \mathbf{e}_x + S_{i}^y \mathbf{e}_y + S_{i}^z \mathbf{e}_z$ ~\cite{janssen19}.

The nearest-neighbor Heisenberg-Kitaev model with $J_{2A} = J_{2B} = 0$ has been extensively studied using a variety of methods~\cite{chaloupka10,price12,price13,chaloupka13,janssen16,janssen17,consoli20,rousochatzakis24}. For $S \le 3/2$, the zero-temperature phase diagram includes spin-liquid phases near the Kitaev points $\phi = \pi/2$ and $3\pi/2$~\cite{koga18,stavropoulos19,dong20,jin22}, in addition to four phases exhibiting collinear magnetic order. These collinear phases appear for all values of $S$ and, listed in order of increasing $\phi \in [0, 2\pi)$, are referred to as Néel, zigzag, ferromagnet, and stripy.
Interestingly, in the classical limit $S \to \infty$, all four ordered phases exhibit an accidental ground-state degeneracy, leading to SU(2)-symmetric ground-state manifolds~\cite{janssen19}. As a consequence, the magnon spectra calculated within linear spin-wave theory exhibit pseudo-Goldstone modes at the ordering wave vectors or at wave vectors related to them by crystal symmetries~\cite{chaloupka15,winter18,rau18,khatua23}.

We have explicitly verified that the inclusion of small $J_{2A}$ and $J_{2B}$ does not alter the overall structure of the classical phase diagram. Furthermore, because these couplings are isotropic, the ordered phases retain their accidental ground-state degeneracy. As in the standard case with $J_{2A} = J_{2B} = 0$, this degeneracy is lifted by thermal or quantum fluctuations through an order-by-disorder mechanism, which favors spin orientations along one of the three cubic axes $\mathbf{e}_x$, $\mathbf{e}_y$, or $\mathbf{e}_z$~\cite{chaloupka10,chaloupka13,price12,price13}.

The model defined in Eq.~\eqref{eq:KH-term} exhibits several symmetries, including combined spin-lattice $C_3^*$ rotations, time reversal $\Theta$, translations by the crystallographic unit vectors $\mathbf{t}_\pm$, and $C_2^{\alpha}$ spin rotations about the $\alpha = x, y, z$ axes~\cite{janssen19}. In addition, the two crystallographic sublattices $A$ and $B$ are related by a reflection through a plane containing any of the three types of honeycomb bonds. When $J_{2A} \ne J_{2B}$, the two sublattices become inequivalent, and the system no longer retains symmetry under these mirror reflections. For magnetically-ordered states with a two-site unit cell, this removes the constraint that the sublattice magnetizations be equal in magnitude, and generically results in ferrimagnetism. However, as we will show below, additional symmetries can prohibit magnetically-ordered states with larger unit cells from developing a net magnetization, irrespective of the values of $J_{2A}$ and $J_{2B}$.

\subsection{Magnetization corrections}
\label{subsec:results}

Applying linear spin-wave theory as described in Sec.~\ref{sec:lsw}, we have studied the effect of quantum fluctuations around the classical ground states of the Heisenberg-Kitaev model $\mathcal{H}_{\mathrm{HK}}$ defined in Eq.~\eqref{eq:KH-term}. 
To compute the magnetization corrections, we employed Eq.~\eqref{eq:magn-corrections}, which applies to collinear ground states and diagonal spin interaction matrices. We have explicitly verified that the more general method based on Eq.~\eqref{eq:magn-external-field}, valid also for noncollinear configurations, yields consistent results.
The numerical calculations presented below were carried out for fixed values $(J_{2A}, J_{2B}) = (0.2, -0.1)$, while varying the angle $\phi$, which parametrizes the ratio between the nearest-neighbor Kitaev ($K$) and Heisenberg ($J$) couplings.
\subsubsection{Ferromagnetic state}
\label{subsubsec:fm-gs}
\begin{figure}
    \centering
        \begin{overpic}[width=0.8\linewidth]{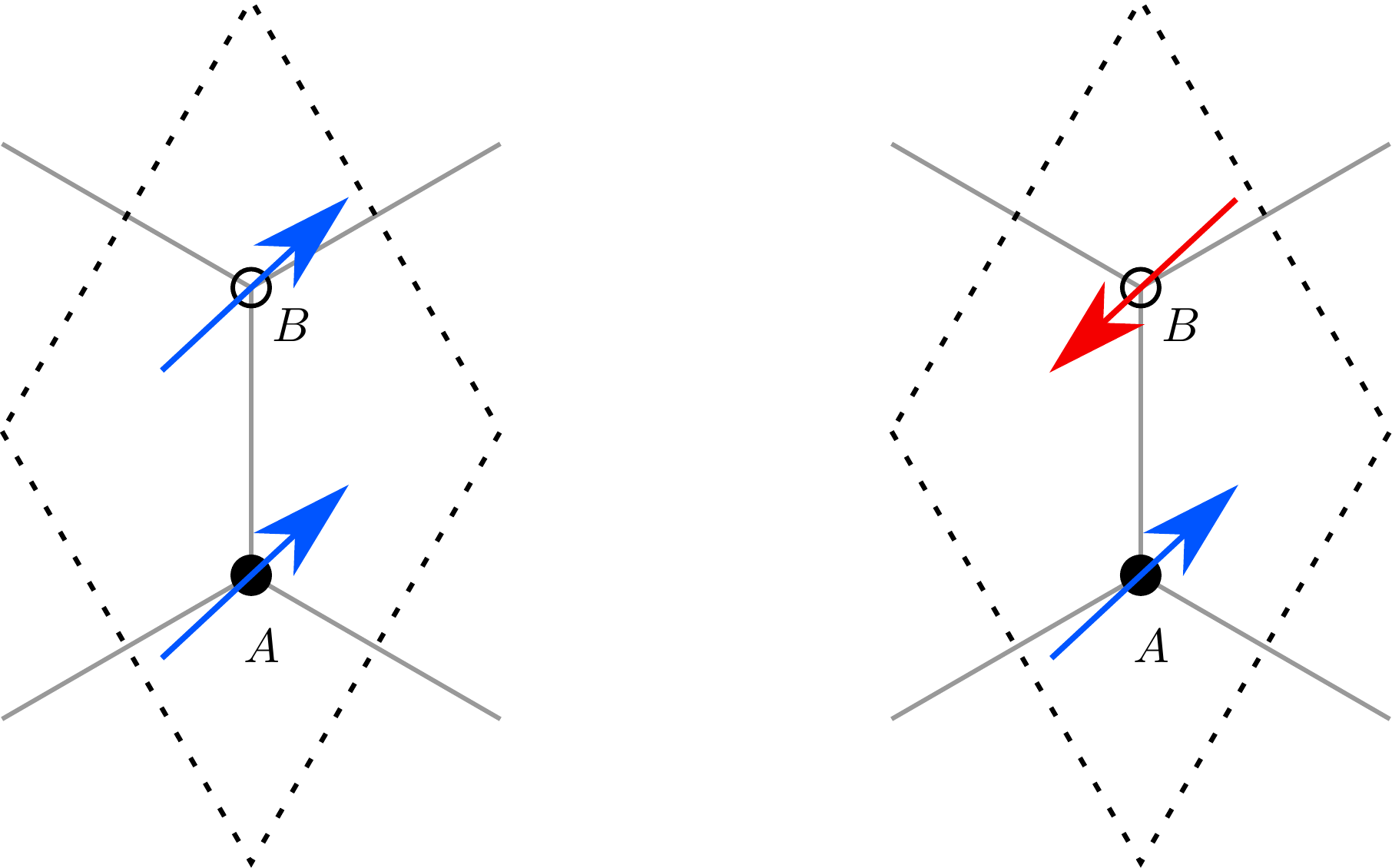}
        \put(0,60){(a)}
        \put(65,60){(b)}
    \end{overpic}
    \caption{Two-site magnetic unit cell in the (a) ferromagnetic and (b) Néel ground states. Here, the magnetic sublattices ($A$ and $B$) coincide with the crystallographic sublattices, represented by the filled and open circles.}
    \label{fig:two-site-cell}
\end{figure}

The ferromagnetic phase of the nearest-neighbor Heisenberg-Kitaev model corresponds to states with the same two-site unit cell as the underlying honeycomb lattice, see Fig.~\ref{fig:two-site-cell}(a). As discussed above, the inclusion of couplings $J_{2A}\ne J_{2B}$ eliminates \emph{all} symmetries that exchange $A \leftrightarrow B$, and thus allows fluctuations to act differently on each sublattice. We consequently expect that, while still oriented along the same direction, the sublattice magnetizations $\mathbf{m}_A$ and $\mathbf{m}_B$ will acquire different magnitudes.
\begin{table}[!tb]
\caption{Sublattice magnetizations $\mathbf{m}_\mu = m_\mu \mathbf{e}_z$ in the Heisenberg-Kitaev model with nearest-neighbor couplings $(J,K)=(\cos \phi,\sin \phi)$ and next-nearest-neighbor couplings $(J_{2A}, J_{2B}) = (0.2, -0.1)$, from linear spin-wave theory for $S=1/2$. The angles $\phi = \pi/4$ and $5\pi/4$ serve as representative points within the Néel and ferromagnetic phases, respectively, with classical ground states assumed as $\mathbf{S}_A = \mathbf{S}_B = S \mathbf{e}_z$ and $\mathbf{S}_A = -\mathbf{S}_B = S \mathbf{e}_z$.
}
    \centering
    \begin{tabularx}{\linewidth}{X >{\centering\arraybackslash}X >{\centering\arraybackslash}X >{\centering\arraybackslash}X>{\centering\arraybackslash}X}
    \hline 
    \hline
        Order & $\phi$ & $m_A$ & $m_B$ & $|\mathbf m_\text{sec}|$ \\
        \hline
        Ferromagnet & $5\pi/4$ & 0.481 & 0.485 & 0.002\\
        N\'eel & $\pi/4$ & 0.137 & $-0.141$ & 0.002 \\
        \hline
        \hline
    \end{tabularx}
    \label{tab:two-site-results}
\end{table}

To validate this picture, we carried out an explicit spin-wave calculation at a representative point in the ferromagnetic phase of the Heisenberg-Kitaev model, namely $\phi = 5\pi/4$. Taking $\mathbf{S}_A = \mathbf{S}_B = S \mathbf{e}_z$ as a classical ground state, we found that the system indeed develops different sublattice magnetizations, given by $\mathbf{m}_A \approx 0.481 \mathbf e_z$ and $\mathbf{m}_B \approx 0.485 \mathbf e_z$ for $S=1/2$, see Table~\ref{tab:two-site-results}. 
The resulting state can thus be viewed as a weak spin-density wave superimposed on a ferromagnetic background, $\mathbf m = (m \pm \Delta m) \mathbf e_z$, with net magnetization $m\approx0.483$ and  an oscillation amplitude of $\Delta m \approx 0.002$.

We observe the same qualitative trend for every $\phi$ in the ferromagnetic phase except at the SU(2)-symmetric point $\phi = \pi$. Here, the ferromagnetic state $\ket{\uparrow\uparrow \cdots\uparrow}$ is an exact ground state of the Hamiltonian, so that the system cannot have a nonzero staggered magnetization at $T=0$.
\subsubsection{Néel state}
The Néel order features a two-site unit cell with antiparallel spins, see Fig.~\ref{fig:two-site-cell}(b). Upon the inclusion of $J_{2A} \ne J_{2B}$, we encounter a behavior analogous to that of the ferromagnetic case: Provided that $K \ne 0$, the inequivalent sublattices experience different magnetization corrections at $T=0$. As a result, the system acquires a nonvanishing net magnetization, $\mathbf{m} = (\mathbf{m}_A + \mathbf{m}_B)/2$, and becomes ferrimagnetic.
(The discussion of the case $K = 0$ is deferred to Sec.~\ref{subsec:ferri-strength}.)
For a representative point $\phi = \pi/4$, a linear spin-wave calculation about the classical state $\mathbf{S}_A = -\mathbf{S}_B = S \mathbf{e}_z$ yields $\mathbf m_A \approx 0.137 \mathbf e_z$ and $\mathbf m_B \approx - 0.141 \mathbf e_z$ for $S=1/2$, as shown in Table~\ref{tab:two-site-results} .

\subsubsection{Zigzag state}
\label{subsubsec:zz-gs}
\begin{figure}[tb!]
\centering
\begin{overpic}[width=0.89135\linewidth]{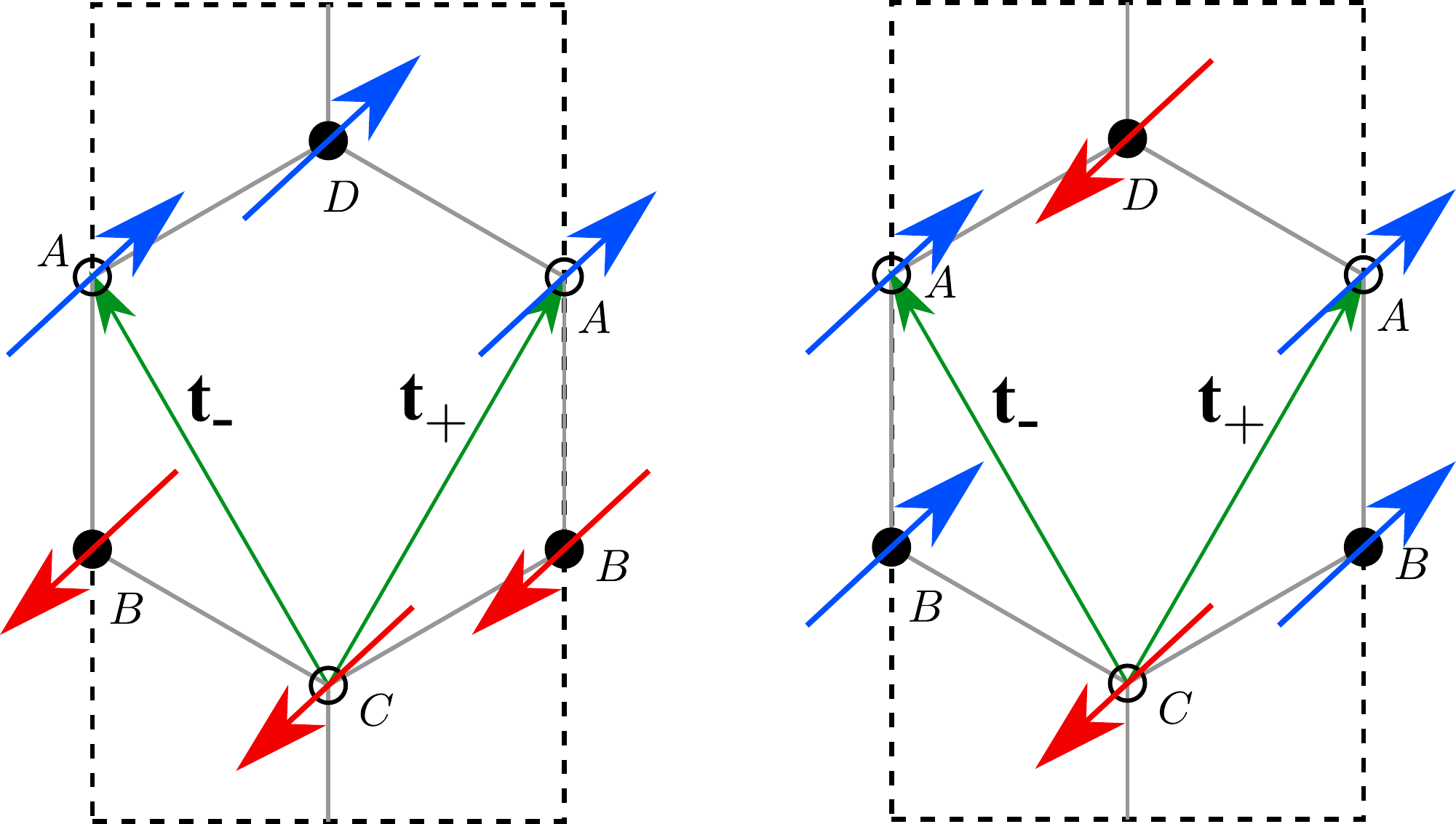}
\put(-3,55){(a)}
\put(52,55){(b)} 
\end{overpic}
\caption{Four-site magnetic unit cell in the (a) zigzag and (b) stripy ground states. $A$, $B$, $C$, and $D$ denote the four magnetic sublattices, with the two crystallographic sublattices indicated by filled and open circles. $\mathbf t_{+}$ and $\mathbf t_-$ are lattice translation vectors.
}
\label{fig:four-site-cell}
\end{figure}

Zigzag ground states can be viewed as a sequence of ferromagnetically ordered chains whose ordering direction alternates across nearest-neighbor bonds of a given type $\gamma=x,y,$ or $z$. This results in a four-site magnetic unit cell as depicted in Fig.~\ref{fig:four-site-cell}(a). Importantly, the transformation $\mathcal{T}=[\Theta ||\{E|\mathbf{t}_{\pm}\}]$, which combines time reversal $\Theta$ with a translation by either $\mathbf{t}_{+}$ or $\mathbf{t}_{-}$, acts on the spin configuration as
\begin{equation}
    \label{eq:four-site-symmetry-action}
    \mathcal{T} : 
    (\mathbf{S}_A, \mathbf{S}_B, \mathbf{S}_C, \mathbf{S}_D)
    \longmapsto
    -(\mathbf{S}_C, \mathbf{S}_D, \mathbf{S}_A, \mathbf{S}_B).
\end{equation}
and thus constitutes a symmetry of the Hamiltonian and the zigzag state. The combined translation and time reversal transformation $\mathcal{T}$ represents a particular instance of the transformations discussed in Sec.~\ref{subsub:translation+time-rev}. As a consequence of Eq.~\eqref{eq:translation+TR-symmetry}, the sublattice magnetizations in the zigzag phase satisfy the relations $\mathbf{m}_A = -\mathbf{m}_C$ and $\mathbf{m}_B = -\mathbf{m}_D$. Note that this result only relies on symmetries, and is thus valid to all orders in $1/S$. We conclude that, even if $J_{2A} \ne J_{2B}$, a zigzag state \emph{cannot} become ferrimagnetic.

\begin{table}[tb!]
\caption{Sublattice magnetizations $\mathbf{m}_\mu = m_\mu \mathbf{e}_z$ in the Heisenberg-Kitaev model with nearest-neighbor couplings $(J,K)=(\cos \phi,\sin \phi)$ and next-nearest-neighbor couplings $(J_{2A}, J_{2B}) = (0.2, -0.1)$, from linear spin-wave theory for $S=1/2$. The angles $\phi=5\pi/8$ and $13\pi/8$ serve as representative points within the zigzag and stripy phases, respectively, with classical ground states assumed as $\mathbf{S}_A = - \mathbf{S}_B = -\mathbf{S}_C = \mathbf{S}_D = S \mathbf{e}_z$ and $\mathbf{S}_A = \mathbf{S}_B = -\mathbf{S}_C = - \mathbf{S}_D = S \mathbf{e}_z$.
}
    \centering
    \begin{tabularx}{\linewidth}{X >{\centering\arraybackslash}X>{\centering\arraybackslash}X>{\centering\arraybackslash}X>{\centering\arraybackslash}X>{\centering\arraybackslash}X>{\centering\arraybackslash}X}
    \hline 
    \hline
        Order & $\phi$ & $m_A$ & $m_B$ & $m_C$ & $m_D$ & $|\mathbf m_\text{sec}|$\\
        \hline
        Zigzag & $5\pi/8$ & 0.238 & $-0.250$ & $-0.238$ & 0.250 & 0\\
        Stripy & $13\pi/8$ & 0.480 & 0.489 & $-0.480$ & $-0.489$ & 0\\
        \hline
        \hline
    \end{tabularx}
    \label{tab:four-site-results}
\end{table}

\begin{figure}[tb!]
\centering
\begin{overpic}[width=0.95\linewidth]{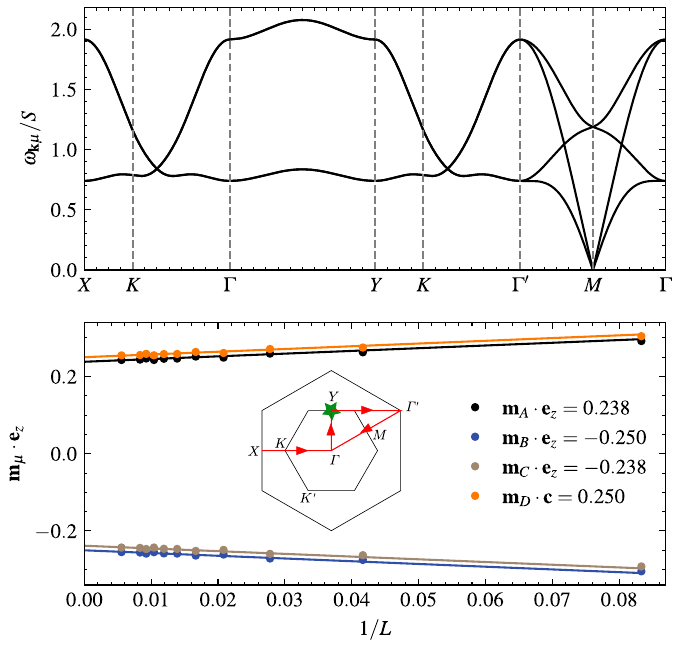}
\put(0,95){(a)}
\put(0,48){(b)}
\end{overpic}
\caption{%
(a)~Magnon spectrum in the $z$-zigzag ground state of the Heisenberg-Kitaev model with nearest-neighbor couplings $(J, K) = (\cos \phi, \sin \phi)$ at $\phi = 5\pi/8$, and next-nearest-neighbor couplings $(J_{2A}, J_{2B}) = (0.2, -0.1)$, from linear spin-wave theory. The momentum path through the extended Brillouin zone is shown in the central inset of (b), with the green star indicating the Bragg peak associated with the $z$-zigzag order. Pseudo-Goldstone modes appear at $\mathbf{M}$ due to an accidental SU(2) degeneracy of the classical ground state.
(b)~Sublattice magnetizations $\mathbf{m}_\mu$ along the [001] direction for $\mu = A, B, C, D$ in the $z$-zigzag ground state for $S=1/2$, using the same model parameters as in~(a). Extrapolated values in the thermodynamic limit $1/L \to 0$ are shown in the right inset.
}
\label{fig:HK-ZZ}
\end{figure}

To support this conclusion, we performed linear spin-wave calculations for $\phi = 5\pi/8$ around a classical $z$-zigzag reference state with $\mathbf{S}_A=\mathbf{S}_D = S \mathbf{e}_z$ and $\mathbf{S}_B=\mathbf{S}_C = -S \mathbf{e}_z$. The resulting spectrum is shown in  Fig.~\ref{fig:HK-ZZ}(a), where linearly dispersing pseudo-Goldstone modes appear due to the accidental SU(2) symmetry of the classical ground-state manifold.
When extrapolated to the thermodynamic limit, the sublattice magnetizations are found to be $\mathbf{m}_A = -\mathbf{m}_C \approx 0.238 \mathbf{e}_z$ and $\mathbf{m}_D = -\mathbf{m}_B \approx 0.250 \mathbf{e}_z$ for $S=1/2$ [see Fig.~\ref{fig:HK-ZZ}(b) and Table~\ref{tab:four-site-results}], yielding a vanishing net moment on each crystallographic sublattice, consistent with the symmetry analysis.

It is worth emphasizing that, in the collinear zigzag state, the constraint on sublattice magnetization corrections arises from the translational symmetry of the underlying honeycomb lattice. This is a fundamental symmetry that holds irrespective of any fine-tuning of the exchange parameters.

\subsubsection{Stripy state}
The remaining ordered phase of the nearest-neighbor Heisenberg-Kitaev model, the stripy phase, is characterized by a collinear spin configuration which is uniform along stripes parallel to one of the $\gamma$ bonds ($x,y$, or $z$), but alternates between adjacent stripes. This results in a four-site unit cell, as illustrated in Fig.\ref{fig:four-site-cell}(b), from which it is evident that the stripy phase also exhibits the symmetry $\mathcal{T} = [\Theta || \{E | \mathbf{t}_{\pm}\}]$ described in Eq.~\eqref{eq:four-site-symmetry-action}. Hence, just as in the zigzag case, symmetry prevents a stripy state from developing ferrimagnetism.

We once more sought to corroborate this symmetry-based argument by performing a linear spin-wave calculation for a representative point of the phase diagram, which we here took as $\phi = 13\pi/8$. Choosing a classical reference state given by $\mathbf{S}_A=\mathbf{S}_B= S\mathbf{e}_z$ and $\mathbf{S}_C=\mathbf{S}_{D} = -S\mathbf{e}_z$, we obtained $\mathbf{m}_A= -\mathbf{m}_C \approx 0.480 \mathbf{e}_z$ and $\mathbf{m}_B=-\mathbf{m}_D\approx 0.489 \mathbf{e}_z$ for $S=1/2$, as shown in Table~\ref{tab:four-site-results}. 
These results are consistent with symmetries and with the expected zero net magnetization.

\subsection{Discussion}
\label{subsec:ferri-strength}

\begin{figure}[tb!]
\centering
\includegraphics[width=\linewidth]{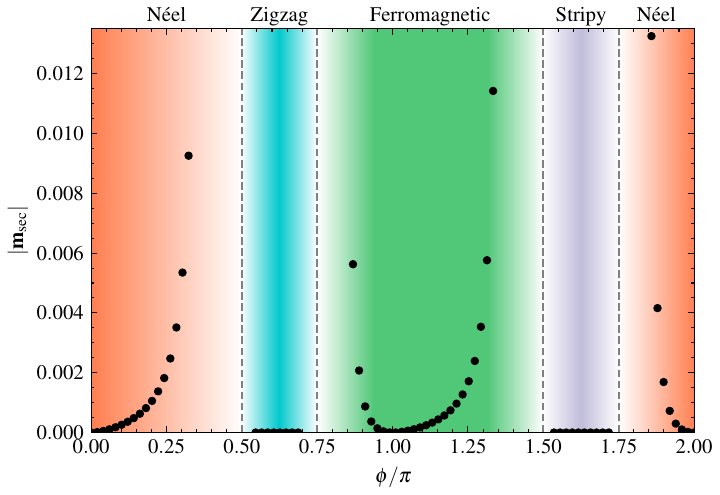}
\caption{%
Magnitude of the secondary order parameter $|\mathbf m_\text{sec}|$, defined in Eq.~\eqref{eq:unsat-magn}, as a function of the Heisenberg-Kitaev angle $\phi$, with nearest-neighbor couplings $(J,K) = (\cos\phi, \sin\phi)$ and fixed next-nearest-neighbor couplings $(J_{2A},J_{2B})=(0.2,-0.1)$ at zero temperature, from linear spin-wave theory for $S=1/2$.
As dictated by symmetry, there is no uncompensated moment in the zigzag and stripy phases, nor at the Heisenberg points ($\phi = 0, \pi$). The colors indicate the four magnetically-ordered phases of the nearest-neighbor model, while the white regions mark potential new phases near its phase boundaries (gray dashed lines), induced by the next-nearest-neighbor couplings.
}
\label{fig:corr-vs-phi-HK}
\end{figure}

In the previous subsection, we discussed conditions under which the reduction of spin-space symmetries by couplings $J_{2A}\neq J_{2B}$ can lead to ferrimagnetism. However, we would like to emphasize that the Kitaev exchange also has significant influence on the qualitative nature of the magnetic properties at $T=0$. In fact, as noted above, even when $J_{2A} \ne J_{2B}$, the \emph{ground state} at the ferromagnetic Heisenberg point $\phi = \pi$ (corresponding to $J = -1$ and $K = 0$) does not exhibit a staggered magnetization. It turns out that an analogous condition holds for the antiferromagnetic Heisenberg point $\phi=0$ ($J=1$ and $K=0$): Marshall's theorem~\cite{marshall55}, along with its extensions by Lieb, Schultz, and Mattis~\cite{lieb61,lieb62}, establishes that the ground state remains a total singlet for $J_{2A}, J_{2B} > -J$. Consequently, as long as this condition is satisfied, introducing unequal couplings $J_{2A} \ne J_{2B}$ does not induce a net magnetization at zero temperature.
At both SU(2)-symmetric points discussed above, the characteristic signature of ferrimagnetism, namely the coexistence of nonzero uniform and staggered magnetizations, emerges only at finite temperature. This arises from unequal thermal population of split magnon bands, which leads to a nonvanishing uniform magnetization at $\phi = 0$ and a staggered magnetization at $\phi = \pi$~\cite{consoli21}.
In light of the above discussion, we aim to quantify the degree of ferrimagnetism across the Heisenberg-Kitaev phase diagram at zero temperature. Ferrimagnetism refers to the coexistence of ferromagnetic and antiferromagnetic order, implying the presence of two distinct order parameters. In cases of weak ferrimagnetism, such as those considered above, one order parameter dominates and is referred to as the \emph{primary} order parameter. The subdominant component, which signals the presence and strength of ferrimagnetic behavior, is referred to as the \emph{secondary} order parameter. We define this secondary order parameter as the uniform magnetization in phases with dominant antiferromagnetic order, and as the staggered magnetization in dominantly ferromagnetic phases,
\begin{align} \label{eq:unsat-magn}
\mathbf m_\text{sec} = 
\begin{cases}
\frac12 (\mathbf{m}_A + \mathbf{m}_B), & \text{N\'eel order},\\
\frac14 (\mathbf{m}_A + \mathbf{m}_B + \mathbf{m}_C + \mathbf{m}_D), & \text{stripy and zigzag order},\\
\frac12 (\mathbf{m}_A - \mathbf{m}_B), & \text{ferromagnetic order}.
\end{cases}
\end{align}
The magnitude of the secondary order parameter $|\mathbf m_\text{sec}|$ is given for representative values of $\phi$ in Tables~\ref{tab:two-site-results} and \ref{tab:four-site-results}.
Figure~\ref{fig:corr-vs-phi-HK} shows $|\mathbf m_\text{sec}|$ as a function of the nearest-neighbor Heisenberg-Kitaev angle $\phi$, with the next-nearest-neighbor couplings held fixed at $(J_{2A}, J_{2B}) = (0.2, -0.1)$.
Consistent with the general argument above, $|\mathbf{m}_\text{sec}|$ vanishes at the Heisenberg points $\phi = 0$ and $\phi = \pi$. Moreover, it vanishes throughout the entire zigzag and stripy phases, in agreement with the analysis presented in the previous subsection.
Finally, we note that the couplings $J_{2A}$ and $J_{2B}$ shift the phase boundaries of the nearest-neighbor model and appear to stabilize narrow regions of noncollinear order, indicated by the white areas in Fig.~\ref{fig:corr-vs-phi-HK}. A detailed characterization of these emergent phases is left for future work.

\section{Heisenberg-Kitaev-Gamma model}
\label{sec:ext-HK}
In this section, we extend the Heisenberg-Kitaev model by adding a nearest-neighbor $\Gamma$ term, which has been argued to be relevant in a number of Kitaev materials~\cite{rousochatzakis24}, including $\alpha$-RuCl$_3$~\cite{winter16, janssen17, janssen19, maksimov20, moeller25}, \BCAO~\cite{maksimov25}, and \NCTO~\cite{krueger23, francini24vestigial, francini24ferri}.
Our aim is to examine the impact of $J_{2A} \ne J_{2B}$ on the zigzag and triple-$\mathbf{q}$ states, and to determine whether these states can support ferrimagnetism, both at zero and finite temperatures. As before, we fix $(J_{2A}, J_{2B}) = (0.2, -0.1)$, though the results remain qualitatively unchanged for other values, provided the sublattice symmetry is explicitly broken.
%

\subsection{Model}
The Heisenberg-Kitaev model defined in Eq.~\eqref{eq:KH-term} can be augmented with further interactions that respect the  $C_3^*$ symmetry. Here, we consider
\begin{align}
\label{eq:gammaKH-term}
    \mathcal H_{\mathrm{HK}\Gamma} = &\sum_{\gamma=x,y,z} \sum_{\langle ij \rangle_{\gamma}} 
    \Bigl[ J \mathbf{S}_i \cdot \mathbf{S}_j + K S_{i}^{\gamma}S_{j}^{\gamma} 
    +\Gamma(S_i^\alpha S_j^\beta + S_i^\beta S_j^\alpha)\Bigr]\, 
    \notag \\
     &+J_{2A}\sum_{\llangle ij \rrangle_A}\mathbf{S}_i \cdot \mathbf{S}_j + J_{2B}\sum_{\llangle ij \rrangle_B}\mathbf{S}_i \cdot \mathbf{S}_j\,,
\end{align}
which includes a nearest-neighbor $\Gamma$ coupling defined such that the spin indices $(\alpha,\beta,\gamma)$ form a cyclic permutation of $(x,y,z)$~\cite{rau14a}. 
We use the parametrization
\begin{equation}
    (J,K,\Gamma) = A (\sin\theta \cos\phi, \sin\theta \sin\phi, \cos\theta),
    \label{eq:JKGamma-param}
\end{equation}
with the overall energy scale set to $A=1$, as before.
The Heisenberg-Kitaev model of Sec.~\ref{sec:model} is recovered by setting $\theta=\pi/2$.

The classical ground-state phase diagram of $\mathcal{H}_{\mathrm{HK}\Gamma}$ for $J_{2A}=J_{2B}=0$ has been discussed in Refs.~\cite{rau14a, rau14b, janssen17, chen23, stavropoulos24}. It exhibits coplanar 120° order and 
a range of noncollinear multi-$\mathbf q$ states,
in addition to the four collinear phases (ferromagnetic, Néel, zigzag, and stripy) that appear in the absence of a $\Gamma$ interaction. 
Our focus here is on the magnetic properties of the single-$\mathbf{q}$ zigzag phase and the noncollinear triple-$\mathbf{q}$ state. The latter can be stabilized by either a six-spin ring-exchange interaction or suitably chosen local field terms~\cite{krueger23, francini24vestigial, francini24ferri}.

An important aspect for the following discussion is that the off-diagonal $\Gamma$ interaction breaks the $C_2^\alpha$ spin-rotation symmetry present in the model when $\Gamma = 0$.

\subsection{Magnetization corrections}

\subsubsection{Zigzag state}
\label{subsub:HKGzz}
Within the zigzag phase, the inclusion of a nonzero $\Gamma$ lifts the accidental ground-state degeneracy that emerges in the classical limit $S\to \infty$, see Sec.~\ref{sec:model}. While the selected zigzag configurations are still collinear, they generally order along an axis that deviates from the cubic directions $\mathbf{e}_x$, $\mathbf{e}_y$, and $\mathbf{e}_z$~\cite{rau14a}. For example, for ferromagnetic $J$ and $K$, large enough $\Gamma$, and $J_{2A}=J_{2B}=0$, the zigzag state depicted in Fig.~\ref{fig:four-site-cell}(a) can be parametrized in the cubic coordinate system as~\cite{janssen17}
\begin{align}
        \mathbf{S}_A & = - \mathbf{S}_{B} = - \mathbf{S}_C = \mathbf{S}_D = S \frac{\mathbf e_x + \mathbf e_y + f(\Gamma/|K|) \mathbf e_z}{\sqrt{2+f(\Gamma/|K|)^2}}\,,
\end{align}
where
\begin{equation}
    f(\zeta)=\frac{2+\zeta-\sqrt{4+4\zeta+9\zeta^2}}{2\zeta}
\end{equation}
characterizes the tilting of the spins as a function of the ratio $\zeta=\Gamma/|K|$. We have confirmed that this parametrization remains valid over an extended range of $J_{2A}$ and $J_{2B}$ values, including the specific choice $(J_{2A}, J_{2B}) = (0.2, -0.1)$ used in the following analysis.

Using analogous symmetry arguments as those outlined in Sec.~\ref{subsubsec:zz-gs}, we can understand the behavior of the uniform magnetization at all temperatures in the zigzag phase.
Although the spins are tilted away from the cubic directions, the state is still collinear, so that the transformation $\mathcal{T}=[\Theta||\{E|\mathbf{t}_{\pm}\}]$ remains a symmetry of both the Heisenberg-Kitaev-Gamma Hamiltonian $\mathcal H_\mathrm{HK\Gamma}$ defined in Eq.~\eqref{eq:gammaKH-term} and the density matrix $\rho=\rme^{-\mathcal{H}_\mathrm{HK\Gamma}/T}/\operatorname{Tr}(\rme^{-\mathcal{H}_\mathrm{HK\Gamma}/T})$. The transformation in Eq.~\eqref{eq:four-site-symmetry-action} maps antiparallel spins onto each other, resulting in constrained corrections to sublattice magnetizations. Due to the zigzag structure, this symmetry prohibits a finite net magnetization at all temperatures $T\ge0$.

\begin{figure*}
    \centering
    \begin{overpic}[width=\textwidth]{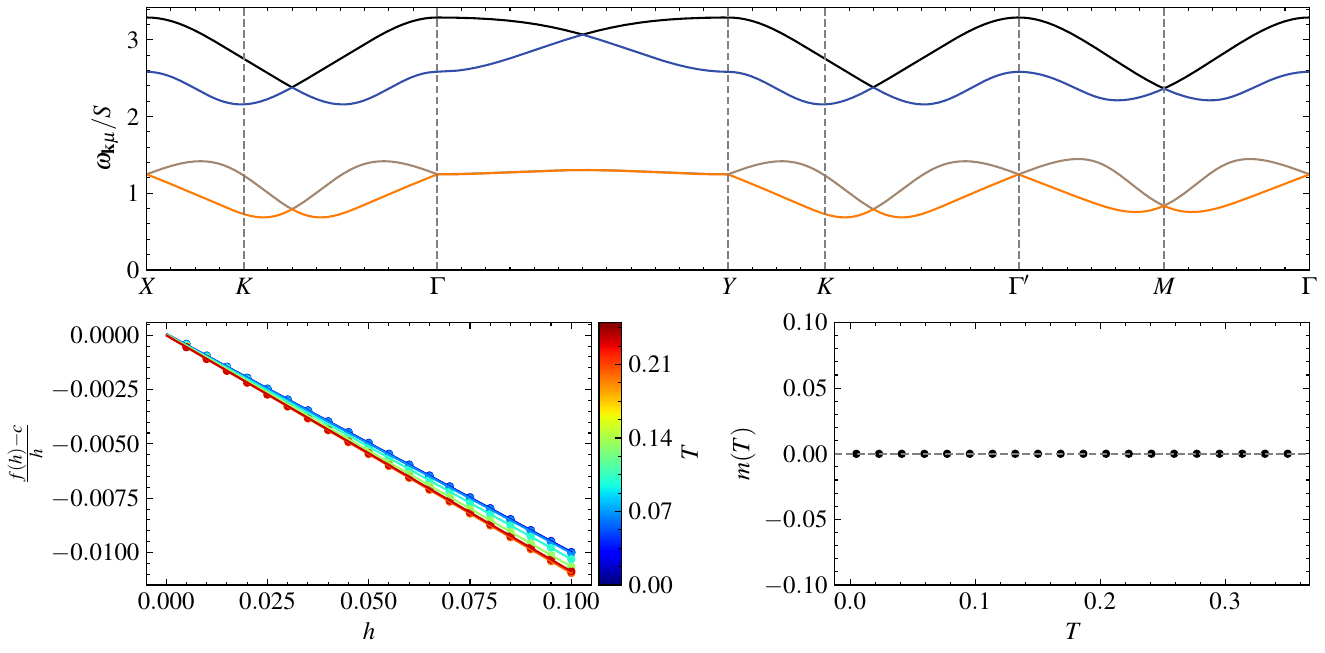}
    \put(0,49){(a)}
    \put(0,25){(b)}
    \put(55,25){(c)}
    \end{overpic}
    \caption{%
(a)~Magnon spectrum in the $z$-zigzag ground state of the Heisenberg-Kitaev-Gamma model with nearest-neighbor couplings $(J,K,\Gamma) = (\sin\theta \cos\phi, \sin\theta \sin\phi, \cos\theta)$ at $\phi = 5\pi/4$ and $\theta = \pi/16$ and next-nearest-neighbor couplings $(J_{2A},J_{2B})=(0.2,-0.1)$, from linear spin-wave theory.
The momentum path through the extended Brillouin zone is shown in the central inset of Fig.~\ref{fig:HK-ZZ}(b). 
(b)~Free-energy density $f(h, T)$ as a function of $h$ at various fixed temperatures, using the same model parameters as in~(a) and $S=1/2$, with an external field applied along the out-of-plane $[111]$ direction. The data are plotted as $[f(h) - c]/h$, where the temperature-dependent coefficient $c$ is extracted from a quadratic fit according to Eq.~\eqref{eq:quadratic-fit}. The absence of a finite vertical intercept for $h \to 0$ demonstrates that the out-of-plane magnetization vanishes in the zero-field limit.
(c)~Out-of-plane magnetization $m(T)$ as a function of temperature, using the same model parameters as in~(a), obtained by extrapolating $[f(h) - c]/h$ to the zero-field limit, confirming the absence of ferrimagnetism at all temperatures within the zigzag phase.
   }
    \label{fig:GHK-ZZ}
\end{figure*}

To illustrate that ferrimagnetism does not arise in the zigzag phase in the Heisenberg-Kitaev-Gamma model with explicit sublattice symmetry breaking introduced by $J_{2A} \ne J_{2B}$, we carried out a linear spin-wave calculation at a representative point in parameter space, $(\phi,\theta) = (5\pi/4, \pi/16)$.
The resulting spectrum, shown in Fig.~\ref{fig:GHK-ZZ}(a), exhibits a finite excitation gap, as the presence of $\Gamma \ne 0$ lifts the accidental ground-state degeneracy that is present in the pure Heisenberg-Kitaev model $\mathcal H_\mathrm{HK}$.
To obtain the magnetization correction arising from both thermal and quantum fluctuations, we add a small external magnetic field and compute the free-energy density $f(T,h) = \mathcal{F}(T,h)/N$ for different temperatures $T$ and fields $h$ using Eq.~\eqref{eq:free-energy}.
Away from phase transitions, the free-energy density is analytic and admits a Taylor expansion for small $h$ as
\begin{equation}
    \label{eq:quadratic-fit}
    f(T,h) = c(T) + b(T) h + a(T)h^2 + \mathcal{O}(h^3)\,,
\end{equation}
with temperature-dependent coefficients $c(T)$, $b(T)$, and $a(T)$.
For a given temperature $T$, the system is ferrimagnetic if $f(T,h)$ features a finite term linear in $h$,
\begin{align}
m(T)=-\frac{\partial f(T,h)}{\partial h} \biggr|_{h\rightarrow0}=-b(T).
\end{align}
Figure~\ref{fig:GHK-ZZ}(b) shows $[f(T,h) - c(T)]/h = b(T) + a(T) h + \mathcal{O}(h^2)$ for $S=1/2$ as a function of $h$ for different fixed temperatures $T$, with the coefficient $c(T)$ obtained in each case from a quadratic fit according to Eq.~\eqref{eq:quadratic-fit}. Here, the field is applied along the out-of-plane $[111]$ direction; however, analogous results are expected for other field orientations as well.
In this plot, the presence of ferrimagnetism would be indicated by a finite intercept of the curve $[f(h) - c]/h$ on the vertical axis. However, the data consistently show a zero intercept at all temperatures, indicating that the uniform magnetization of the zigzag state vanishes identically, in agreement with the general symmetry analysis.
This result is summarized in Fig.~\ref{fig:GHK-ZZ}(c), which shows $m(T)$ for various temperatures $T$, where each data point is obtained from extrapolating $[f(h) - c]/h$ at fixed temperature $T$ to the limit $h \to 0$.

\subsubsection{Triple-q state}
\label{subsec:HKG-triple-q}
\begin{figure}
    \centering
    \includegraphics[width=0.6\linewidth]{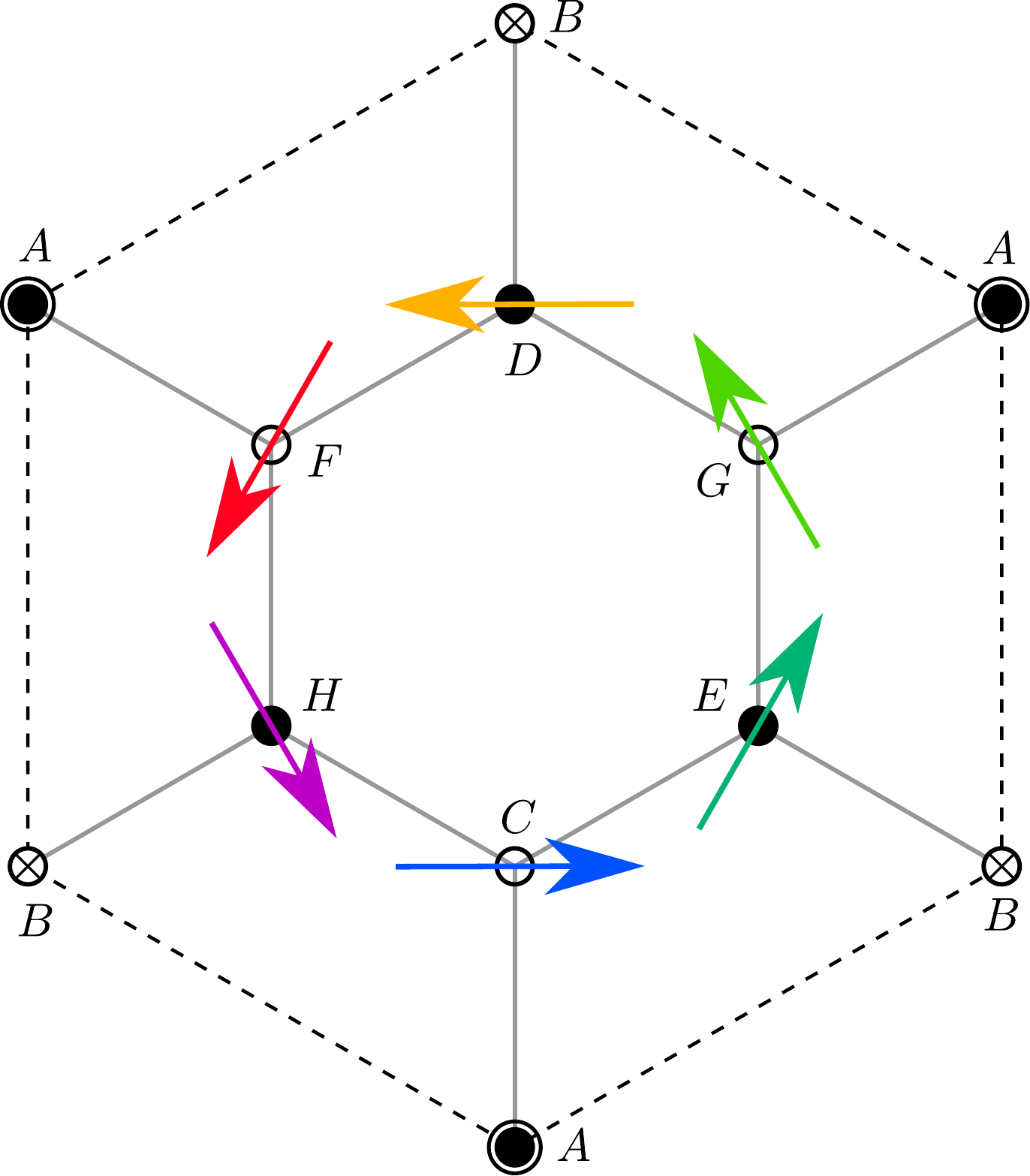}
    \caption{Eight-site magnetic unit cell in the triple-$\mathbf{q}$ state. $A$, $B$, $C$, $D$, $E$, $F$, $G$, and $H$ denote the eight magnetic sublattices, with the crystallographic sublattices indicated by filled and open circles. The spins on the $A$ and $B$ magnetic sublattices point along the out-of-plane $[111]$ and $[\bar{1} \bar{1} \bar{1}]$ directions, respectively, as indicated by the circled and crossed sites. The spins on the remaining six magnetic sublattices, $C$, $D$, $E$, $F$, $G$, and $H$, have finite in-plane components forming a vortex-like pattern around the hexagonal plaquette, while their out-of-plane components alternate in sign.}
    \label{fig:eight-site-cell}
\end{figure}

The triple-$\mathbf{q}$ order proposed as the ground state of {\NCTO} is a noncoplanar magnetic configuration with an eight-site unit cell~\cite{krueger23}, as illustrated in Fig.~\ref{fig:eight-site-cell}.
In this configuration, the spins on magnetic sublattices $A$ and $B$ align along the $[111]$ and $[\bar{1}\bar{1}\bar{1}]$ directions of the cubic coordinate system, respectively, which are perpendicular to the honeycomb plane spanned by the cubic $[11\bar{2}]$ and $[\bar{1}10]$ directions~\cite{janssen19}. The spins on the remaining six magnetic sublattices, $C$, $D$, $E$, $F$, $G$, and $H$, exhibit finite in-plane components arranged in a vortex-like pattern around the hexagonal plaquette. Their out-of-plane components alternate in sign; see Appendix~\ref{appendix:3q-ansatz-details} for further details.
Although the triple-$\mathbf{q}$ order appears as a ground state of the bilinear Heisenberg-Kitaev-$\Gamma$ model $\mathcal{H}_\mathrm{HK\Gamma}$ [Eq.~\eqref{eq:gammaKH-term}] only at isolated points in parameter space, it can be stabilized by nonbilinear spin exchange interactions~\cite{krueger23}.
In particular, different forms of six-spin ring exchange, which constitute the leading correction to the nearest-neighbor Heisenberg interaction in the strong-coupling expansion of the Hubbard model on the honeycomb lattice~\cite{yang12}, have been proposed as a possible mechanism for stabilizing this order in \NCTO~\cite{krueger23, francini24vestigial, francini24ferri, wang23, gu25}.
At the level of linear spin-wave theory, the ring exchange interaction has been shown to effectively generate a local-field term alongside a renormalization of the bilinear couplings~\cite{krueger23}.
Here, the local field is aligned with the classical spin configuration.
In this work, we neglect both the renormalization of the bilinear couplings and the dependence of the local field directions on bilinear couplings. This simplification is not expected to qualitatively affect our results, as the emergence of ferrimagnetism is determined solely by the symmetries of the model and its ground state, rather than by the specific microscopic mechanism that stabilizes the ground-state order.

Concretely, we consider the Hamiltonian
\begin{equation}
    \label{eq:3q-model}
    \mathcal{H}_{3\mathbf{q}}=\mathcal{H}_{\mathrm{HK\Gamma}} + \mathcal{H}_{\mathrm{loc}}\,,
\end{equation}
where the local field term given by
\begin{equation}
    \label{eq:local-field}
    \mathcal{H}_{\mathrm{loc}}=-h_{\mathrm{loc}}\sum_{i\mu}\hat{\mathbf{n}}_\mu \cdot \mathbf{S}_{i\mu}\,
\end{equation}
with the field directions chosen as
\begin{align}
    \label{eq:local-field-directions}
        \hat{\mathbf{n}}_A & =-\hat{\mathbf{n}}_B= \frac{\mathbf e_x + \mathbf e_y + \mathbf e_z}{\sqrt{3}} \,, \\
        \hat{\mathbf{n}}_C & =-\hat{\mathbf{n}}_D= \frac{\mathbf e_x + \mathbf e_y - \mathbf e_z}{\sqrt{3}} \,, \\
        \hat{\mathbf{n}}_E & =-\hat{\mathbf{n}}_F= \frac{- \mathbf e_x + \mathbf e_y - \mathbf e_z}{\sqrt{3}} \,, \\
        \hat{\mathbf{n}}_G & =-\hat{\mathbf{n}}_H= \frac{ - \mathbf e_x + \mathbf e_y + \mathbf e_z}{\sqrt{3}} \,.
\end{align}
In addition to being invariant under 
\begin{align}
    \label{eq:c3-symmetry}
    C_3^* 
    &: & 
        \mathbf{S}_{i\mu} & \mapsto \mathsf{O}_{\mathbf{c}}(2\pi/3)\, \mathbf{S}_{j\nu},
        & 
        \unitvec{n}_{\mu\phantom{i}} & \mapsto \mathsf{O}_{\mathbf{c}}(2\pi/3)\, \unitvec{n}_\mu,
\end{align}
with $\mathbf{R}_{j\nu} = \mathsf{O}_{\mathbf{c}}(2\pi/3) \mathbf{R}_{i\mu}$,
which implements simultaneous $2\pi/3$-rotations around the $[111]$ out-of-plane direction in real and in spin space, $\mathcal{H}_\mathrm{loc}$ also exhibits three spin-space symmetries given by
\begin{equation}
    \label{eq:HK-c2-symmetry}
    \mathcal{T}^{\alpha}=[C_2^\alpha || \{E|\mathbf{t}^{(\alpha)}\}]
    =
    \begin{cases}
        [C_2^x || \{E|\mathbf{t}_{-}\}],  &\alpha=x\,, \\
        [C_2^y || \{E|\mathbf{t}_{+}\}], &\alpha=y\,, \\
        [C_2^z || \{E|\mathbf{t}_{+} - \mathbf{t}_{-}\}], &\alpha=z\,,
    \end{cases}
\end{equation}
consisting of a lattice translation by a vector $\mathbf{t}^{(\alpha)} \in \{\mathbf t_\mp,\mathbf t_+ - \mathbf t_-\}$ combined with a $\pi$-rotation about the cubic axis $\mathbf{e}_\alpha \in \{\mathbf e_x, \mathbf e_y, \mathbf e_z\}$ in spin space.
In the absence of the $\Gamma$ term, Eqs.~\eqref{eq:c3-symmetry} and \eqref{eq:HK-c2-symmetry} are symmetries not only of $\mathcal{H}_\mathrm{loc}$, but also of the full Hamiltonian $\mathcal{H}_{3\mathbf{q}}$. As in previous sections, this fact alone imposes severe constraints on the different sublattice magnetizations. For example, given that $\mathcal{T}^x$ permutes the sublattices according to $(A,B,C,D) \leftrightarrow (H,G,F,E)$,
the symmetry derived from the composite transformation $C_3^* \mathcal{T}^x$ requires that the sublattice magnetizations satisfy $\mathbf m_A + \mathbf m_D + \mathbf m_E + \mathbf m_H = 0$ and $\mathbf m_B + \mathbf m_C + \mathbf m_F + \mathbf m_G = 0$.
As a result, the total magnetization $\mathbf{m} = \sum_{\mu} \mathbf m_\mu$
of a triple-$\mathbf{q}$ state must vanish by symmetry when $\Gamma=0$. 
However, the situation changes when $\Gamma \ne 0$, as the transformations in Eq.~\eqref{eq:HK-c2-symmetry} no longer remain symmetries of $\mathcal{H}_{3\mathbf{q}}$. As a result, the constraints on the sublattice magnetizations are lifted, enabling the system to develop ferrimagnetism. The $C_3^*$ symmetry in Eq.~\eqref{eq:c3-symmetry} only constrains the in-plane components of $\mathbf{m}_E + \mathbf{m}_D + \mathbf{m}_H$ and $\mathbf{m}_C + \mathbf{m}_G + \mathbf{m}_F$ to vanish, leaving the out-of-plane components and the magnetizations on sublattices $A$ and $B$ unconstrained.

To substantiate our symmetry-based argument with an explicit spin-wave calculation, we first identified a parameter regime in which a classical triple-$\mathbf{q}$ configuration serves as a suitable reference state for a $1/S$ expansion. For this condition to hold, the triple-$\mathbf{q}$ state must minimize $\mathcal{H}_{3\mathbf{q}}$ within the space of all classical spin configurations whose magnetic unit cells share the size and shape depicted in Fig.~\ref{fig:eight-site-cell}. Since each spin configuration is parametrized by two angles, the minimization involves a function of 16 variables. We addressed this problem using a two-step numerical procedure.
First, for a given set of parameters $(\phi,\theta,J_{2A},J_{2B},h_\mathrm{loc},h)$, where $h$ denotes the magnitude of the external field $\mathbf{h}$, we imposed a triple-$\mathbf{q}$ ansatz on $\mathcal{H}_{3\mathbf{q}}$ (see Appendix~\ref{appendix:3q-ansatz-details} for details) and determined the optimal configuration within this restricted two-dimensional subspace.
Second, we computed the Hessian of $\mathcal{H}_{3\mathbf{q}}$ to verify whether the optimal triple-$\mathbf{q}$ solution also corresponds to a local minimum in the full 16-dimensional configuration space.
An example of a parameter set that yields a stable triple-$\mathbf{q}$ ground state, as defined above, is 
\begin{gather}
(\phi, \theta) = (\pi - \arctan{2}, \pi/3), \label{eq:parameters-3q-A}\\
(J_{2A}, J_{2B}, h_{\mathrm{loc}}/S) = A ( 
0.2, -0.1, 0.8), \label{eq:parameters-3q-B}
\end{gather}
with small $h \to 0$ and the overall energy scale set to $A=1$, as before. We use this parameter set for the explicit calculations presented in the remainder of this section.
Previous work on the classical model~\cite{francini24ferri} demonstrated that finite-temperature fluctuations generically induce a uniform magnetization along the $[111]$ direction in the classical triple-$\mathbf{q}$ state. In the low-temperature limit, however, this classical uniform magnetization vanishes unless sublattice-dependent $g$-factors are introduced.
In the following, we investigate the interplay between quantum and thermal fluctuations, focusing in particular on whether quantum fluctuations can induce a uniform magnetization in the triple-$\mathbf{q}$ state already at zero temperature.
For the explicit calculations, we again apply the external field $\mathbf{h}$ along the [111] direction, enabling direct comparison with both the classical analysis in Ref.~\cite{francini24ferri} and experimental results on \NCTO~\cite{yao20}.

\begin{figure*}
    \centering
    \begin{overpic}[width=\textwidth]{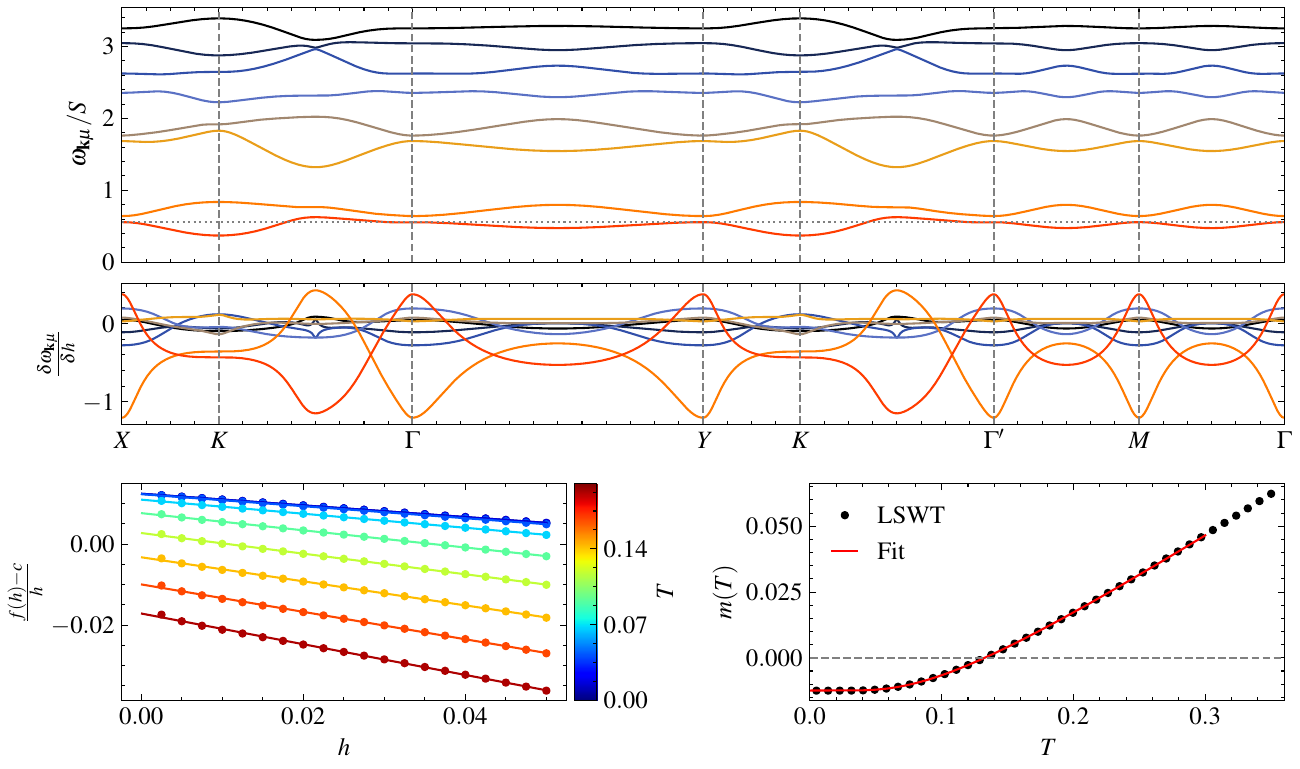}
        \put(0,59){(a)}
        \put(0,37){(b)}
        \put(0,22){(c)}
        \put(55,22){(d)}
    \end{overpic}
    \caption{%
(a)~Magnon spectrum in the triple-$\mathbf q$ ground state of the Heisenberg-Kitaev-Gamma model with nearest-neighbor couplings $(J,K,\Gamma) = (\sin\theta \cos\phi, \sin\theta \sin\phi, \cos\theta)$ at $\phi = \pi-\arctan2$ and $\theta= \pi/3$, next-nearest-neighbor couplings $(J_{2A},J_{2B})=(0.2,-0.1)$, and local field $h_\mathrm{loc}/S = 0.8$, from linear spin-wave theory.
The momentum path through the extended Brillouin zone is shown in the central inset of Fig.~\ref{fig:HK-ZZ}(b). 
The dashed horizontal line indicates the activation energy $\Delta/S = 0.566(1)$ for $S=1/2$, which is obtained by fitting Eq.~\eqref{eq:temp-fit} to the out-of-plane magnetization $m(T)$ from Eq.~\eqref{eq:explicit-magn-correction}. Note that $\Delta/S$ is slightly larger than the true magnon excitation gap $\Omega/S = 0.3714$, as a result of the reduced weight $\partial \omega_{\mathbf k\mu}/\partial h$ at the momentum $\mathbf k = K$, for which the excitation energy becomes minimal.
(b)~Weights $\partial \omega_{\mathbf k\mu}/\partial h \simeq \delta \omega_{\mathbf k\mu}/\delta h$, obtained from finite-difference derivative, taking an increment of $\delta h = 2.5\times 10^{-3}$, using the same model parameters as in~(a), with the external field applied along the out-of-plane $[111]$ direction. The largest weights occur at the $X$, $\Gamma$, $\Gamma'$, $Y$, and $M$ points, but not at the $K$ point, at which the excitation energy becomes minimal.
(c)~Free-energy density $f(h, T)$ as a function of $h$ at various fixed temperatures, using the same model parameters as in~(a) and $S=1/2$. The data are plotted as $[f(h) - c]/h$, where the temperature-dependent coefficient $c$ is extracted from a quadratic fit according to Eq.~\eqref{eq:quadratic-fit}. The presence of a finite vertical intercept for $h\to 0$ demonstrates the existence of a finite out-of-plane magnetization in the zero-field limit.
(d)~Out-of-plane magnetization $m(T)$ as a function of temperature, using the same model parameters as in~(a), obtained by extrapolating $[f(h) - c]/h$ to the zero-field limit, confirming the ferrimagnetic behavior of the triple-$\mathbf q$ state and the emergence of a compensation point at a finite temperature $T_\star \simeq 0.133$.
}
\label{fig:GHK-3q}
\end{figure*}
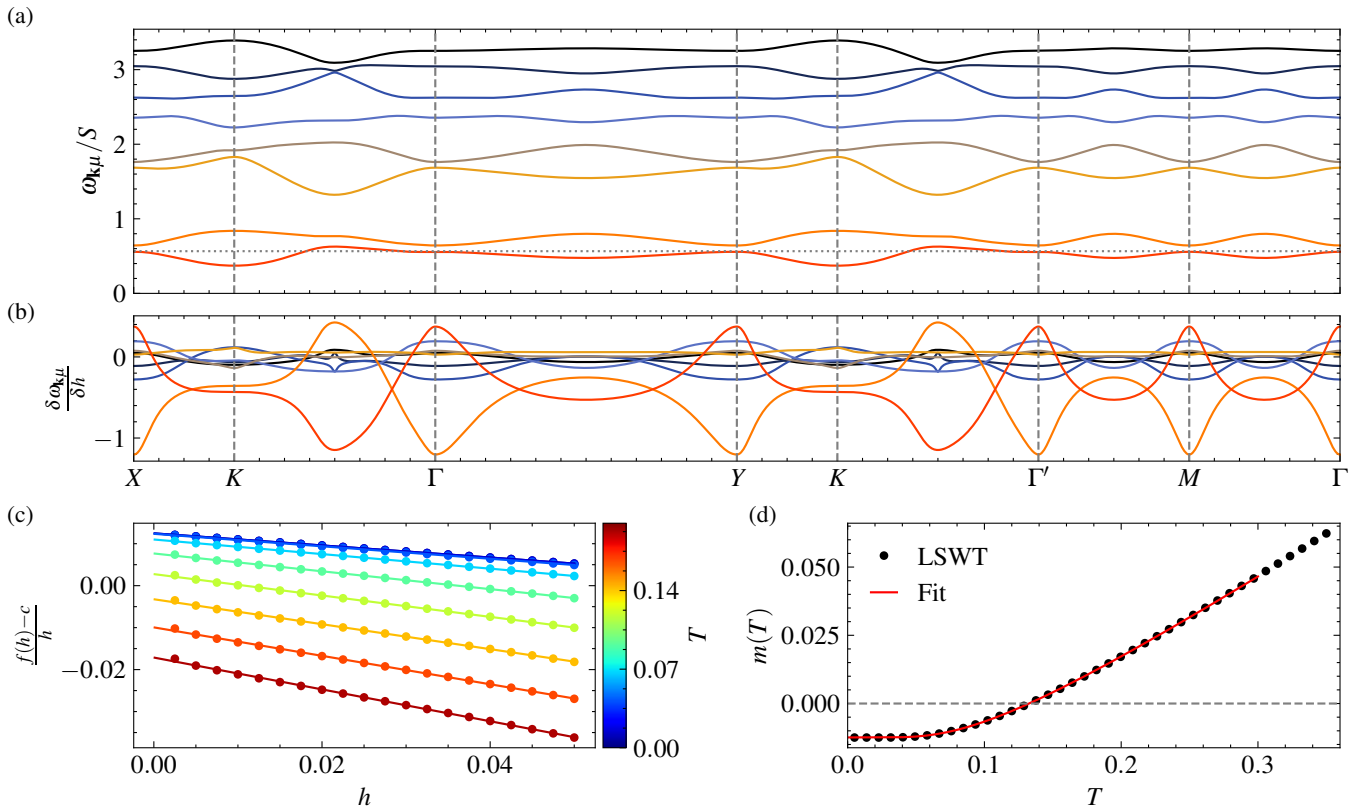

Figure~\ref{fig:GHK-3q}(a) displays the magnon excitation spectrum obtained from linear spin-wave theory using the parameter set specified in Eqs.~\eqref{eq:parameters-3q-A}--\eqref{eq:parameters-3q-B}. As a result of the finite $\Gamma$ interaction and the presence of the local field, the spectrum is gapped, as before. To obtain the magnetization correction, we again compute the free-energy density $f(T,h)$ at various temperatures $T$ and external fields $h$, and extrapolate to the low-field limit.
Figure~\ref{fig:GHK-3q}(c) shows $[f(T,h) - c(T)]/h$ for $S=1/2$ as a function of $h$ for several fixed temperatures $T$, with the coefficient $c(T)$ extracted in each case via a quadratic fit according to Eq.~\eqref{eq:quadratic-fit}. In contrast to the zigzag case, the curves $[f(h) - c]/h$ now exhibit a clear finite intercept as $h \to 0$, with the intercept value varying as a function of temperature $T$. This result is summarized in Fig.~\ref{fig:GHK-3q}(d), which displays the out-of-plane component of the residual magnetization $m(T)$ in the limit $h \to 0$ as a function of temperature $T$. Note that $m(T)$ changes sign at a characteristic temperature $T_\star \simeq 0.133$, indicating the presence of a compensation point, as observed in {\NCTO}~\cite{yao20}. In our calculation, this behavior occurs because the magnetic sublattices with larger zero-temperature magnetizations turn out to be more susceptible to thermal fluctuations than those with smaller ones.
Importantly, this result does not depend on any sublattice-dependent $g$-factors, thereby avoiding the fine-tuning issue present in the classical model~\cite{francini24ferri}.
Since the magnetization corrections are small, we expect $T_\star$ to lie within the temperature range where linear spin-wave theory remains valid.

Furthermore, the magnetization curve $m(T)$ displays activated behavior at low temperatures, which may be associated with the presence of a finite gap in the magnon excitation spectrum, in full qualitative agreement with experimental observations in \NCTO~\cite{yao20}.
The red curve in Fig.~\ref{fig:GHK-3q}(d) indicates a fit of the low-temperature portion of $m(T)$ using the simplified function
\begin{equation}
    \label{eq:temp-fit}
    m(T) = A + \frac{B}{\rme^{\Delta/T}-1}\,,
\end{equation}
with fitting parameters $A$, $B$, and $\Delta$.
Excellent agreement, as shown in Fig.~\ref{fig:GHK-3q}(d), is achieved using the best-fit values $A =-0.01242(1)$, $B =0.0924(3)$, and $\Delta = 0.2829(6)$ for $S=1/2$.
We have explicitly verified that the value of $A$ obtained from the fitting procedure agrees, within numerical uncertainty, with the zero-temperature magnetization $m(0)$ calculated directly from Eq.~\eqref{eq:magn-external-field}, lending further credibility to the fitting approach. However, caution is required when interpreting the activation energy $\Delta$. In particular, we emphasize that this energy scale should not be directly identified with the gap $\Omega = \min(\omega_{\mathbf{k}\mu}) = 0.1857$ in the magnon excitation spectrum. This is reflected in the difference between the gap value $\Omega$ and the optimal fit parameter $\Delta = 0.2829(6)$. The latter is indicated by the horizontal dotted line in Fig.~\ref{fig:GHK-3q}(a).
To understand the value of $\Delta$, it is helpful to examine Eq.~\eqref{eq:explicit-magn-correction}. Given that the summand in the temperature-dependent part of $m(T)$ is weighted by the partial derivatives $\partial \omega_{\mathbf{k}\mu} / \partial h$, $\Delta$ is shifted from $\Omega$ toward the eigenenergies for which the weights are largest. Figure~\ref{fig:GHK-3q}(b) shows the weights $\partial \omega_{\mathbf{k}\mu} / \partial h$, obtained using a finite-difference approximation. Among the high-symmetry points, the two low-energy bands exhibit the largest weights at the $X$, $\Gamma$, $\Gamma’$, $Y$, and $M$ points, as well as at the midpoint between $\Gamma$ and $K$. Notably, the corresponding excitation energies are in reasonable agreement with the fitted value of $\Delta$. In contrast, the minimal gap $\Omega$ occurs at the $K$ point, which carries significantly less weight and thus contributes less to the temperature-induced increase in magnetization and to the activation energy $\Delta$.
%
%

\section{Conclusion}
\label{sec:conclusion}
In this paper, we highlighted the crucial role of symmetries in linear spin-wave calculations. We focused on Heisenberg-Kitaev-Gamma systems with explicit sublattice symmetry breaking, a scenario relevant for \NCTO\ and related cobaltate honeycomb magnets. The nature of the ground state in these materials remains unsettled, with both zigzag and multi-$\mathbf{q}$ states under consideration.

Linear spin-wave theory can be used to quantify the effects of quantum fluctuations around classical long-range ordered states. We studied various models that stabilize either the collinear single-$\mathbf{q}$ zigzag or the triple-$\mathbf{q}$ noncollinear states. The symmetries strongly constrain how quantum fluctuations affect the ordered states, leading to markedly different outcomes for the two ground states.

First, we demonstrated that a collinear zigzag phase cannot exhibit ferrimagnetism, neither at zero nor at finite temperature, as long as the combined symmetry of crystallographic translations with time-reversal symmetry is preserved.
Importantly, both the Heisenberg-Kitaev-Gamma Hamiltonian and the collinear zigzag state respect this symmetry, regardless of the specific microscopic parameters. Therefore, ferrimagnetism is forbidden in Heisenberg-Kitaev-Gamma models with a zigzag ground state, provided the combined translation and time-reversal symmetry is preserved. We are not aware of an extended Heisenberg-Kitaev-Gamma model on the honeycomb lattice realizing a zigzag-type ground state that spontaneously breaks this symmetry.

In contrast, the noncollinear triple-$\mathbf{q}$ ground state emerging in the presence of a ring exchange interaction naturally exhibits ferrimagnetic behavior. The $C_3^*$ symmetry of the Hamiltonian and ground state imposes constraints on the spins but permits a nonzero magnetization once the $\mathcal{T}^\alpha$ symmetry of the Heisenberg-Kitaev model is broken by the $\Gamma$ term. In our calculations, we employed a local-field term to stabilize the triple-$\mathbf{q}$ state; however, on symmetry grounds, similar results are expected when using the full model with ring exchange interactions. This is because the local-field model and the full model share the same symmetries.
The key ingredients for ferrimagnetic behavior are a triple-$\mathbf{q}$ ground state, a finite off-diagonal interaction such as the $\Gamma$ interaction, and a sublattice imbalance promoted, e.g., by different next-nearest-neighbor couplings $J_{2A}$ and $J_{2B}$.
These conditions are realistic in systems like \NCTO\ and do not require fine-tuning~\cite{yao20,krueger23}.
These general results are expected to hold at all orders of the spin-wave expansion since they are rooted in symmetry. Consequently, although higher-order terms in the $1/S$ expansion may modify quantitative details beyond our linear spin-wave theory results, the qualitative behavior should remain unchanged.

The finite-temperature magnetization curve $m(T)$ of the triple-$\mathbf{q}$ state as a function of temperature $T$ exhibits notable features. 
First, $m(T)$ shows an activated behavior at low temperatures, which may be associated with the presence of a finite gap in the magnon excitation spectrum. We have shown, however, that the activation energy $\Delta$ extracted from fitting $m(T)$ to an activated form does not necessarily correspond to the actual magnon gap $\Omega$. This is because the contribution of a given excitation to the magnetization correction in Eq.~\eqref{eq:explicit-magn-correction} is weighted by $\partial \omega_{\mathbf{k} \mu} / \partial h$, which can cause the fitted value of $\Delta$ to deviate from the actual gap $\Omega$. This result is independent of the specific model or ground state and may have broader significance for interpreting experimental magnetization curves as a function of temperature.
Second, specific to our model with a triple-$\mathbf{q}$ ground state, for temperatures $T$ above the activation energy $\Delta$, the initially negative residual out-of-plane magnetization increases and crosses zero at a compensation point $T_\star$, where it changes sign. Overall, the calculated magnetization curve qualitatively reproduces the experimental observations in \NCTO\ at low temperatures~\cite{yao20}. Deviations appear only at elevated temperatures, where the spin-wave approximation ceases to be valid.
It follows that the triple-$\mathbf{q}$ state not only supports zero-temperature ferrimagnetism, but also permits a finite-temperature compensation point---without relying on sublattice-dependent $g$-factors---thus avoiding the fine-tuning required in the classical model~\cite{francini24ferri}.

Since our general results are rooted purely in symmetry considerations, we expect them to remain qualitatively valid across the entire family of extended Heisenberg-Kitaev-Gamma models that share the same symmetries. In particular, a triple-$\mathbf{q}$ ground state is expected to exhibit ferrimagnetism in any extended Heisenberg-Kitaev-Gamma model with off-diagonal interactions. This also holds in the presence of other bond-dependent interactions, such as Dzyaloshinskii-Moriya terms, provided the ground state maintains noncollinear triple-$\mathbf{q}$ order and the crystallographic sublattice symmetry is explicitly broken.
For the same reason, zigzag order cannot exhibit ferrimagnetism in more complex models as long as the combined time-reversal and translational symmetry is preserved.

\begin{acknowledgments}
We thank M.\ Fornoville, Y.\ Li, and M.\ Vojta for insightful comments and collaboration on related projects.
This work has been supported by the Deutsche Forschungsgemeinschaft through 
Project No.\ 247310070 (SFB 1143, A02 {\&} A07), 
Project No.\ 390858490 (W\"urzburg-Dresden Cluster of Excellence \textit{ct.qmat}, EXC 2147), and 
Project No.\ 411750675 (Emmy Noether program, JA2306/4-1).

\end{acknowledgments}

\section*{Data availability}
%
The data that support the findings of this article are openly
available~\cite{data-availability}.
%

\appendix

\section{Details of spin-wave calculation}
\label{appendix:LSW-details}
In this appendix, we provide details about the linear spin-wave calculations, following Ref.~\cite{colpa78}. 

The Holstein-Primakoff bosons in Eq.~\eqref{eq:HP-bosons} are expressed in momentum space by means of a Fourier transformation,
\begin{equation}
    \label{eq:HP-fourier}
    \begin{split}
        a_{i\mu} &=\frac{1}{\sqrt{N/p}} \sum_{\mathbf{k}} \rme^{\rmi\mathbf{k}\cdot \mathbf{R}_{i\mu}} a_{\mu}(\mathbf{k}), \\
    \end{split}
\end{equation}
where $\mathbf{R}_{i\mu}=\mathbf{R}_i + \mathbf{d}_\mu$ and $\mathbf{R}_i$ being the position vector of the magnetic unit cell $i$ and $\mathbf{d}_{\mu}$ the position vector of the magnetic sublattice $\mu$ inside the magnetic unit cell. Considering only bilinear boson-boson terms, the Hamiltonian acquires the expression in Eq.~\eqref{eq:momentum-hamiltonian} with the $2p$-component vector satisfying the commutation relations
\begin{equation}
    \label{eq:commutation-relations}
    [\mathbf{x}(\mathbf{k}), \mathbf{x}^{\dagger}(\mathbf{k})]=\Sigma= \begin{pmatrix} \mathsf{I}_{p} & 0 \\
    0 &  -\mathsf{I}_{p}
    \end{pmatrix},
\end{equation}
with $\mathsf{I}_{p}$ being a $p\times p$ identity matrix.
The Bogoliubov transformation $\mathbf{x}(\mathbf{k})=\mathsf{T}(\mathbf{k})\boldsymbol{\psi}(\mathbf{k})$ introduces new bosonic modes $\boldsymbol{\psi}$ diagonalizing the quadratic Hamiltonian and fulfilling the commutation relations $[\boldsymbol{\psi}(\mathbf{k}),\boldsymbol{\psi}^\dagger(\mathbf{k})]=\Sigma$. Thus, the problem reduces to finding the matrix $\mathsf{T}$. First, the Cholesky decomposition is applied to the Hermitian matrix $\mathsf{M}(\mathbf{k})$ of Eq.~\eqref{eq:momentum-hamiltonian} to find the complex matrix $\mathsf{K}$ such that
\begin{equation}
    \label{eq:cholesky-decomposition}
    \mathsf{M}=\mathsf{K}^{\dagger}\mathsf{K}.
\end{equation}
After solving the eigenvalue problem of the Hermitian matrix $\mathsf{K}\Sigma\mathsf{K}^{\dagger}$, the eigenvectors are arranged into the matrix $\mathsf{U}$ diagonalizing the matrix $\mathsf{L}=\mathsf{U}^{\dagger}\mathsf{K}\Sigma\mathsf{K}^{\dagger}\mathsf{U}$. Finally, the spectrum is obtained as $\mathsf{E}=\Sigma\mathsf{L}$ and the Bogoliubov transformation matrix is given by $\mathsf{T}=\mathsf{K}^{-1}\mathsf{U}\mathsf{E}^{1/2}$. The matrix $\mathsf{T}$ of the Bogoliubov transformation has the structure
\begin{equation}
    \label{eq:bogoliubov-matrix}
    \mathsf{T}(\mathbf{k})=\begin{pmatrix}
        \mathsf{U}(\mathbf{k}) & \mathsf{V}(\mathbf{k}) \\
        \mathsf{V}^*(-\mathbf{k}) & \mathsf{U}^*(-\mathbf{k}) 
    \end{pmatrix},
\end{equation}
with $\mathsf{U}$ and $\mathsf{V}$ being $p\times p$ matrices, and satisfies the orthogonality relation $\mathsf{T}\Sigma\mathsf{T}^\dagger=\mathsf{T}^\dagger \Sigma \mathsf{T}$.

The knowledge of the Bogoliubov transformation matrix $\mathsf{T}$ allows for the computation of different physical quantities such as magnetization corrections. If the classical ground state is collinear and the Hamiltonian does not mix different spin components, as in the case of Heisenberg or Kitaev interactions, then the sublattice magnetizations are given by
\begin{align}
        m_{\mu}^3 
        &= S-\frac{p}{N} \sum_{\mathbf{k}}\langle 0 | a_{\mu}^\dagger (\mathbf{k}) a_{\mu}(\mathbf{k}) | 0 \rangle  + \mathcal{O} (1/S)
	\nonumber \\ \label{eq:full-magn-corrections} 
        &= S-\frac{p}{N} \sum_{\mathbf{k}}\langle 0 | \boldsymbol{\psi}^\dagger(\mathbf{k}) \mathsf{T}(\mathbf{k})^\dagger \Lambda^{(\mu)} \mathsf{T}(\mathbf{k})\boldsymbol{\psi}(\mathbf{k}) | 0 \rangle + \mathcal{O} (1/S)
\end{align}
with $\Lambda_{\alpha\beta}^{(\mu)}=\delta_{\alpha\mu}\delta_{\beta\mu}$. Inserting Eq.~\eqref{eq:bogoliubov-matrix}, this reduces to
\begin{equation}
    \label{eq:magn-correction}
    m^3_\mu = S-\frac{p}{N}\sum_{\mathbf{k}}\sum_{\alpha=1}^p |\mathsf{V}_{\mu,\alpha}(\mathbf{k})|^2 + \mathcal O(1/S)\,,
\end{equation}
However, if the classical ground state is not collinear and/or the Hamiltonian mixes different spin components, such as the $\Gamma$ term in Eq.~\eqref{eq:gammaKH-term}, then the calculation of the sublattice magnetizations also need to take into account corrections to the directions of the ordered moments. In these cases, we introduce an external magnetic field and calculate the total magnetization according to Eq.~\eqref{eq:magn-external-field}, with a ground-state energy explicitly given by
\begin{align}
    E_{\mathrm{gs}} & = S(S+1)\sum_{i\mu,j\nu}\mathbf{e}_{i\mu}^{(3)} \mathsf{J}_{\mu\nu}(\mathbf{R}_i - \mathbf{R}_j) \mathbf{e}_{j\nu}^{(3)}
    \nonumber \\ & \quad
    - ( S+1/2) \sum_{i\mu} \mathbf{h}_{i\mu}\cdot \mathbf{e}_{i\mu}^{(3)} 
    + \frac{1}{2}\sum_{\mathbf{k}\mu}\omega_{\mathbf{k}\mu}\, + \mathcal{O}(1/S^0).
\label{eq:LSW-ground-state}
\end{align}
where $\mathbf{e}_{i\mu}^{(3)}$ is the third axis in the local reference frame.  The field $\mathbf{h}_{i\mu}$ encodes both the presence of an external field and a site-dependent term. Despite losing the information of individual sublattice magnetizations, Eqs.~\eqref{eq:magn-external-field} and \eqref{eq:LSW-ground-state} yield a result that is fully consistent up to order $1/S^{0}$. Since Eq.~\eqref{eq:LSW-ground-state} accounts for external field terms, it is more general than Eq.~\eqref{eq:bilinear-hamiltonian} and is the expression we effectively used to compute the magnetization corrections in Eqs.~\eqref{eq:magn-external-field} and \eqref{eq:explicit-magn-correction}.
%
%

\section{Triple-q ansatz}
\label{appendix:3q-ansatz-details}

In this appendix, we provide details on the triple-$\mathbf{q}$ ansatz used in Sec.~\ref{subsec:HKG-triple-q}. There are several ways of describing the spins on the lattice~\cite{janssen16}. In the following, we express each spin in a basis defined by the crystallographic axes $\{\mathbf{a},\mathbf{b},\mathbf{c}\}$ as
\begin{equation}
    \label{eq:spin-parametrization}
    \mathbf{S}_i = S(\sin{\beta_i}\cos{\alpha_i}\mathbf{a}+\sin{\beta_i}\sin{\alpha_i}\mathbf{b}+\cos{\beta_i}\mathbf{c})\,,
\end{equation}
where $S$ is the spin magnitude, and the crystallographic axes are given by
\begin{equation}
    \label{eq:crystal-axis}
    \mathbf{a}=\frac{\mathbf{e}_x+\mathbf{e}_y-2\mathbf{e}_z}{\sqrt{6}},\quad \mathbf{b}=\frac{-\mathbf{e}_x+\mathbf{e}_y}{\sqrt{2}},\quad \mathbf{c}=\frac{\mathbf{e_x+\mathbf{e}_y}+\mathbf{e}_z}{\sqrt{3}}.
\end{equation}
For simplicity, consider an eight-site magnetic unit cell as in Fig.~\ref{fig:eight-site-cell}. Since the triple-$\mathbf{q}$ state is characterized by three-fold rotation symmetry around the $\mathbf{c}$ axis, namely the $C_3^*$ symmetry, the set of angles $\{\alpha_i,\beta_i\}$ is constrained. Under the symmetry, the in-plane components of the spins rotate, while the out-of-plane components remain invariant. As a consequence, the $C_3^*$-invariant spins of sublattices $A$ and $B$ in Fig.~\ref{fig:eight-site-cell} are fixed to $\mathbf{S}_A=S\mathbf{c}$ and $\mathbf{S}_B=-S\mathbf{c}$.
Following similar symmetry arguments, one can show that the initial set of 16 angles $\{\alpha_i,\beta_i\}$ is reduced to a set of only two angles $\beta_<$ and $\beta_>$ for the two different crystallographic sublattices of the vortex structure, corresponding to the ($C$, $F$, $G$) and ($D$, $E$, $H$) sites, respectively, in Fig.~\ref{fig:eight-site-cell}. These angles describe the out-of-plane tilting of these two sets of spins, which generate the in-plane vortex structure. Moreover, the angles are responsible for a nonzero out-of-plane magnetization if $\beta_>\neq\pi-\beta_<$, already at the classical level. Table~\ref{tab:triple-q-ansatz} reports the angles for the spins in the eight-site magnetic unit cell.

\begin{table}[!b]
\caption{Angle configuration for the triple-$\mathbf{q}$ state shown in Fig.~\ref{fig:eight-site-cell}. The $C_3^*$ symmetry fixes all the $\alpha_i$ angles, and relates different $\beta_i$, reducing the degrees of freedom necessary to describe this state from $16$ to 2, namely $\beta_>$ and $\beta_<$.}
    \centering
    \begin{tabularx}{\linewidth}{>{\centering\arraybackslash}X >{\centering\arraybackslash}X >{\centering\arraybackslash}X}
    \hline
    \hline
        Sublattice & $\alpha$ & $\beta$  \\
        \hline
        $A$ & n/a & 0 \\
        $B$ & n/a & $\pi$ \\
        $C$ & 0 & $\beta_<$ \\
        $D$ & $\pi$ & $\beta_>$\\
        $E$ & $\pi/3$ & $\beta_>$\\
        $F$ & $4\pi/3$ & $\beta_<$ \\
        $G$ & $2\pi/3$ & $\beta_<$ \\
        $H$ &  $5\pi/3$ & $\beta_>$ \\
        \hline
        \hline
    \end{tabularx}
    \label{tab:triple-q-ansatz}
\end{table}
The triple-$\mathbf{q}$ ansatz is of great help in the energy minimization of the Hamiltonian in Eq.~\eqref{eq:3q-model}, reducing the minimum space to a two-dimensional manifold spanned by $\beta_{\gtrless}$. For the set of parameters used in Sec.~\ref{subsec:HKG-triple-q}, $(\phi,\theta,J_{2A},J_{2B},h_{\mathrm{loc}}/S)=(\pi-\arctan2, \pi/3,0.2,-0.1,0.8)$, energy minimization leads to $\beta_< \approx 0.489\pi$ and $ \beta_> \approx0.553\pi$, with a classical out-of-plane classical magnetization of $\mathbf{M}\cdot\mathbf{c}/S \approx-0.05$.
%
%

\bibliographystyle{longapsrev4-2}
\bibliography{LSW-HK-ferri}

\begin{thebibliography}{93}%
\makeatletter
\providecommand \@ifxundefined [1]{%
 \@ifx{#1\undefined}
}%
\providecommand \@ifnum [1]{%
 \ifnum #1\expandafter \@firstoftwo
 \else \expandafter \@secondoftwo
 \fi
}%
\providecommand \@ifx [1]{%
 \ifx #1\expandafter \@firstoftwo
 \else \expandafter \@secondoftwo
 \fi
}%
\providecommand \natexlab [1]{#1}%
\providecommand \enquote  [1]{``#1''}%
\providecommand \bibnamefont  [1]{#1}%
\providecommand \bibfnamefont [1]{#1}%
\providecommand \citenamefont [1]{#1}%
\providecommand \href@noop [0]{\@secondoftwo}%
\providecommand \href [0]{\begingroup \@sanitize@url \@href}%
\providecommand \@href[1]{\@@startlink{#1}\@@href}%
\providecommand \@@href[1]{\endgroup#1\@@endlink}%
\providecommand \@sanitize@url [0]{\catcode `\\12\catcode `\$12\catcode
  `\&12\catcode `\#12\catcode `\^12\catcode `\_12\catcode `\%12\relax}%
\providecommand \@@startlink[1]{}%
\providecommand \@@endlink[0]{}%
\providecommand \url  [0]{\begingroup\@sanitize@url \@url }%
\providecommand \@url [1]{\endgroup\@href {#1}{\urlprefix }}%
\providecommand \urlprefix  [0]{URL }%
\providecommand \Eprint [0]{\href }%
\providecommand \doibase [0]{https://doi.org/}%
\providecommand \selectlanguage [0]{\@gobble}%
\providecommand \bibinfo  [0]{\@secondoftwo}%
\providecommand \bibfield  [0]{\@secondoftwo}%
\providecommand \translation [1]{[#1]}%
\providecommand \BibitemOpen [0]{}%
\providecommand \bibitemStop [0]{}%
\providecommand \bibitemNoStop [0]{.\EOS\space}%
\providecommand \EOS [0]{\spacefactor3000\relax}%
\providecommand \BibitemShut  [1]{\csname bibitem#1\endcsname}%
\let\auto@bib@innerbib\@empty
\bibitem [{\citenamefont {Kitaev}(2006)}]{kitaev06}%
  \BibitemOpen
  \bibfield  {author} {\bibinfo {author} {\bibfnamefont {A.}~\bibnamefont
  {Kitaev}},\ }\bibfield  {title} {\bibinfo {title} {Anyons in an exactly
  solved model and beyond},\ }\href
  {https://doi.org/https://doi.org/10.1016/j.aop.2005.10.005} {\bibfield
  {journal} {\bibinfo  {journal} {Ann. Phys. (N. Y.)}\ }\textbf {\bibinfo
  {volume} {321}},\ \bibinfo {pages} {2} (\bibinfo {year} {2006})}\BibitemShut
  {NoStop}%
\bibitem [{\citenamefont {Savary}\ and\ \citenamefont
  {Balents}(2016)}]{savary16}%
  \BibitemOpen
  \bibfield  {author} {\bibinfo {author} {\bibfnamefont {L.}~\bibnamefont
  {Savary}}\ and\ \bibinfo {author} {\bibfnamefont {L.}~\bibnamefont
  {Balents}},\ }\bibfield  {title} {\bibinfo {title} {{Quantum spin liquids: a
  review}},\ }\href {https://doi.org/10.1088/0034-4885/80/1/016502} {\bibfield
  {journal} {\bibinfo  {journal} {Rep. Prog. Phys.}\ }\textbf {\bibinfo
  {volume} {80}},\ \bibinfo {pages} {016502} (\bibinfo {year}
  {2016})}\BibitemShut {NoStop}%
\bibitem [{\citenamefont {Zhou}\ \emph {et~al.}(2017)\citenamefont {Zhou},
  \citenamefont {Kanoda},\ and\ \citenamefont {Ng}}]{zhou17}%
  \BibitemOpen
  \bibfield  {author} {\bibinfo {author} {\bibfnamefont {Y.}~\bibnamefont
  {Zhou}}, \bibinfo {author} {\bibfnamefont {K.}~\bibnamefont {Kanoda}},\ and\
  \bibinfo {author} {\bibfnamefont {T.-K.}\ \bibnamefont {Ng}},\ }\bibfield
  {title} {\bibinfo {title} {{Quantum spin liquid states}},\ }\href
  {https://doi.org/10.1103/RevModPhys.89.025003} {\bibfield  {journal}
  {\bibinfo  {journal} {Rev. Mod. Phys.}\ }\textbf {\bibinfo {volume} {89}},\
  \bibinfo {pages} {025003} (\bibinfo {year} {2017})}\BibitemShut {NoStop}%
\bibitem [{\citenamefont {Knolle}\ and\ \citenamefont
  {Moessner}(2019)}]{knolle19}%
  \BibitemOpen
  \bibfield  {author} {\bibinfo {author} {\bibfnamefont {J.}~\bibnamefont
  {Knolle}}\ and\ \bibinfo {author} {\bibfnamefont {R.}~\bibnamefont
  {Moessner}},\ }\bibfield  {title} {\bibinfo {title} {{A Field Guide to Spin
  Liquids}},\ }\href {https://doi.org/10.1146/annurev-conmatphys-031218-013401}
  {\bibfield  {journal} {\bibinfo  {journal} {Annu. Rev. Condens. Matter
  Phys.}\ }\textbf {\bibinfo {volume} {10}},\ \bibinfo {pages} {451} (\bibinfo
  {year} {2019})}\BibitemShut {NoStop}%
\bibitem [{\citenamefont {Broholm}\ \emph {et~al.}(2020)\citenamefont
  {Broholm}, \citenamefont {Cava}, \citenamefont {Kivelson}, \citenamefont
  {Nocera}, \citenamefont {Norman},\ and\ \citenamefont {Senthil}}]{broholm20}%
  \BibitemOpen
  \bibfield  {author} {\bibinfo {author} {\bibfnamefont {C.}~\bibnamefont
  {Broholm}}, \bibinfo {author} {\bibfnamefont {R.~J.}\ \bibnamefont {Cava}},
  \bibinfo {author} {\bibfnamefont {S.~A.}\ \bibnamefont {Kivelson}}, \bibinfo
  {author} {\bibfnamefont {D.~G.}\ \bibnamefont {Nocera}}, \bibinfo {author}
  {\bibfnamefont {M.~R.}\ \bibnamefont {Norman}},\ and\ \bibinfo {author}
  {\bibfnamefont {T.}~\bibnamefont {Senthil}},\ }\bibfield  {title} {\bibinfo
  {title} {{Quantum spin liquids}},\ }\href
  {https://doi.org/10.1126/science.aay0668} {\bibfield  {journal} {\bibinfo
  {journal} {Science}\ }\textbf {\bibinfo {volume} {367}},\ \bibinfo {pages}
  {263} (\bibinfo {year} {2020})}\BibitemShut {NoStop}%
\bibitem [{\citenamefont {Jackeli}\ and\ \citenamefont
  {Khaliullin}(2009)}]{jackeli09}%
  \BibitemOpen
  \bibfield  {author} {\bibinfo {author} {\bibfnamefont {G.}~\bibnamefont
  {Jackeli}}\ and\ \bibinfo {author} {\bibfnamefont {G.}~\bibnamefont
  {Khaliullin}},\ }\bibfield  {title} {\bibinfo {title} {Mott Insulators in the
  Strong Spin-Orbit Coupling Limit: From Heisenberg to a Quantum Compass and
  Kitaev Models},\ }\href {https://doi.org/10.1103/PhysRevLett.102.017205}
  {\bibfield  {journal} {\bibinfo  {journal} {Phys. Rev. Lett.}\ }\textbf
  {\bibinfo {volume} {102}},\ \bibinfo {pages} {017205} (\bibinfo {year}
  {2009})}\BibitemShut {NoStop}%
\bibitem [{\citenamefont {Chaloupka}\ \emph {et~al.}(2010)\citenamefont
  {Chaloupka}, \citenamefont {Jackeli},\ and\ \citenamefont
  {Khaliullin}}]{chaloupka10}%
  \BibitemOpen
  \bibfield  {author} {\bibinfo {author} {\bibfnamefont {J.}~\bibnamefont
  {Chaloupka}}, \bibinfo {author} {\bibfnamefont {G.}~\bibnamefont {Jackeli}},\
  and\ \bibinfo {author} {\bibfnamefont {G.}~\bibnamefont {Khaliullin}},\
  }\bibfield  {title} {\bibinfo {title} {Kitaev-Heisenberg Model on a Honeycomb
  Lattice: Possible Exotic Phases in Iridium Oxides
  ${A}_{2}{\mathrm{IrO}}_{3}$},\ }\href
  {https://doi.org/10.1103/PhysRevLett.105.027204} {\bibfield  {journal}
  {\bibinfo  {journal} {Phys. Rev. Lett.}\ }\textbf {\bibinfo {volume} {105}},\
  \bibinfo {pages} {027204} (\bibinfo {year} {2010})}\BibitemShut {NoStop}%
\bibitem [{\citenamefont {Chaloupka}\ \emph {et~al.}(2013)\citenamefont
  {Chaloupka}, \citenamefont {Jackeli},\ and\ \citenamefont
  {Khaliullin}}]{chaloupka13}%
  \BibitemOpen
  \bibfield  {author} {\bibinfo {author} {\bibfnamefont {J.}~\bibnamefont
  {Chaloupka}}, \bibinfo {author} {\bibfnamefont {G.}~\bibnamefont {Jackeli}},\
  and\ \bibinfo {author} {\bibfnamefont {G.}~\bibnamefont {Khaliullin}},\
  }\bibfield  {title} {\bibinfo {title} {Zigzag Magnetic Order in the Iridium
  Oxide ${\mathrm{Na}}_{2}{\mathrm{IrO}}_{3}$},\ }\href
  {https://doi.org/10.1103/PhysRevLett.110.097204} {\bibfield  {journal}
  {\bibinfo  {journal} {Phys. Rev. Lett.}\ }\textbf {\bibinfo {volume} {110}},\
  \bibinfo {pages} {097204} (\bibinfo {year} {2013})}\BibitemShut {NoStop}%
\bibitem [{\citenamefont {Yao}\ and\ \citenamefont {Li}(2020)}]{yao20}%
  \BibitemOpen
  \bibfield  {author} {\bibinfo {author} {\bibfnamefont {W.}~\bibnamefont
  {Yao}}\ and\ \bibinfo {author} {\bibfnamefont {Y.}~\bibnamefont {Li}},\
  }\bibfield  {title} {\bibinfo {title} {Ferrimagnetism and anisotropic phase
  tunability by magnetic fields in
  ${\mathrm{Na}}_{2}{\mathrm{Co}}_{2}{\mathrm{TeO}}_{6}$},\ }\href
  {https://doi.org/10.1103/PhysRevB.101.085120} {\bibfield  {journal} {\bibinfo
   {journal} {Phys. Rev. B}\ }\textbf {\bibinfo {volume} {101}},\ \bibinfo
  {pages} {085120} (\bibinfo {year} {2020})}\BibitemShut {NoStop}%
\bibitem [{\citenamefont {Songvilay}\ \emph {et~al.}(2020)\citenamefont
  {Songvilay}, \citenamefont {Robert}, \citenamefont {Petit}, \citenamefont
  {Rodriguez-Rivera}, \citenamefont {Ratcliff}, \citenamefont {Damay},
  \citenamefont {Bal\'edent}, \citenamefont {Jim\'enez-Ruiz}, \citenamefont
  {Lejay}, \citenamefont {Pachoud}, \citenamefont {Hadj-Azzem}, \citenamefont
  {Simonet},\ and\ \citenamefont {Stock}}]{songvilay20}%
  \BibitemOpen
  \bibfield  {author} {\bibinfo {author} {\bibfnamefont {M.}~\bibnamefont
  {Songvilay}}, \bibinfo {author} {\bibfnamefont {J.}~\bibnamefont {Robert}},
  \bibinfo {author} {\bibfnamefont {S.}~\bibnamefont {Petit}}, \bibinfo
  {author} {\bibfnamefont {J.~A.}\ \bibnamefont {Rodriguez-Rivera}}, \bibinfo
  {author} {\bibfnamefont {W.~D.}\ \bibnamefont {Ratcliff}}, \bibinfo {author}
  {\bibfnamefont {F.}~\bibnamefont {Damay}}, \bibinfo {author} {\bibfnamefont
  {V.}~\bibnamefont {Bal\'edent}}, \bibinfo {author} {\bibfnamefont
  {M.}~\bibnamefont {Jim\'enez-Ruiz}}, \bibinfo {author} {\bibfnamefont
  {P.}~\bibnamefont {Lejay}}, \bibinfo {author} {\bibfnamefont
  {E.}~\bibnamefont {Pachoud}}, \bibinfo {author} {\bibfnamefont
  {A.}~\bibnamefont {Hadj-Azzem}}, \bibinfo {author} {\bibfnamefont
  {V.}~\bibnamefont {Simonet}},\ and\ \bibinfo {author} {\bibfnamefont
  {C.}~\bibnamefont {Stock}},\ }\bibfield  {title} {\bibinfo {title} {Kitaev
  interactions in the Co honeycomb antiferromagnets
  ${\mathrm{Na}}_{3}{\mathrm{Co}}_{2}{\mathrm{SbO}}_{6}$ and
  ${\mathrm{Na}}_{2}{\mathrm{Co}}_{2}{\mathrm{TeO}}_{6}$},\ }\href
  {https://doi.org/10.1103/PhysRevB.102.224429} {\bibfield  {journal} {\bibinfo
   {journal} {Phys. Rev. B}\ }\textbf {\bibinfo {volume} {102}},\ \bibinfo
  {pages} {224429} (\bibinfo {year} {2020})}\BibitemShut {NoStop}%
\bibitem [{\citenamefont {Lin}\ \emph {et~al.}(2021)\citenamefont {Lin},
  \citenamefont {Jeong}, \citenamefont {Kim}, \citenamefont {Wang},
  \citenamefont {Huang}, \citenamefont {Masuda}, \citenamefont {Asai},
  \citenamefont {Itoh}, \citenamefont {G{\"u}nther}, \citenamefont {Russina},
  \citenamefont {Lu}, \citenamefont {Sheng}, \citenamefont {Wang},
  \citenamefont {Wang}, \citenamefont {Wang}, \citenamefont {Ren},
  \citenamefont {Xi}, \citenamefont {Tong}, \citenamefont {Ling}, \citenamefont
  {Liu}, \citenamefont {Wu}, \citenamefont {Mei}, \citenamefont {Qu},
  \citenamefont {Zhou}, \citenamefont {Wang}, \citenamefont {Park},
  \citenamefont {Wan},\ and\ \citenamefont {Ma}}]{lin21}%
  \BibitemOpen
  \bibfield  {author} {\bibinfo {author} {\bibfnamefont {G.}~\bibnamefont
  {Lin}}, \bibinfo {author} {\bibfnamefont {J.}~\bibnamefont {Jeong}}, \bibinfo
  {author} {\bibfnamefont {C.}~\bibnamefont {Kim}}, \bibinfo {author}
  {\bibfnamefont {Y.}~\bibnamefont {Wang}}, \bibinfo {author} {\bibfnamefont
  {Q.}~\bibnamefont {Huang}}, \bibinfo {author} {\bibfnamefont
  {T.}~\bibnamefont {Masuda}}, \bibinfo {author} {\bibfnamefont
  {S.}~\bibnamefont {Asai}}, \bibinfo {author} {\bibfnamefont {S.}~\bibnamefont
  {Itoh}}, \bibinfo {author} {\bibfnamefont {G.}~\bibnamefont {G{\"u}nther}},
  \bibinfo {author} {\bibfnamefont {M.}~\bibnamefont {Russina}}, \bibinfo
  {author} {\bibfnamefont {Z.}~\bibnamefont {Lu}}, \bibinfo {author}
  {\bibfnamefont {J.}~\bibnamefont {Sheng}}, \bibinfo {author} {\bibfnamefont
  {L.}~\bibnamefont {Wang}}, \bibinfo {author} {\bibfnamefont {J.}~\bibnamefont
  {Wang}}, \bibinfo {author} {\bibfnamefont {G.}~\bibnamefont {Wang}}, \bibinfo
  {author} {\bibfnamefont {Q.}~\bibnamefont {Ren}}, \bibinfo {author}
  {\bibfnamefont {C.}~\bibnamefont {Xi}}, \bibinfo {author} {\bibfnamefont
  {W.}~\bibnamefont {Tong}}, \bibinfo {author} {\bibfnamefont {L.}~\bibnamefont
  {Ling}}, \bibinfo {author} {\bibfnamefont {Z.}~\bibnamefont {Liu}}, \bibinfo
  {author} {\bibfnamefont {L.}~\bibnamefont {Wu}}, \bibinfo {author}
  {\bibfnamefont {J.}~\bibnamefont {Mei}}, \bibinfo {author} {\bibfnamefont
  {Z.}~\bibnamefont {Qu}}, \bibinfo {author} {\bibfnamefont {H.}~\bibnamefont
  {Zhou}}, \bibinfo {author} {\bibfnamefont {X.}~\bibnamefont {Wang}}, \bibinfo
  {author} {\bibfnamefont {J.-G.}\ \bibnamefont {Park}}, \bibinfo {author}
  {\bibfnamefont {Y.}~\bibnamefont {Wan}},\ and\ \bibinfo {author}
  {\bibfnamefont {J.}~\bibnamefont {Ma}},\ }\bibfield  {title} {\bibinfo
  {title} {Field-induced quantum spin disordered state in spin-1/2 honeycomb
  magnet Na$_2$Co$_2$TeO$_6$},\ }\href
  {https://doi.org/10.1038/s41467-021-25567-7} {\bibfield  {journal} {\bibinfo
  {journal} {Nat. Commun.}\ }\textbf {\bibinfo {volume} {12}},\ \bibinfo
  {pages} {5559} (\bibinfo {year} {2021})}\BibitemShut {NoStop}%
\bibitem [{\citenamefont {Chen}\ \emph {et~al.}(2021)\citenamefont {Chen},
  \citenamefont {Li}, \citenamefont {Hu}, \citenamefont {Hu}, \citenamefont
  {Yue}, \citenamefont {Sutarto}, \citenamefont {He}, \citenamefont {Iida},
  \citenamefont {Kamazawa}, \citenamefont {Yu}, \citenamefont {Lin},\ and\
  \citenamefont {Li}}]{chen21}%
  \BibitemOpen
  \bibfield  {author} {\bibinfo {author} {\bibfnamefont {W.}~\bibnamefont
  {Chen}}, \bibinfo {author} {\bibfnamefont {X.}~\bibnamefont {Li}}, \bibinfo
  {author} {\bibfnamefont {Z.}~\bibnamefont {Hu}}, \bibinfo {author}
  {\bibfnamefont {Z.}~\bibnamefont {Hu}}, \bibinfo {author} {\bibfnamefont
  {L.}~\bibnamefont {Yue}}, \bibinfo {author} {\bibfnamefont {R.}~\bibnamefont
  {Sutarto}}, \bibinfo {author} {\bibfnamefont {F.}~\bibnamefont {He}},
  \bibinfo {author} {\bibfnamefont {K.}~\bibnamefont {Iida}}, \bibinfo {author}
  {\bibfnamefont {K.}~\bibnamefont {Kamazawa}}, \bibinfo {author}
  {\bibfnamefont {W.}~\bibnamefont {Yu}}, \bibinfo {author} {\bibfnamefont
  {X.}~\bibnamefont {Lin}},\ and\ \bibinfo {author} {\bibfnamefont
  {Y.}~\bibnamefont {Li}},\ }\bibfield  {title} {\bibinfo {title} {Spin-orbit
  phase behavior of ${\mathrm{Na}}_{2}{\mathrm{Co}}_{2}{\mathrm{TeO}}_{6}$ at
  low temperatures},\ }\href {https://doi.org/10.1103/PhysRevB.103.L180404}
  {\bibfield  {journal} {\bibinfo  {journal} {Phys. Rev. B}\ }\textbf {\bibinfo
  {volume} {103}},\ \bibinfo {pages} {L180404} (\bibinfo {year}
  {2021})}\BibitemShut {NoStop}%
\bibitem [{\citenamefont {Lee}\ \emph {et~al.}(2021)\citenamefont {Lee},
  \citenamefont {Lee}, \citenamefont {Choi}, \citenamefont {Jang},
  \citenamefont {Kalaivanan}, \citenamefont {Sankar},\ and\ \citenamefont
  {Choi}}]{lee21}%
  \BibitemOpen
  \bibfield  {author} {\bibinfo {author} {\bibfnamefont {C.~H.}\ \bibnamefont
  {Lee}}, \bibinfo {author} {\bibfnamefont {S.}~\bibnamefont {Lee}}, \bibinfo
  {author} {\bibfnamefont {Y.~S.}\ \bibnamefont {Choi}}, \bibinfo {author}
  {\bibfnamefont {Z.~H.}\ \bibnamefont {Jang}}, \bibinfo {author}
  {\bibfnamefont {R.}~\bibnamefont {Kalaivanan}}, \bibinfo {author}
  {\bibfnamefont {R.}~\bibnamefont {Sankar}},\ and\ \bibinfo {author}
  {\bibfnamefont {K.-Y.}\ \bibnamefont {Choi}},\ }\bibfield  {title} {\bibinfo
  {title} {Multistage development of anisotropic magnetic correlations in the
  Co-based honeycomb lattice
  ${\mathrm{Na}}_{2}{\mathrm{Co}}_{2}{\mathrm{TeO}}_{6}$},\ }\href
  {https://doi.org/10.1103/PhysRevB.103.214447} {\bibfield  {journal} {\bibinfo
   {journal} {Phys. Rev. B}\ }\textbf {\bibinfo {volume} {103}},\ \bibinfo
  {pages} {214447} (\bibinfo {year} {2021})}\BibitemShut {NoStop}%
\bibitem [{\citenamefont {Hong}\ \emph {et~al.}(2021)\citenamefont {Hong},
  \citenamefont {Gillig}, \citenamefont {Hentrich}, \citenamefont {Yao},
  \citenamefont {Kocsis}, \citenamefont {Witte}, \citenamefont {Schreiner},
  \citenamefont {Baumann}, \citenamefont {P\'erez}, \citenamefont {Wolter},
  \citenamefont {Li}, \citenamefont {B\"uchner},\ and\ \citenamefont
  {Hess}}]{hong21}%
  \BibitemOpen
  \bibfield  {author} {\bibinfo {author} {\bibfnamefont {X.}~\bibnamefont
  {Hong}}, \bibinfo {author} {\bibfnamefont {M.}~\bibnamefont {Gillig}},
  \bibinfo {author} {\bibfnamefont {R.}~\bibnamefont {Hentrich}}, \bibinfo
  {author} {\bibfnamefont {W.}~\bibnamefont {Yao}}, \bibinfo {author}
  {\bibfnamefont {V.}~\bibnamefont {Kocsis}}, \bibinfo {author} {\bibfnamefont
  {A.~R.}\ \bibnamefont {Witte}}, \bibinfo {author} {\bibfnamefont
  {T.}~\bibnamefont {Schreiner}}, \bibinfo {author} {\bibfnamefont
  {D.}~\bibnamefont {Baumann}}, \bibinfo {author} {\bibfnamefont
  {N.}~\bibnamefont {P\'erez}}, \bibinfo {author} {\bibfnamefont {A.~U.~B.}\
  \bibnamefont {Wolter}}, \bibinfo {author} {\bibfnamefont {Y.}~\bibnamefont
  {Li}}, \bibinfo {author} {\bibfnamefont {B.}~\bibnamefont {B\"uchner}},\ and\
  \bibinfo {author} {\bibfnamefont {C.}~\bibnamefont {Hess}},\ }\bibfield
  {title} {\bibinfo {title} {Strongly scattered phonon heat transport of the
  candidate Kitaev material
  ${\mathrm{Na}}_{2}{\mathrm{Co}}_{2}{\mathrm{TeO}}_{6}$},\ }\href
  {https://doi.org/10.1103/PhysRevB.104.144426} {\bibfield  {journal} {\bibinfo
   {journal} {Phys. Rev. B}\ }\textbf {\bibinfo {volume} {104}},\ \bibinfo
  {pages} {144426} (\bibinfo {year} {2021})}\BibitemShut {NoStop}%
\bibitem [{\citenamefont {Samarakoon}\ \emph {et~al.}(2021)\citenamefont
  {Samarakoon}, \citenamefont {Chen}, \citenamefont {Zhou},\ and\ \citenamefont
  {Garlea}}]{samarakoon21}%
  \BibitemOpen
  \bibfield  {author} {\bibinfo {author} {\bibfnamefont {A.~M.}\ \bibnamefont
  {Samarakoon}}, \bibinfo {author} {\bibfnamefont {Q.}~\bibnamefont {Chen}},
  \bibinfo {author} {\bibfnamefont {H.}~\bibnamefont {Zhou}},\ and\ \bibinfo
  {author} {\bibfnamefont {V.~O.}\ \bibnamefont {Garlea}},\ }\bibfield  {title}
  {\bibinfo {title} {Static and dynamic magnetic properties of honeycomb
  lattice antiferromagnets ${\mathrm{Na}}_{2}{M}_{2}{\mathrm{TeO}}_{6}$,
  $M=\mathrm{Co}$ and Ni},\ }\href
  {https://doi.org/10.1103/PhysRevB.104.184415} {\bibfield  {journal} {\bibinfo
   {journal} {Phys. Rev. B}\ }\textbf {\bibinfo {volume} {104}},\ \bibinfo
  {pages} {184415} (\bibinfo {year} {2021})}\BibitemShut {NoStop}%
\bibitem [{\citenamefont {Kim}\ \emph {et~al.}(2022)\citenamefont {Kim},
  \citenamefont {Jeong}, \citenamefont {Lin}, \citenamefont {Park},
  \citenamefont {Masuda}, \citenamefont {Asai}, \citenamefont {Itoh},
  \citenamefont {Kim}, \citenamefont {Zhou}, \citenamefont {Ma},\ and\
  \citenamefont {Park}}]{kim22}%
  \BibitemOpen
  \bibfield  {author} {\bibinfo {author} {\bibfnamefont {C.}~\bibnamefont
  {Kim}}, \bibinfo {author} {\bibfnamefont {J.}~\bibnamefont {Jeong}}, \bibinfo
  {author} {\bibfnamefont {G.}~\bibnamefont {Lin}}, \bibinfo {author}
  {\bibfnamefont {P.}~\bibnamefont {Park}}, \bibinfo {author} {\bibfnamefont
  {T.}~\bibnamefont {Masuda}}, \bibinfo {author} {\bibfnamefont
  {S.}~\bibnamefont {Asai}}, \bibinfo {author} {\bibfnamefont {S.}~\bibnamefont
  {Itoh}}, \bibinfo {author} {\bibfnamefont {H.-S.}\ \bibnamefont {Kim}},
  \bibinfo {author} {\bibfnamefont {H.}~\bibnamefont {Zhou}}, \bibinfo {author}
  {\bibfnamefont {J.}~\bibnamefont {Ma}},\ and\ \bibinfo {author}
  {\bibfnamefont {J.-G.}\ \bibnamefont {Park}},\ }\bibfield  {title} {\bibinfo
  {title} {Antiferromagnetic Kitaev interaction in $J_\text{eff} = 1/2$ cobalt
  honeycomb materials Na$_3$Co$_2$SbO$_6$ and Na$_2$Co$_2$TeO$_6$},\ }\href
  {https://doi.org/10.1088/1361-648x/ac2644} {\bibfield  {journal} {\bibinfo
  {journal} {J. Phys. Condens. Matter}\ }\textbf {\bibinfo {volume} {34}},\
  \bibinfo {pages} {045802} (\bibinfo {year} {2022})}\BibitemShut {NoStop}%
\bibitem [{\citenamefont {Mukherjee}\ \emph {et~al.}(2022)\citenamefont
  {Mukherjee}, \citenamefont {Manna}, \citenamefont {Saha}, \citenamefont
  {Majumdar},\ and\ \citenamefont {Giri}}]{mukherjee22}%
  \BibitemOpen
  \bibfield  {author} {\bibinfo {author} {\bibfnamefont {S.}~\bibnamefont
  {Mukherjee}}, \bibinfo {author} {\bibfnamefont {G.}~\bibnamefont {Manna}},
  \bibinfo {author} {\bibfnamefont {P.}~\bibnamefont {Saha}}, \bibinfo {author}
  {\bibfnamefont {S.}~\bibnamefont {Majumdar}},\ and\ \bibinfo {author}
  {\bibfnamefont {S.}~\bibnamefont {Giri}},\ }\bibfield  {title} {\bibinfo
  {title} {Ferroelectric order with a linear high-field magnetoelectric
  coupling in ${\mathrm{Na}}_{2}{\mathrm{Co}}_{2}{\mathrm{TeO}}_{6}$: A
  proposed Kitaev compound},\ }\href
  {https://doi.org/10.1103/PhysRevMaterials.6.054407} {\bibfield  {journal}
  {\bibinfo  {journal} {Phys. Rev. Materials}\ }\textbf {\bibinfo {volume}
  {6}},\ \bibinfo {pages} {054407} (\bibinfo {year} {2022})}\BibitemShut
  {NoStop}%
\bibitem [{\citenamefont {Sanders}\ \emph {et~al.}(2022)\citenamefont
  {Sanders}, \citenamefont {Mole}, \citenamefont {Liu}, \citenamefont {Brown},
  \citenamefont {Yu}, \citenamefont {Ling},\ and\ \citenamefont
  {Rachel}}]{sanders22}%
  \BibitemOpen
  \bibfield  {author} {\bibinfo {author} {\bibfnamefont {A.~L.}\ \bibnamefont
  {Sanders}}, \bibinfo {author} {\bibfnamefont {R.~A.}\ \bibnamefont {Mole}},
  \bibinfo {author} {\bibfnamefont {J.}~\bibnamefont {Liu}}, \bibinfo {author}
  {\bibfnamefont {A.~J.}\ \bibnamefont {Brown}}, \bibinfo {author}
  {\bibfnamefont {D.}~\bibnamefont {Yu}}, \bibinfo {author} {\bibfnamefont
  {C.~D.}\ \bibnamefont {Ling}},\ and\ \bibinfo {author} {\bibfnamefont
  {S.}~\bibnamefont {Rachel}},\ }\bibfield  {title} {\bibinfo {title} {Dominant
  Kitaev interactions in the honeycomb materials
  ${\mathrm{Na}}_{3}{\mathrm{Co}}_{2}{\mathrm{SbO}}_{6}$ and
  ${\mathrm{Na}}_{2}{\mathrm{Co}}_{2}{\mathrm{TeO}}_{6}$},\ }\href
  {https://doi.org/10.1103/PhysRevB.106.014413} {\bibfield  {journal} {\bibinfo
   {journal} {Phys. Rev. B}\ }\textbf {\bibinfo {volume} {106}},\ \bibinfo
  {pages} {014413} (\bibinfo {year} {2022})}\BibitemShut {NoStop}%
\bibitem [{\citenamefont {Yang}\ \emph {et~al.}(2022)\citenamefont {Yang},
  \citenamefont {Kim}, \citenamefont {Choi}, \citenamefont {Lee}, \citenamefont
  {Lin}, \citenamefont {Ma}, \citenamefont {Kratochv\'{\i}lov\'a},
  \citenamefont {Proschek}, \citenamefont {Moon}, \citenamefont {Lee},
  \citenamefont {Oh},\ and\ \citenamefont {Park}}]{yang22}%
  \BibitemOpen
  \bibfield  {author} {\bibinfo {author} {\bibfnamefont {H.}~\bibnamefont
  {Yang}}, \bibinfo {author} {\bibfnamefont {C.}~\bibnamefont {Kim}}, \bibinfo
  {author} {\bibfnamefont {Y.}~\bibnamefont {Choi}}, \bibinfo {author}
  {\bibfnamefont {J.~H.}\ \bibnamefont {Lee}}, \bibinfo {author} {\bibfnamefont
  {G.}~\bibnamefont {Lin}}, \bibinfo {author} {\bibfnamefont {J.}~\bibnamefont
  {Ma}}, \bibinfo {author} {\bibfnamefont {M.}~\bibnamefont
  {Kratochv\'{\i}lov\'a}}, \bibinfo {author} {\bibfnamefont {P.}~\bibnamefont
  {Proschek}}, \bibinfo {author} {\bibfnamefont {E.-G.}\ \bibnamefont {Moon}},
  \bibinfo {author} {\bibfnamefont {K.~H.}\ \bibnamefont {Lee}}, \bibinfo
  {author} {\bibfnamefont {Y.~S.}\ \bibnamefont {Oh}},\ and\ \bibinfo {author}
  {\bibfnamefont {J.-G.}\ \bibnamefont {Park}},\ }\bibfield  {title} {\bibinfo
  {title} {Significant thermal Hall effect in the $3d$ cobalt Kitaev system
  ${\mathrm{Na}}_{2}{\mathrm{Co}}_{2}\mathrm{Te}{\mathrm{O}}_{6}$},\ }\href
  {https://doi.org/10.1103/PhysRevB.106.L081116} {\bibfield  {journal}
  {\bibinfo  {journal} {Phys. Rev. B}\ }\textbf {\bibinfo {volume} {106}},\
  \bibinfo {pages} {L081116} (\bibinfo {year} {2022})}\BibitemShut {NoStop}%
\bibitem [{\citenamefont {Yao}\ \emph {et~al.}(2022)\citenamefont {Yao},
  \citenamefont {Iida}, \citenamefont {Kamazawa},\ and\ \citenamefont
  {Li}}]{yao22}%
  \BibitemOpen
  \bibfield  {author} {\bibinfo {author} {\bibfnamefont {W.}~\bibnamefont
  {Yao}}, \bibinfo {author} {\bibfnamefont {K.}~\bibnamefont {Iida}}, \bibinfo
  {author} {\bibfnamefont {K.}~\bibnamefont {Kamazawa}},\ and\ \bibinfo
  {author} {\bibfnamefont {Y.}~\bibnamefont {Li}},\ }\bibfield  {title}
  {\bibinfo {title} {Excitations in the Ordered and Paramagnetic States of
  Honeycomb Magnet ${\mathrm{Na}}_{2}{\mathrm{Co}}_{2}{\mathrm{TeO}}_{6}$},\
  }\href {https://doi.org/10.1103/PhysRevLett.129.147202} {\bibfield  {journal}
  {\bibinfo  {journal} {Phys. Rev. Lett.}\ }\textbf {\bibinfo {volume} {129}},\
  \bibinfo {pages} {147202} (\bibinfo {year} {2022})}\BibitemShut {NoStop}%
\bibitem [{\citenamefont {Kr\"uger}\ \emph {et~al.}(2023)\citenamefont
  {Kr\"uger}, \citenamefont {Chen}, \citenamefont {Jin}, \citenamefont {Li},\
  and\ \citenamefont {Janssen}}]{krueger23}%
  \BibitemOpen
  \bibfield  {author} {\bibinfo {author} {\bibfnamefont {W.~G.~F.}\
  \bibnamefont {Kr\"uger}}, \bibinfo {author} {\bibfnamefont {W.}~\bibnamefont
  {Chen}}, \bibinfo {author} {\bibfnamefont {X.}~\bibnamefont {Jin}}, \bibinfo
  {author} {\bibfnamefont {Y.}~\bibnamefont {Li}},\ and\ \bibinfo {author}
  {\bibfnamefont {L.}~\bibnamefont {Janssen}},\ }\bibfield  {title} {\bibinfo
  {title} {Triple-q Order in
  ${\mathrm{Na}}_{2}{\mathrm{Co}}_{2}{\mathrm{TeO}}_{6}$ from Proximity to
  Hidden-SU(2)-Symmetric Point},\ }\href
  {https://doi.org/10.1103/PhysRevLett.131.146702} {\bibfield  {journal}
  {\bibinfo  {journal} {Phys. Rev. Lett.}\ }\textbf {\bibinfo {volume} {131}},\
  \bibinfo {pages} {146702} (\bibinfo {year} {2023})}\BibitemShut {NoStop}%
\bibitem [{\citenamefont {Yao}\ \emph {et~al.}(2023)\citenamefont {Yao},
  \citenamefont {Zhao}, \citenamefont {Qiu}, \citenamefont {Balz},
  \citenamefont {Stewart}, \citenamefont {Lynn},\ and\ \citenamefont
  {Li}}]{yao23}%
  \BibitemOpen
  \bibfield  {author} {\bibinfo {author} {\bibfnamefont {W.}~\bibnamefont
  {Yao}}, \bibinfo {author} {\bibfnamefont {Y.}~\bibnamefont {Zhao}}, \bibinfo
  {author} {\bibfnamefont {Y.}~\bibnamefont {Qiu}}, \bibinfo {author}
  {\bibfnamefont {C.}~\bibnamefont {Balz}}, \bibinfo {author} {\bibfnamefont
  {J.~R.}\ \bibnamefont {Stewart}}, \bibinfo {author} {\bibfnamefont {J.~W.}\
  \bibnamefont {Lynn}},\ and\ \bibinfo {author} {\bibfnamefont
  {Y.}~\bibnamefont {Li}},\ }\bibfield  {title} {\bibinfo {title} {Magnetic
  ground state of the Kitaev
  ${\mathrm{Na}}_{2}{\mathrm{Co}}_{2}{\mathrm{TeO}}_{6}$ spin liquid
  candidate},\ }\href {https://doi.org/10.1103/PhysRevResearch.5.L022045}
  {\bibfield  {journal} {\bibinfo  {journal} {Phys. Rev. Res.}\ }\textbf
  {\bibinfo {volume} {5}},\ \bibinfo {pages} {L022045} (\bibinfo {year}
  {2023})}\BibitemShut {NoStop}%
\bibitem [{\citenamefont {Xiang}\ \emph {et~al.}(2023)\citenamefont {Xiang},
  \citenamefont {Dhakal}, \citenamefont {Ozerov}, \citenamefont {Jiang},
  \citenamefont {Mou}, \citenamefont {Ozarowski}, \citenamefont {Huang},
  \citenamefont {Zhou}, \citenamefont {Fang}, \citenamefont {Winter},
  \citenamefont {Jiang},\ and\ \citenamefont {Smirnov}}]{xiang23}%
  \BibitemOpen
  \bibfield  {author} {\bibinfo {author} {\bibfnamefont {L.}~\bibnamefont
  {Xiang}}, \bibinfo {author} {\bibfnamefont {R.}~\bibnamefont {Dhakal}},
  \bibinfo {author} {\bibfnamefont {M.}~\bibnamefont {Ozerov}}, \bibinfo
  {author} {\bibfnamefont {Y.}~\bibnamefont {Jiang}}, \bibinfo {author}
  {\bibfnamefont {B.~S.}\ \bibnamefont {Mou}}, \bibinfo {author} {\bibfnamefont
  {A.}~\bibnamefont {Ozarowski}}, \bibinfo {author} {\bibfnamefont
  {Q.}~\bibnamefont {Huang}}, \bibinfo {author} {\bibfnamefont
  {H.}~\bibnamefont {Zhou}}, \bibinfo {author} {\bibfnamefont {J.}~\bibnamefont
  {Fang}}, \bibinfo {author} {\bibfnamefont {S.~M.}\ \bibnamefont {Winter}},
  \bibinfo {author} {\bibfnamefont {Z.}~\bibnamefont {Jiang}},\ and\ \bibinfo
  {author} {\bibfnamefont {D.}~\bibnamefont {Smirnov}},\ }\bibfield  {title}
  {\bibinfo {title} {Disorder-Enriched Magnetic Excitations in a
  Heisenberg-Kitaev Quantum Magnet
  ${\mathrm{Na}}_{2}{\mathrm{Co}}_{2}{\mathrm{TeO}}_{6}$},\ }\href
  {https://doi.org/10.1103/PhysRevLett.131.076701} {\bibfield  {journal}
  {\bibinfo  {journal} {Phys. Rev. Lett.}\ }\textbf {\bibinfo {volume} {131}},\
  \bibinfo {pages} {076701} (\bibinfo {year} {2023})}\BibitemShut {NoStop}%
\bibitem [{\citenamefont {Zhang}\ \emph
  {et~al.}(2023{\natexlab{a}})\citenamefont {Zhang}, \citenamefont {Lee},
  \citenamefont {Woods}, \citenamefont {Peria}, \citenamefont {Thomas},
  \citenamefont {Movshovich}, \citenamefont {Brosha}, \citenamefont {Huang},
  \citenamefont {Zhou}, \citenamefont {Zapf},\ and\ \citenamefont
  {Lee}}]{zhang23}%
  \BibitemOpen
  \bibfield  {author} {\bibinfo {author} {\bibfnamefont {S.}~\bibnamefont
  {Zhang}}, \bibinfo {author} {\bibfnamefont {S.}~\bibnamefont {Lee}}, \bibinfo
  {author} {\bibfnamefont {A.~J.}\ \bibnamefont {Woods}}, \bibinfo {author}
  {\bibfnamefont {W.~K.}\ \bibnamefont {Peria}}, \bibinfo {author}
  {\bibfnamefont {S.~M.}\ \bibnamefont {Thomas}}, \bibinfo {author}
  {\bibfnamefont {R.}~\bibnamefont {Movshovich}}, \bibinfo {author}
  {\bibfnamefont {E.}~\bibnamefont {Brosha}}, \bibinfo {author} {\bibfnamefont
  {Q.}~\bibnamefont {Huang}}, \bibinfo {author} {\bibfnamefont
  {H.}~\bibnamefont {Zhou}}, \bibinfo {author} {\bibfnamefont {V.~S.}\
  \bibnamefont {Zapf}},\ and\ \bibinfo {author} {\bibfnamefont
  {M.}~\bibnamefont {Lee}},\ }\bibfield  {title} {\bibinfo {title} {Electronic
  and magnetic phase diagrams of the Kitaev quantum spin liquid candidate
  ${\mathrm{Na}}_{2}{\mathrm{Co}}_{2}{\mathrm{TeO}}_{6}$},\ }\href
  {https://doi.org/10.1103/PhysRevB.108.064421} {\bibfield  {journal} {\bibinfo
   {journal} {Phys. Rev. B}\ }\textbf {\bibinfo {volume} {108}},\ \bibinfo
  {pages} {064421} (\bibinfo {year} {2023}{\natexlab{a}})}\BibitemShut
  {NoStop}%
\bibitem [{\citenamefont {Hong}\ \emph {et~al.}(2023)\citenamefont {Hong},
  \citenamefont {Gillig}, \citenamefont {Yao}, \citenamefont {Janssen},
  \citenamefont {Kocsis}, \citenamefont {Gass}, \citenamefont {Li},
  \citenamefont {Wolter}, \citenamefont {Büchner},\ and\ \citenamefont
  {Hess}}]{hong23}%
  \BibitemOpen
  \bibfield  {author} {\bibinfo {author} {\bibfnamefont {X.}~\bibnamefont
  {Hong}}, \bibinfo {author} {\bibfnamefont {M.}~\bibnamefont {Gillig}},
  \bibinfo {author} {\bibfnamefont {W.}~\bibnamefont {Yao}}, \bibinfo {author}
  {\bibfnamefont {L.}~\bibnamefont {Janssen}}, \bibinfo {author} {\bibfnamefont
  {V.}~\bibnamefont {Kocsis}}, \bibinfo {author} {\bibfnamefont
  {S.}~\bibnamefont {Gass}}, \bibinfo {author} {\bibfnamefont {Y.}~\bibnamefont
  {Li}}, \bibinfo {author} {\bibfnamefont {A.~U.~B.}\ \bibnamefont {Wolter}},
  \bibinfo {author} {\bibfnamefont {B.}~\bibnamefont {Büchner}},\ and\
  \bibinfo {author} {\bibfnamefont {C.}~\bibnamefont {Hess}},\ }\bibinfo
  {title} {Phonon thermal transport shaped by strong spin-phonon scattering in
  a Kitaev material Na$_2$Co$_2$TeO$_6$},\ \Eprint
  {https://arxiv.org/abs/2306.16963} {arXiv:2306.16963}\BibitemShut {NoStop}%
\bibitem [{\citenamefont {Pilch}\ \emph {et~al.}(2023)\citenamefont {Pilch},
  \citenamefont {Peedu}, \citenamefont {Bera}, \citenamefont {Yusuf},
  \citenamefont {Nagel}, \citenamefont {R\~{o}\~{o}m},\ and\ \citenamefont
  {Wang}}]{pilch23}%
  \BibitemOpen
  \bibfield  {author} {\bibinfo {author} {\bibfnamefont {P.}~\bibnamefont
  {Pilch}}, \bibinfo {author} {\bibfnamefont {L.}~\bibnamefont {Peedu}},
  \bibinfo {author} {\bibfnamefont {A.~K.}\ \bibnamefont {Bera}}, \bibinfo
  {author} {\bibfnamefont {S.~M.}\ \bibnamefont {Yusuf}}, \bibinfo {author}
  {\bibfnamefont {U.}~\bibnamefont {Nagel}}, \bibinfo {author} {\bibfnamefont
  {T.}~\bibnamefont {R\~{o}\~{o}m}},\ and\ \bibinfo {author} {\bibfnamefont
  {Z.}~\bibnamefont {Wang}},\ }\bibfield  {title} {\bibinfo {title} {Field- and
  polarization-dependent quantum spin dynamics in the honeycomb magnet
  ${\mathrm{Na}}_{2}{\mathrm{Co}}_{2}{\mathrm{TeO}}_{6}$: Magnetic excitations
  and continuum},\ }\href {https://doi.org/10.1103/PhysRevB.108.L140406}
  {\bibfield  {journal} {\bibinfo  {journal} {Phys. Rev. B}\ }\textbf {\bibinfo
  {volume} {108}},\ \bibinfo {pages} {L140406} (\bibinfo {year}
  {2023})}\BibitemShut {NoStop}%
\bibitem [{\citenamefont {Bera}\ \emph {et~al.}(2023)\citenamefont {Bera},
  \citenamefont {Yusuf}, \citenamefont {Orlandi}, \citenamefont {Manuel},
  \citenamefont {Bhaskaran},\ and\ \citenamefont {Zvyagin}}]{bera23}%
  \BibitemOpen
  \bibfield  {author} {\bibinfo {author} {\bibfnamefont {A.~K.}\ \bibnamefont
  {Bera}}, \bibinfo {author} {\bibfnamefont {S.~M.}\ \bibnamefont {Yusuf}},
  \bibinfo {author} {\bibfnamefont {F.}~\bibnamefont {Orlandi}}, \bibinfo
  {author} {\bibfnamefont {P.}~\bibnamefont {Manuel}}, \bibinfo {author}
  {\bibfnamefont {L.}~\bibnamefont {Bhaskaran}},\ and\ \bibinfo {author}
  {\bibfnamefont {S.~A.}\ \bibnamefont {Zvyagin}},\ }\bibfield  {title}
  {\bibinfo {title} {Field-induced phase transitions and anisotropic magnetic
  properties of the Kitaev-Heisenberg compound
  ${\mathrm{Na}}_{2}{\mathrm{Co}}_{2}\mathrm{Te}{\mathrm{O}}_{6}$},\ }\href
  {https://doi.org/10.1103/PhysRevB.108.214419} {\bibfield  {journal} {\bibinfo
   {journal} {Phys. Rev. B}\ }\textbf {\bibinfo {volume} {108}},\ \bibinfo
  {pages} {214419} (\bibinfo {year} {2023})}\BibitemShut {NoStop}%
\bibitem [{\citenamefont {Gillig}\ \emph {et~al.}(2023)\citenamefont {Gillig},
  \citenamefont {Hong}, \citenamefont {Wellm}, \citenamefont {Kataev},
  \citenamefont {Yao}, \citenamefont {Li}, \citenamefont {B\"uchner},\ and\
  \citenamefont {Hess}}]{gillig23}%
  \BibitemOpen
  \bibfield  {author} {\bibinfo {author} {\bibfnamefont {M.}~\bibnamefont
  {Gillig}}, \bibinfo {author} {\bibfnamefont {X.}~\bibnamefont {Hong}},
  \bibinfo {author} {\bibfnamefont {C.}~\bibnamefont {Wellm}}, \bibinfo
  {author} {\bibfnamefont {V.}~\bibnamefont {Kataev}}, \bibinfo {author}
  {\bibfnamefont {W.}~\bibnamefont {Yao}}, \bibinfo {author} {\bibfnamefont
  {Y.}~\bibnamefont {Li}}, \bibinfo {author} {\bibfnamefont {B.}~\bibnamefont
  {B\"uchner}},\ and\ \bibinfo {author} {\bibfnamefont {C.}~\bibnamefont
  {Hess}},\ }\bibfield  {title} {\bibinfo {title} {Phononic-magnetic dichotomy
  of the thermal Hall effect in the Kitaev material
  ${\mathrm{Na}}_{2}{\mathrm{Co}}_{2}{\mathrm{TeO}}_{6}$},\ }\href
  {https://doi.org/10.1103/PhysRevResearch.5.043110} {\bibfield  {journal}
  {\bibinfo  {journal} {Phys. Rev. Res.}\ }\textbf {\bibinfo {volume} {5}},\
  \bibinfo {pages} {043110} (\bibinfo {year} {2023})}\BibitemShut {NoStop}%
\bibitem [{\citenamefont {Miao}\ \emph {et~al.}(2024)\citenamefont {Miao},
  \citenamefont {Jin}, \citenamefont {Yao}, \citenamefont {Chen}, \citenamefont
  {Koda}, \citenamefont {Tan}, \citenamefont {Xie}, \citenamefont {Ji},
  \citenamefont {Kamiyama},\ and\ \citenamefont {Li}}]{miao24}%
  \BibitemOpen
  \bibfield  {author} {\bibinfo {author} {\bibfnamefont {P.}~\bibnamefont
  {Miao}}, \bibinfo {author} {\bibfnamefont {X.}~\bibnamefont {Jin}}, \bibinfo
  {author} {\bibfnamefont {W.}~\bibnamefont {Yao}}, \bibinfo {author}
  {\bibfnamefont {Y.}~\bibnamefont {Chen}}, \bibinfo {author} {\bibfnamefont
  {A.}~\bibnamefont {Koda}}, \bibinfo {author} {\bibfnamefont {Z.}~\bibnamefont
  {Tan}}, \bibinfo {author} {\bibfnamefont {W.}~\bibnamefont {Xie}}, \bibinfo
  {author} {\bibfnamefont {W.}~\bibnamefont {Ji}}, \bibinfo {author}
  {\bibfnamefont {T.}~\bibnamefont {Kamiyama}},\ and\ \bibinfo {author}
  {\bibfnamefont {Y.}~\bibnamefont {Li}},\ }\bibfield  {title} {\bibinfo
  {title} {Persistent spin dynamics in magnetically ordered honeycomb-lattice
  cobalt oxides},\ }\href {https://doi.org/10.1103/PhysRevB.109.134431}
  {\bibfield  {journal} {\bibinfo  {journal} {Phys. Rev. B}\ }\textbf {\bibinfo
  {volume} {109}},\ \bibinfo {pages} {134431} (\bibinfo {year}
  {2024})}\BibitemShut {NoStop}%
\bibitem [{\citenamefont {Zhou}\ \emph {et~al.}(2024)\citenamefont {Zhou},
  \citenamefont {Li}, \citenamefont {Kim}, \citenamefont {Matsuo},
  \citenamefont {Mehlawat}, \citenamefont {Matsui}, \citenamefont {Yang},
  \citenamefont {Miyata}, \citenamefont {Su}, \citenamefont {Kindo},
  \citenamefont {Park}, \citenamefont {Kohama}, \citenamefont {Li},\ and\
  \citenamefont {Matsuda}}]{zhou24}%
  \BibitemOpen
  \bibfield  {author} {\bibinfo {author} {\bibfnamefont {X.-G.}\ \bibnamefont
  {Zhou}}, \bibinfo {author} {\bibfnamefont {H.}~\bibnamefont {Li}}, \bibinfo
  {author} {\bibfnamefont {C.}~\bibnamefont {Kim}}, \bibinfo {author}
  {\bibfnamefont {A.}~\bibnamefont {Matsuo}}, \bibinfo {author} {\bibfnamefont
  {K.}~\bibnamefont {Mehlawat}}, \bibinfo {author} {\bibfnamefont
  {K.}~\bibnamefont {Matsui}}, \bibinfo {author} {\bibfnamefont
  {Z.}~\bibnamefont {Yang}}, \bibinfo {author} {\bibfnamefont {A.}~\bibnamefont
  {Miyata}}, \bibinfo {author} {\bibfnamefont {G.}~\bibnamefont {Su}}, \bibinfo
  {author} {\bibfnamefont {K.}~\bibnamefont {Kindo}}, \bibinfo {author}
  {\bibfnamefont {J.-G.}\ \bibnamefont {Park}}, \bibinfo {author}
  {\bibfnamefont {Y.}~\bibnamefont {Kohama}}, \bibinfo {author} {\bibfnamefont
  {W.}~\bibnamefont {Li}},\ and\ \bibinfo {author} {\bibfnamefont {Y.~H.}\
  \bibnamefont {Matsuda}},\ }\bibinfo {title} {Emergent quantum disordered
  phase in Na$_2$Co$_2$TeO$_6$ under intermediate magnetic field along $c$
  axis},\ \Eprint {https://arxiv.org/abs/2408.01957}
  {arXiv:2408.01957}\BibitemShut {NoStop}%
\bibitem [{\citenamefont {Lin}\ \emph {et~al.}(2024)\citenamefont {Lin},
  \citenamefont {Shu}, \citenamefont {Zhao}, \citenamefont {Li}, \citenamefont
  {Ma}, \citenamefont {Jiao}, \citenamefont {Li}, \citenamefont {Duan},
  \citenamefont {Huang}, \citenamefont {Sheng}, \citenamefont {Kolesnikov},
  \citenamefont {Li}, \citenamefont {Wu}, \citenamefont {Chen}, \citenamefont
  {Yu}, \citenamefont {Wang}, \citenamefont {Liu}, \citenamefont {Zhou},\ and\
  \citenamefont {Ma}}]{lin24}%
  \BibitemOpen
  \bibfield  {author} {\bibinfo {author} {\bibfnamefont {G.}~\bibnamefont
  {Lin}}, \bibinfo {author} {\bibfnamefont {M.}~\bibnamefont {Shu}}, \bibinfo
  {author} {\bibfnamefont {Q.}~\bibnamefont {Zhao}}, \bibinfo {author}
  {\bibfnamefont {G.}~\bibnamefont {Li}}, \bibinfo {author} {\bibfnamefont
  {Y.}~\bibnamefont {Ma}}, \bibinfo {author} {\bibfnamefont {J.}~\bibnamefont
  {Jiao}}, \bibinfo {author} {\bibfnamefont {Y.}~\bibnamefont {Li}}, \bibinfo
  {author} {\bibfnamefont {G.}~\bibnamefont {Duan}}, \bibinfo {author}
  {\bibfnamefont {Q.}~\bibnamefont {Huang}}, \bibinfo {author} {\bibfnamefont
  {J.}~\bibnamefont {Sheng}}, \bibinfo {author} {\bibfnamefont {A.~I.}\
  \bibnamefont {Kolesnikov}}, \bibinfo {author} {\bibfnamefont
  {L.}~\bibnamefont {Li}}, \bibinfo {author} {\bibfnamefont {L.}~\bibnamefont
  {Wu}}, \bibinfo {author} {\bibfnamefont {H.}~\bibnamefont {Chen}}, \bibinfo
  {author} {\bibfnamefont {R.}~\bibnamefont {Yu}}, \bibinfo {author}
  {\bibfnamefont {X.}~\bibnamefont {Wang}}, \bibinfo {author} {\bibfnamefont
  {Z.}~\bibnamefont {Liu}}, \bibinfo {author} {\bibfnamefont {H.}~\bibnamefont
  {Zhou}},\ and\ \bibinfo {author} {\bibfnamefont {J.}~\bibnamefont {Ma}},\
  }\bibinfo {title} {Evidence for field induced quantum spin liquid behavior in
  a spin-1/2 honeycomb magnet},\ \Eprint {https://arxiv.org/abs/2409.07959}
  {arXiv:2409.07959}\BibitemShut {NoStop}%
\bibitem [{\citenamefont {Arneth}\ \emph {et~al.}(2024)\citenamefont {Arneth},
  \citenamefont {Choi}, \citenamefont {Kalaivanan}, \citenamefont {Sankar},\
  and\ \citenamefont {Klingeler}}]{arneth24}%
  \BibitemOpen
  \bibfield  {author} {\bibinfo {author} {\bibfnamefont {J.}~\bibnamefont
  {Arneth}}, \bibinfo {author} {\bibfnamefont {K.~Y.}\ \bibnamefont {Choi}},
  \bibinfo {author} {\bibfnamefont {R.}~\bibnamefont {Kalaivanan}}, \bibinfo
  {author} {\bibfnamefont {R.}~\bibnamefont {Sankar}},\ and\ \bibinfo {author}
  {\bibfnamefont {R.}~\bibnamefont {Klingeler}},\ }\bibinfo {title} {Signatures
  of a Quantum Critical Endpoint in the Kitaev Candidate Na$_2$Co$_2$TeO$_6$},\
  \Eprint {https://arxiv.org/abs/2409.05661} {arXiv:2409.05661}\BibitemShut
  {NoStop}%
\bibitem [{\citenamefont {Bischof}\ \emph {et~al.}(2025)\citenamefont
  {Bischof}, \citenamefont {Arneth}, \citenamefont {Kalaivanan}, \citenamefont
  {Sankar}, \citenamefont {Choi},\ and\ \citenamefont {Klingeler}}]{bischof25}%
  \BibitemOpen
  \bibfield  {author} {\bibinfo {author} {\bibfnamefont {L.}~\bibnamefont
  {Bischof}}, \bibinfo {author} {\bibfnamefont {J.}~\bibnamefont {Arneth}},
  \bibinfo {author} {\bibfnamefont {R.}~\bibnamefont {Kalaivanan}}, \bibinfo
  {author} {\bibfnamefont {R.}~\bibnamefont {Sankar}}, \bibinfo {author}
  {\bibfnamefont {K.-Y.}\ \bibnamefont {Choi}},\ and\ \bibinfo {author}
  {\bibfnamefont {R.}~\bibnamefont {Klingeler}},\ }\bibinfo {title} {Spin waves
  in Na$_2$Co$_2$TeO$_6$ studied by high-frequency/high-field ESR: Successes
  and failures of the triple-$\mathbf{q}$ model},\ \Eprint
  {https://arxiv.org/abs/2506.03789} {arXiv:2506.03789}\BibitemShut {NoStop}%
\bibitem [{\citenamefont {Yan}\ \emph {et~al.}(2019)\citenamefont {Yan},
  \citenamefont {Okamoto}, \citenamefont {Wu}, \citenamefont {Zheng},
  \citenamefont {Zhou}, \citenamefont {Cao},\ and\ \citenamefont
  {McGuire}}]{yan19}%
  \BibitemOpen
  \bibfield  {author} {\bibinfo {author} {\bibfnamefont {J.-Q.}\ \bibnamefont
  {Yan}}, \bibinfo {author} {\bibfnamefont {S.}~\bibnamefont {Okamoto}},
  \bibinfo {author} {\bibfnamefont {Y.}~\bibnamefont {Wu}}, \bibinfo {author}
  {\bibfnamefont {Q.}~\bibnamefont {Zheng}}, \bibinfo {author} {\bibfnamefont
  {H.~D.}\ \bibnamefont {Zhou}}, \bibinfo {author} {\bibfnamefont {H.~B.}\
  \bibnamefont {Cao}},\ and\ \bibinfo {author} {\bibfnamefont {M.~A.}\
  \bibnamefont {McGuire}},\ }\bibfield  {title} {\bibinfo {title} {Magnetic
  order in single crystals of Na$_3$Co$_2$SbO$_6$ with a honeycomb arrangement
  of $3d^7$ Co$^{2+}$ ions},\ }\href
  {https://doi.org/10.1103/PhysRevMaterials.3.074405} {\bibfield  {journal}
  {\bibinfo  {journal} {Phys. Rev. Materials}\ }\textbf {\bibinfo {volume}
  {3}},\ \bibinfo {pages} {074405} (\bibinfo {year} {2019})}\BibitemShut
  {NoStop}%
\bibitem [{\citenamefont {Li}\ \emph {et~al.}(2022)\citenamefont {Li},
  \citenamefont {Gu}, \citenamefont {Chen}, \citenamefont {Garlea},
  \citenamefont {Iida}, \citenamefont {Kamazawa}, \citenamefont {Li},
  \citenamefont {Deng}, \citenamefont {Xiao}, \citenamefont {Zheng},
  \citenamefont {Ye}, \citenamefont {Peng}, \citenamefont {Zaliznyak},
  \citenamefont {Tranquada},\ and\ \citenamefont {Li}}]{li22}%
  \BibitemOpen
  \bibfield  {author} {\bibinfo {author} {\bibfnamefont {X.}~\bibnamefont
  {Li}}, \bibinfo {author} {\bibfnamefont {Y.}~\bibnamefont {Gu}}, \bibinfo
  {author} {\bibfnamefont {Y.}~\bibnamefont {Chen}}, \bibinfo {author}
  {\bibfnamefont {V.~O.}\ \bibnamefont {Garlea}}, \bibinfo {author}
  {\bibfnamefont {K.}~\bibnamefont {Iida}}, \bibinfo {author} {\bibfnamefont
  {K.}~\bibnamefont {Kamazawa}}, \bibinfo {author} {\bibfnamefont
  {Y.}~\bibnamefont {Li}}, \bibinfo {author} {\bibfnamefont {G.}~\bibnamefont
  {Deng}}, \bibinfo {author} {\bibfnamefont {Q.}~\bibnamefont {Xiao}}, \bibinfo
  {author} {\bibfnamefont {X.}~\bibnamefont {Zheng}}, \bibinfo {author}
  {\bibfnamefont {Z.}~\bibnamefont {Ye}}, \bibinfo {author} {\bibfnamefont
  {Y.}~\bibnamefont {Peng}}, \bibinfo {author} {\bibfnamefont {I.~A.}\
  \bibnamefont {Zaliznyak}}, \bibinfo {author} {\bibfnamefont {J.~M.}\
  \bibnamefont {Tranquada}},\ and\ \bibinfo {author} {\bibfnamefont
  {Y.}~\bibnamefont {Li}},\ }\bibfield  {title} {\bibinfo {title} {Giant
  Magnetic In-Plane Anisotropy and Competing Instabilities in
  ${\mathrm{Na}}_{3}{\mathrm{Co}}_{2}{\mathrm{SbO}}_{6}$},\ }\href
  {https://doi.org/10.1103/PhysRevX.12.041024} {\bibfield  {journal} {\bibinfo
  {journal} {Phys. Rev. X}\ }\textbf {\bibinfo {volume} {12}},\ \bibinfo
  {pages} {041024} (\bibinfo {year} {2022})}\BibitemShut {NoStop}%
\bibitem [{\citenamefont {Gu}\ \emph {et~al.}(2024)\citenamefont {Gu},
  \citenamefont {Li}, \citenamefont {Chen}, \citenamefont {Iida}, \citenamefont
  {Nakao}, \citenamefont {Munakata}, \citenamefont {Garlea}, \citenamefont
  {Li}, \citenamefont {Deng}, \citenamefont {Zaliznyak}, \citenamefont
  {Tranquada},\ and\ \citenamefont {Li}}]{gu24}%
  \BibitemOpen
  \bibfield  {author} {\bibinfo {author} {\bibfnamefont {Y.}~\bibnamefont
  {Gu}}, \bibinfo {author} {\bibfnamefont {X.}~\bibnamefont {Li}}, \bibinfo
  {author} {\bibfnamefont {Y.}~\bibnamefont {Chen}}, \bibinfo {author}
  {\bibfnamefont {K.}~\bibnamefont {Iida}}, \bibinfo {author} {\bibfnamefont
  {A.}~\bibnamefont {Nakao}}, \bibinfo {author} {\bibfnamefont
  {K.}~\bibnamefont {Munakata}}, \bibinfo {author} {\bibfnamefont {V.~O.}\
  \bibnamefont {Garlea}}, \bibinfo {author} {\bibfnamefont {Y.}~\bibnamefont
  {Li}}, \bibinfo {author} {\bibfnamefont {G.}~\bibnamefont {Deng}}, \bibinfo
  {author} {\bibfnamefont {I.~A.}\ \bibnamefont {Zaliznyak}}, \bibinfo {author}
  {\bibfnamefont {J.~M.}\ \bibnamefont {Tranquada}},\ and\ \bibinfo {author}
  {\bibfnamefont {Y.}~\bibnamefont {Li}},\ }\bibfield  {title} {\bibinfo
  {title} {In-plane multi-$\mathrm{q}$ magnetic ground state of
  ${\mathrm{Na}}_{3}{\mathrm{Co}}_{2}{\mathrm{SbO}}_{6}$},\ }\href
  {https://doi.org/10.1103/PhysRevB.109.L060410} {\bibfield  {journal}
  {\bibinfo  {journal} {Phys. Rev. B}\ }\textbf {\bibinfo {volume} {109}},\
  \bibinfo {pages} {L060410} (\bibinfo {year} {2024})}\BibitemShut {NoStop}%
\bibitem [{\citenamefont {Hu}\ \emph {et~al.}(2024)\citenamefont {Hu},
  \citenamefont {Chen}, \citenamefont {Cui}, \citenamefont {Li}, \citenamefont
  {Li}, \citenamefont {Xu}, \citenamefont {Chen}, \citenamefont {Li},
  \citenamefont {Gu}, \citenamefont {Yu}, \citenamefont {Zhou}, \citenamefont
  {Li},\ and\ \citenamefont {Yu}}]{hu24}%
  \BibitemOpen
  \bibfield  {author} {\bibinfo {author} {\bibfnamefont {Z.}~\bibnamefont
  {Hu}}, \bibinfo {author} {\bibfnamefont {Y.}~\bibnamefont {Chen}}, \bibinfo
  {author} {\bibfnamefont {Y.}~\bibnamefont {Cui}}, \bibinfo {author}
  {\bibfnamefont {S.}~\bibnamefont {Li}}, \bibinfo {author} {\bibfnamefont
  {C.}~\bibnamefont {Li}}, \bibinfo {author} {\bibfnamefont {X.}~\bibnamefont
  {Xu}}, \bibinfo {author} {\bibfnamefont {Y.}~\bibnamefont {Chen}}, \bibinfo
  {author} {\bibfnamefont {X.}~\bibnamefont {Li}}, \bibinfo {author}
  {\bibfnamefont {Y.}~\bibnamefont {Gu}}, \bibinfo {author} {\bibfnamefont
  {R.}~\bibnamefont {Yu}}, \bibinfo {author} {\bibfnamefont {R.}~\bibnamefont
  {Zhou}}, \bibinfo {author} {\bibfnamefont {Y.}~\bibnamefont {Li}},\ and\
  \bibinfo {author} {\bibfnamefont {W.}~\bibnamefont {Yu}},\ }\bibfield
  {title} {\bibinfo {title} {Field-induced phase transitions and quantum
  criticality in the honeycomb antiferromagnet
  ${\mathrm{Na}}_{3}{\mathrm{Co}}_{2}{\mathrm{SbO}}_{6}$},\ }\href
  {https://doi.org/10.1103/PhysRevB.109.054411} {\bibfield  {journal} {\bibinfo
   {journal} {Phys. Rev. B}\ }\textbf {\bibinfo {volume} {109}},\ \bibinfo
  {pages} {054411} (\bibinfo {year} {2024})}\BibitemShut {NoStop}%
\bibitem [{\citenamefont {Zhong}\ \emph {et~al.}(2020)\citenamefont {Zhong},
  \citenamefont {Gao}, \citenamefont {Ong},\ and\ \citenamefont
  {Cava}}]{zhong20}%
  \BibitemOpen
  \bibfield  {author} {\bibinfo {author} {\bibfnamefont {R.}~\bibnamefont
  {Zhong}}, \bibinfo {author} {\bibfnamefont {T.}~\bibnamefont {Gao}}, \bibinfo
  {author} {\bibfnamefont {N.~P.}\ \bibnamefont {Ong}},\ and\ \bibinfo {author}
  {\bibfnamefont {R.~J.}\ \bibnamefont {Cava}},\ }\bibfield  {title} {\bibinfo
  {title} {Weak-field induced nonmagnetic state in a Co-based honeycomb},\
  }\href {https://doi.org/10.1126/sciadv.aay6953} {\bibfield  {journal}
  {\bibinfo  {journal} {Sci. Adv.}\ }\textbf {\bibinfo {volume} {6}},\ \bibinfo
  {pages} {eaay6953} (\bibinfo {year} {2020})}\BibitemShut {NoStop}%
\bibitem [{\citenamefont {Shi}\ \emph {et~al.}(2021)\citenamefont {Shi},
  \citenamefont {Wang}, \citenamefont {Zhong}, \citenamefont {Wang},
  \citenamefont {Hu}, \citenamefont {Zhang}, \citenamefont {Liu}, \citenamefont
  {Dong}, \citenamefont {Wang},\ and\ \citenamefont {Wang}}]{shi21}%
  \BibitemOpen
  \bibfield  {author} {\bibinfo {author} {\bibfnamefont {L.~Y.}\ \bibnamefont
  {Shi}}, \bibinfo {author} {\bibfnamefont {X.~M.}\ \bibnamefont {Wang}},
  \bibinfo {author} {\bibfnamefont {R.~D.}\ \bibnamefont {Zhong}}, \bibinfo
  {author} {\bibfnamefont {Z.~X.}\ \bibnamefont {Wang}}, \bibinfo {author}
  {\bibfnamefont {T.~C.}\ \bibnamefont {Hu}}, \bibinfo {author} {\bibfnamefont
  {S.~J.}\ \bibnamefont {Zhang}}, \bibinfo {author} {\bibfnamefont {Q.~M.}\
  \bibnamefont {Liu}}, \bibinfo {author} {\bibfnamefont {T.}~\bibnamefont
  {Dong}}, \bibinfo {author} {\bibfnamefont {F.}~\bibnamefont {Wang}},\ and\
  \bibinfo {author} {\bibfnamefont {N.~L.}\ \bibnamefont {Wang}},\ }\bibfield
  {title} {\bibinfo {title} {Magnetic excitations of the field-induced states
  in ${\mathrm{BaCo}}_{2}{({\mathrm{AsO}}_{4})}_{2}$ probed by time-domain
  terahertz spectroscopy},\ }\href
  {https://doi.org/10.1103/PhysRevB.104.144408} {\bibfield  {journal} {\bibinfo
   {journal} {Phys. Rev. B}\ }\textbf {\bibinfo {volume} {104}},\ \bibinfo
  {pages} {144408} (\bibinfo {year} {2021})}\BibitemShut {NoStop}%
\bibitem [{\citenamefont {Zhang}\ \emph
  {et~al.}(2023{\natexlab{b}})\citenamefont {Zhang}, \citenamefont {Xu},
  \citenamefont {Halloran}, \citenamefont {Zhong}, \citenamefont {Broholm},
  \citenamefont {Cava}, \citenamefont {Drichko},\ and\ \citenamefont
  {Armitage}}]{zhang22}%
  \BibitemOpen
  \bibfield  {author} {\bibinfo {author} {\bibfnamefont {X.}~\bibnamefont
  {Zhang}}, \bibinfo {author} {\bibfnamefont {Y.}~\bibnamefont {Xu}}, \bibinfo
  {author} {\bibfnamefont {T.}~\bibnamefont {Halloran}}, \bibinfo {author}
  {\bibfnamefont {R.}~\bibnamefont {Zhong}}, \bibinfo {author} {\bibfnamefont
  {C.}~\bibnamefont {Broholm}}, \bibinfo {author} {\bibfnamefont {R.~J.}\
  \bibnamefont {Cava}}, \bibinfo {author} {\bibfnamefont {N.}~\bibnamefont
  {Drichko}},\ and\ \bibinfo {author} {\bibfnamefont {N.~P.}\ \bibnamefont
  {Armitage}},\ }\bibfield  {title} {\bibinfo {title} {A magnetic continuum in
  the cobalt-based honeycomb magnet BaCo$_2$(AsO$_4$)$_2$},\ }\href
  {https://doi.org/10.1038/s41563-022-01403-1} {\bibfield  {journal} {\bibinfo
  {journal} {Nat. Mat.}\ }\textbf {\bibinfo {volume} {22}},\ \bibinfo {pages}
  {58} (\bibinfo {year} {2023}{\natexlab{b}})}\BibitemShut {NoStop}%
\bibitem [{\citenamefont {Maksimov}\ \emph {et~al.}(2022)\citenamefont
  {Maksimov}, \citenamefont {Ushakov}, \citenamefont {Pchelkina}, \citenamefont
  {Li}, \citenamefont {Winter},\ and\ \citenamefont {Streltsov}}]{maksimov22b}%
  \BibitemOpen
  \bibfield  {author} {\bibinfo {author} {\bibfnamefont {P.~A.}\ \bibnamefont
  {Maksimov}}, \bibinfo {author} {\bibfnamefont {A.~V.}\ \bibnamefont
  {Ushakov}}, \bibinfo {author} {\bibfnamefont {Z.~V.}\ \bibnamefont
  {Pchelkina}}, \bibinfo {author} {\bibfnamefont {Y.}~\bibnamefont {Li}},
  \bibinfo {author} {\bibfnamefont {S.~M.}\ \bibnamefont {Winter}},\ and\
  \bibinfo {author} {\bibfnamefont {S.~V.}\ \bibnamefont {Streltsov}},\
  }\bibfield  {title} {\bibinfo {title} {Ab initio guided minimal model for the
  ``Kitaev'' material ${\mathrm{BaCo}}_{2}$(${\mathrm{AsO}}_{4}{)}_{2}$:
  Importance of direct hopping, third-neighbor exchange, and quantum
  fluctuations},\ }\href {https://doi.org/10.1103/PhysRevB.106.165131}
  {\bibfield  {journal} {\bibinfo  {journal} {Phys. Rev. B}\ }\textbf {\bibinfo
  {volume} {106}},\ \bibinfo {pages} {165131} (\bibinfo {year}
  {2022})}\BibitemShut {NoStop}%
\bibitem [{\citenamefont {Maksimov}\ \emph {et~al.}(2025)\citenamefont
  {Maksimov}, \citenamefont {Jiang}, \citenamefont {Regnault},\ and\
  \citenamefont {Chernyshev}}]{maksimov25}%
  \BibitemOpen
  \bibfield  {author} {\bibinfo {author} {\bibfnamefont {P.~A.}\ \bibnamefont
  {Maksimov}}, \bibinfo {author} {\bibfnamefont {S.}~\bibnamefont {Jiang}},
  \bibinfo {author} {\bibfnamefont {L.~P.}\ \bibnamefont {Regnault}},\ and\
  \bibinfo {author} {\bibfnamefont {A.~L.}\ \bibnamefont {Chernyshev}},\
  }\bibinfo {title} {BaCo$_2$(AsO$_4$)$_2$: Strong Kitaev, After All},\ \Eprint
  {https://arxiv.org/abs/2503.20859} {arXiv:2503.20859}\BibitemShut {NoStop}%
\bibitem [{\citenamefont {Liu}\ and\ \citenamefont {Khaliullin}(2018)}]{liu18}%
  \BibitemOpen
  \bibfield  {author} {\bibinfo {author} {\bibfnamefont {H.}~\bibnamefont
  {Liu}}\ and\ \bibinfo {author} {\bibfnamefont {G.}~\bibnamefont
  {Khaliullin}},\ }\bibfield  {title} {\bibinfo {title} {Pseudospin exchange
  interactions in ${d}^{7}$ cobalt compounds: Possible realization of the
  Kitaev model},\ }\href {https://doi.org/10.1103/PhysRevB.97.014407}
  {\bibfield  {journal} {\bibinfo  {journal} {Phys. Rev. B}\ }\textbf {\bibinfo
  {volume} {97}},\ \bibinfo {pages} {014407} (\bibinfo {year}
  {2018})}\BibitemShut {NoStop}%
\bibitem [{\citenamefont {Sano}\ \emph {et~al.}(2018)\citenamefont {Sano},
  \citenamefont {Kato},\ and\ \citenamefont {Motome}}]{sano18}%
  \BibitemOpen
  \bibfield  {author} {\bibinfo {author} {\bibfnamefont {R.}~\bibnamefont
  {Sano}}, \bibinfo {author} {\bibfnamefont {Y.}~\bibnamefont {Kato}},\ and\
  \bibinfo {author} {\bibfnamefont {Y.}~\bibnamefont {Motome}},\ }\bibfield
  {title} {\bibinfo {title} {Kitaev-Heisenberg Hamiltonian for high-spin
  ${d}^{7}$ Mott insulators},\ }\href
  {https://doi.org/10.1103/PhysRevB.97.014408} {\bibfield  {journal} {\bibinfo
  {journal} {Phys. Rev. B}\ }\textbf {\bibinfo {volume} {97}},\ \bibinfo
  {pages} {014408} (\bibinfo {year} {2018})}\BibitemShut {NoStop}%
\bibitem [{\citenamefont {Liu}\ \emph {et~al.}(2020)\citenamefont {Liu},
  \citenamefont {Chaloupka},\ and\ \citenamefont {Khaliullin}}]{liu20}%
  \BibitemOpen
  \bibfield  {author} {\bibinfo {author} {\bibfnamefont {H.}~\bibnamefont
  {Liu}}, \bibinfo {author} {\bibfnamefont {J.}~\bibnamefont {Chaloupka}},\
  and\ \bibinfo {author} {\bibfnamefont {G.}~\bibnamefont {Khaliullin}},\
  }\bibfield  {title} {\bibinfo {title} {Kitaev Spin Liquid in $3d$ Transition
  Metal Compounds},\ }\href {https://doi.org/10.1103/PhysRevLett.125.047201}
  {\bibfield  {journal} {\bibinfo  {journal} {Phys. Rev. Lett.}\ }\textbf
  {\bibinfo {volume} {125}},\ \bibinfo {pages} {047201} (\bibinfo {year}
  {2020})}\BibitemShut {NoStop}%
\bibitem [{\citenamefont {Winter}(2022)}]{winter22}%
  \BibitemOpen
  \bibfield  {author} {\bibinfo {author} {\bibfnamefont {S.~M.}\ \bibnamefont
  {Winter}},\ }\bibfield  {title} {\bibinfo {title} {Magnetic couplings in
  edge-sharing high-spin $d^7$ compounds},\ }\href
  {https://doi.org/10.1088/2515-7639/ac94f8} {\bibfield  {journal} {\bibinfo
  {journal} {J. Phys. Mat.}\ }\textbf {\bibinfo {volume} {5}},\ \bibinfo
  {pages} {045003} (\bibinfo {year} {2022})}\BibitemShut {NoStop}%
\bibitem [{\citenamefont {Rousochatzakis}\ \emph {et~al.}(2024)\citenamefont
  {Rousochatzakis}, \citenamefont {Perkins}, \citenamefont {Luo},\ and\
  \citenamefont {Kee}}]{rousochatzakis24}%
  \BibitemOpen
  \bibfield  {author} {\bibinfo {author} {\bibfnamefont {I.}~\bibnamefont
  {Rousochatzakis}}, \bibinfo {author} {\bibfnamefont {N.~B.}\ \bibnamefont
  {Perkins}}, \bibinfo {author} {\bibfnamefont {Q.}~\bibnamefont {Luo}},\ and\
  \bibinfo {author} {\bibfnamefont {H.-Y.}\ \bibnamefont {Kee}},\ }\bibfield
  {title} {\bibinfo {title} {Beyond Kitaev physics in strong spin-orbit coupled
  magnets},\ }\href {https://doi.org/10.1088/1361-6633/ad208d} {\bibfield
  {journal} {\bibinfo  {journal} {Rep. Prog. Phys.}\ }\textbf {\bibinfo
  {volume} {87}},\ \bibinfo {pages} {026502} (\bibinfo {year}
  {2024})}\BibitemShut {NoStop}%
\bibitem [{\citenamefont {Trebst}\ and\ \citenamefont
  {Hickey}(2022)}]{trebst22}%
  \BibitemOpen
  \bibfield  {author} {\bibinfo {author} {\bibfnamefont {S.}~\bibnamefont
  {Trebst}}\ and\ \bibinfo {author} {\bibfnamefont {C.}~\bibnamefont
  {Hickey}},\ }\bibfield  {title} {\bibinfo {title} {Kitaev materials},\ }\href
  {https://doi.org/https://doi.org/10.1016/j.physrep.2021.11.003} {\bibfield
  {journal} {\bibinfo  {journal} {Phys. Rep.}\ }\textbf {\bibinfo {volume}
  {950}},\ \bibinfo {pages} {1} (\bibinfo {year} {2022})}\BibitemShut {NoStop}%
\bibitem [{\citenamefont {Gu}\ \emph {et~al.}(2025)\citenamefont {Gu},
  \citenamefont {Jin},\ and\ \citenamefont {Li}}]{gu25}%
  \BibitemOpen
  \bibfield  {author} {\bibinfo {author} {\bibfnamefont {Y.}~\bibnamefont
  {Gu}}, \bibinfo {author} {\bibfnamefont {X.}~\bibnamefont {Jin}},\ and\
  \bibinfo {author} {\bibfnamefont {Y.}~\bibnamefont {Li}},\ }\bibfield
  {title} {\bibinfo {title} {On the multi-q characteristics of magnetic ground
  states of honeycomb cobalt oxides},\ }\href
  {http://iopscience.iop.org/article/10.1088/0256-307X/42/2/027303} {\bibfield
  {journal} {\bibinfo  {journal} {Chinese Physics Letters}\ } (\bibinfo {year}
  {2025})}\BibitemShut {NoStop}%
\bibitem [{\citenamefont {Kocsis}\ \emph {et~al.}()\citenamefont {Kocsis},
  \citenamefont {Luther}, \citenamefont {Pérez}, \citenamefont {Yao},
  \citenamefont {Kühne}, \citenamefont {Wolter}, \citenamefont {Li},\ and\
  \citenamefont {Büchner}}]{kocsis24}%
  \BibitemOpen
  \bibfield  {author} {\bibinfo {author} {\bibfnamefont {V.}~\bibnamefont
  {Kocsis}}, \bibinfo {author} {\bibfnamefont {S.}~\bibnamefont {Luther}},
  \bibinfo {author} {\bibfnamefont {N.}~\bibnamefont {Pérez}}, \bibinfo
  {author} {\bibfnamefont {W.}~\bibnamefont {Yao}}, \bibinfo {author}
  {\bibfnamefont {H.}~\bibnamefont {Kühne}}, \bibinfo {author} {\bibfnamefont
  {A.~U.}\ \bibnamefont {Wolter}}, \bibinfo {author} {\bibfnamefont
  {Y.}~\bibnamefont {Li}},\ and\ \bibinfo {author} {\bibfnamefont
  {B.}~\bibnamefont {Büchner}},\ }\bibinfo {title} {Magnetoelectric and
  magnetoelastic couplings in the quantum spin liquid candidate
  Na$_2$Co$_2$TeO$_6$},\ \bibinfo {note} {{Talk at APS March Meeting 2024,
  Minneapolis,
  \href{https://meetings.aps.org/Meeting/MAR24/Session/F22.3}{https://meetings.aps.org/Meeting/MAR24/Session/F22.3}}}\BibitemShut
  {NoStop}%
\bibitem [{\citenamefont {Jin}\ \emph {et~al.}(2025)\citenamefont {Jin},
  \citenamefont {Geng}, \citenamefont {Orlandi}, \citenamefont {Khalyavin},
  \citenamefont {Manuel}, \citenamefont {Liu},\ and\ \citenamefont
  {Li}}]{jin25}%
  \BibitemOpen
  \bibfield  {author} {\bibinfo {author} {\bibfnamefont {X.}~\bibnamefont
  {Jin}}, \bibinfo {author} {\bibfnamefont {M.}~\bibnamefont {Geng}}, \bibinfo
  {author} {\bibfnamefont {F.}~\bibnamefont {Orlandi}}, \bibinfo {author}
  {\bibfnamefont {D.}~\bibnamefont {Khalyavin}}, \bibinfo {author}
  {\bibfnamefont {P.}~\bibnamefont {Manuel}}, \bibinfo {author} {\bibfnamefont
  {Y.}~\bibnamefont {Liu}},\ and\ \bibinfo {author} {\bibfnamefont
  {Y.}~\bibnamefont {Li}},\ }\bibinfo {title} {Robust triple-q magnetic order
  with trainable spin vorticity in Na$_2$Co$_2$TeO$_6$},\ \Eprint
  {https://arxiv.org/abs/2501.07843} {arXiv:2501.07843}\BibitemShut {NoStop}%
\bibitem [{\citenamefont {Lefran\ifmmode~\mbox{\c{c}}\else \c{c}\fi{}ois}\
  \emph {et~al.}(2016)\citenamefont {Lefran\ifmmode~\mbox{\c{c}}\else
  \c{c}\fi{}ois}, \citenamefont {Songvilay}, \citenamefont {Robert},
  \citenamefont {Nataf}, \citenamefont {Jordan}, \citenamefont {Chaix},
  \citenamefont {Colin}, \citenamefont {Lejay}, \citenamefont {Hadj-Azzem},
  \citenamefont {Ballou},\ and\ \citenamefont {Simonet}}]{lefrancois16}%
  \BibitemOpen
  \bibfield  {author} {\bibinfo {author} {\bibfnamefont {E.}~\bibnamefont
  {Lefran\ifmmode~\mbox{\c{c}}\else \c{c}\fi{}ois}}, \bibinfo {author}
  {\bibfnamefont {M.}~\bibnamefont {Songvilay}}, \bibinfo {author}
  {\bibfnamefont {J.}~\bibnamefont {Robert}}, \bibinfo {author} {\bibfnamefont
  {G.}~\bibnamefont {Nataf}}, \bibinfo {author} {\bibfnamefont
  {E.}~\bibnamefont {Jordan}}, \bibinfo {author} {\bibfnamefont
  {L.}~\bibnamefont {Chaix}}, \bibinfo {author} {\bibfnamefont {C.~V.}\
  \bibnamefont {Colin}}, \bibinfo {author} {\bibfnamefont {P.}~\bibnamefont
  {Lejay}}, \bibinfo {author} {\bibfnamefont {A.}~\bibnamefont {Hadj-Azzem}},
  \bibinfo {author} {\bibfnamefont {R.}~\bibnamefont {Ballou}},\ and\ \bibinfo
  {author} {\bibfnamefont {V.}~\bibnamefont {Simonet}},\ }\bibfield  {title}
  {\bibinfo {title} {Magnetic properties of the honeycomb oxide
  ${\mathrm{Na}}_{2}{\mathrm{Co}}_{2}{\mathrm{TeO}}_{6}$},\ }\href
  {https://doi.org/10.1103/PhysRevB.94.214416} {\bibfield  {journal} {\bibinfo
  {journal} {Phys. Rev. B}\ }\textbf {\bibinfo {volume} {94}},\ \bibinfo
  {pages} {214416} (\bibinfo {year} {2016})}\BibitemShut {NoStop}%
\bibitem [{\citenamefont {Francini}\ and\ \citenamefont
  {Janssen}(2024{\natexlab{a}})}]{francini24ferri}%
  \BibitemOpen
  \bibfield  {author} {\bibinfo {author} {\bibfnamefont {N.}~\bibnamefont
  {Francini}}\ and\ \bibinfo {author} {\bibfnamefont {L.}~\bibnamefont
  {Janssen}},\ }\bibfield  {title} {\bibinfo {title} {Ferrimagnetism from
  triple-$\mathrm{q}$ order in
  ${\mathrm{Na}}_{2}{\mathrm{Co}}_{2}{\mathrm{TeO}}_{6}$},\ }\href
  {https://doi.org/10.1103/PhysRevB.110.235118} {\bibfield  {journal} {\bibinfo
   {journal} {Phys. Rev. B}\ }\textbf {\bibinfo {volume} {110}},\ \bibinfo
  {pages} {235118} (\bibinfo {year} {2024}{\natexlab{a}})}\BibitemShut
  {NoStop}%
\bibitem [{\citenamefont {Bradley}\ \emph {et~al.}(1972)\citenamefont
  {Bradley}, \citenamefont {Bradley},\ and\ \citenamefont
  {Cracknell}}]{bradley_book}%
  \BibitemOpen
  \bibfield  {author} {\bibinfo {author} {\bibfnamefont {T.~W.}\ \bibnamefont
  {Bradley}}, \bibinfo {author} {\bibfnamefont {C.~J.}\ \bibnamefont
  {Bradley}},\ and\ \bibinfo {author} {\bibfnamefont {A.~P.}\ \bibnamefont
  {Cracknell}},\ }\href@noop {} {\emph {\bibinfo {title} {{The Mathematical
  Theory of Symmetry in Solids: Representation Theory for Point Groups and
  Space Groups}}}}\ (\bibinfo  {publisher} {Clarendon Press},\ \bibinfo
  {address} {Oxford},\ \bibinfo {year} {1972})\BibitemShut {NoStop}%
\bibitem [{\citenamefont {Corticelli}\ \emph {et~al.}(2022)\citenamefont
  {Corticelli}, \citenamefont {Moessner},\ and\ \citenamefont
  {McClarty}}]{corticelli22}%
  \BibitemOpen
  \bibfield  {author} {\bibinfo {author} {\bibfnamefont {A.}~\bibnamefont
  {Corticelli}}, \bibinfo {author} {\bibfnamefont {R.}~\bibnamefont
  {Moessner}},\ and\ \bibinfo {author} {\bibfnamefont {P.~A.}\ \bibnamefont
  {McClarty}},\ }\bibfield  {title} {\bibinfo {title} {{Spin-space groups and
  magnon band topology}},\ }\href {https://doi.org/10.1103/PhysRevB.105.064430}
  {\bibfield  {journal} {\bibinfo  {journal} {Phys. Rev. B}\ }\textbf {\bibinfo
  {volume} {105}},\ \bibinfo {pages} {064430} (\bibinfo {year}
  {2022})}\BibitemShut {NoStop}%
\bibitem [{\citenamefont {\ifmmode~\check{S}\else \v{S}\fi{}mejkal}\ \emph
  {et~al.}(2022)\citenamefont {\ifmmode~\check{S}\else \v{S}\fi{}mejkal},
  \citenamefont {Sinova},\ and\ \citenamefont {Jungwirth}}]{smejkal22}%
  \BibitemOpen
  \bibfield  {author} {\bibinfo {author} {\bibfnamefont {L.}~\bibnamefont
  {\ifmmode~\check{S}\else \v{S}\fi{}mejkal}}, \bibinfo {author} {\bibfnamefont
  {J.}~\bibnamefont {Sinova}},\ and\ \bibinfo {author} {\bibfnamefont
  {T.}~\bibnamefont {Jungwirth}},\ }\bibfield  {title} {\bibinfo {title}
  {{Beyond Conventional Ferromagnetism and Antiferromagnetism: A Phase with
  Nonrelativistic Spin and Crystal Rotation Symmetry}},\ }\href
  {https://doi.org/10.1103/PhysRevX.12.031042} {\bibfield  {journal} {\bibinfo
  {journal} {Phys. Rev. X}\ }\textbf {\bibinfo {volume} {12}},\ \bibinfo
  {pages} {031042} (\bibinfo {year} {2022})}\BibitemShut {NoStop}%
\bibitem [{\citenamefont {Brinkman}\ and\ \citenamefont
  {Elliott}(1966)}]{brinkman66}%
  \BibitemOpen
  \bibfield  {author} {\bibinfo {author} {\bibfnamefont {W.~F.}\ \bibnamefont
  {Brinkman}}\ and\ \bibinfo {author} {\bibfnamefont {R.~J.}\ \bibnamefont
  {Elliott}},\ }\bibfield  {title} {\bibinfo {title} {{Theory of spin-space
  groups}},\ }\href {https://doi.org/10.1098/rspa.1966.0211} {\bibfield
  {journal} {\bibinfo  {journal} {Proc. R. Soc. London A}\ }\textbf {\bibinfo
  {volume} {294}},\ \bibinfo {pages} {343} (\bibinfo {year}
  {1966})}\BibitemShut {NoStop}%
\bibitem [{\citenamefont {Litvin}\ and\ \citenamefont
  {Opechowski}(1974)}]{litvin74}%
  \BibitemOpen
  \bibfield  {author} {\bibinfo {author} {\bibfnamefont {D.}~\bibnamefont
  {Litvin}}\ and\ \bibinfo {author} {\bibfnamefont {W.}~\bibnamefont
  {Opechowski}},\ }\bibfield  {title} {\bibinfo {title} {{Spin groups}},\
  }\href {https://doi.org/https://doi.org/10.1016/0031-8914(74)90157-8}
  {\bibfield  {journal} {\bibinfo  {journal} {Physica}\ }\textbf {\bibinfo
  {volume} {76}},\ \bibinfo {pages} {538} (\bibinfo {year} {1974})}\BibitemShut
  {NoStop}%
\bibitem [{\citenamefont {Litvin}(1977)}]{litvin77}%
  \BibitemOpen
  \bibfield  {author} {\bibinfo {author} {\bibfnamefont {D.~B.}\ \bibnamefont
  {Litvin}},\ }\bibfield  {title} {\bibinfo {title} {{Spin point groups}},\
  }\href {https://doi.org/10.1107/S0567739477000709} {\bibfield  {journal}
  {\bibinfo  {journal} {Acta Cryst.}\ }\textbf {\bibinfo {volume} {A33}},\
  \bibinfo {pages} {279} (\bibinfo {year} {1977})}\BibitemShut {NoStop}%
\bibitem [{\citenamefont {Francini}\ and\ \citenamefont
  {Janssen}(2024{\natexlab{b}})}]{francini24vestigial}%
  \BibitemOpen
  \bibfield  {author} {\bibinfo {author} {\bibfnamefont {N.}~\bibnamefont
  {Francini}}\ and\ \bibinfo {author} {\bibfnamefont {L.}~\bibnamefont
  {Janssen}},\ }\bibfield  {title} {\bibinfo {title} {Spin vestigial orders in
  extended Heisenberg-Kitaev models near hidden SU(2) points: Application to
  ${\mathrm{Na}}_{2}{\mathrm{Co}}_{2}{\mathrm{TeO}}_{6}$},\ }\href
  {https://doi.org/10.1103/PhysRevB.109.075104} {\bibfield  {journal} {\bibinfo
   {journal} {Phys. Rev. B}\ }\textbf {\bibinfo {volume} {109}},\ \bibinfo
  {pages} {075104} (\bibinfo {year} {2024}{\natexlab{b}})}\BibitemShut
  {NoStop}%
\bibitem [{\citenamefont {Holstein}\ and\ \citenamefont
  {Primakoff}(1940)}]{holstein40}%
  \BibitemOpen
  \bibfield  {author} {\bibinfo {author} {\bibfnamefont {T.}~\bibnamefont
  {Holstein}}\ and\ \bibinfo {author} {\bibfnamefont {H.}~\bibnamefont
  {Primakoff}},\ }\bibfield  {title} {\bibinfo {title} {Field Dependence of the
  Intrinsic Domain Magnetization of a Ferromagnet},\ }\href
  {https://doi.org/10.1103/PhysRev.58.1098} {\bibfield  {journal} {\bibinfo
  {journal} {Phys. Rev.}\ }\textbf {\bibinfo {volume} {58}},\ \bibinfo {pages}
  {1098} (\bibinfo {year} {1940})}\BibitemShut {NoStop}%
\bibitem [{\citenamefont {Rau}\ \emph {et~al.}(2014)\citenamefont {Rau},
  \citenamefont {Lee},\ and\ \citenamefont {Kee}}]{rau14a}%
  \BibitemOpen
  \bibfield  {author} {\bibinfo {author} {\bibfnamefont {J.~G.}\ \bibnamefont
  {Rau}}, \bibinfo {author} {\bibfnamefont {E.~K.-H.}\ \bibnamefont {Lee}},\
  and\ \bibinfo {author} {\bibfnamefont {H.-Y.}\ \bibnamefont {Kee}},\
  }\bibfield  {title} {\bibinfo {title} {Generic Spin Model for the Honeycomb
  Iridates beyond the Kitaev Limit},\ }\href
  {https://doi.org/10.1103/PhysRevLett.112.077204} {\bibfield  {journal}
  {\bibinfo  {journal} {Phys. Rev. Lett.}\ }\textbf {\bibinfo {volume} {112}},\
  \bibinfo {pages} {077204} (\bibinfo {year} {2014})}\BibitemShut {NoStop}%
\bibitem [{\citenamefont {Zhitomirsky}\ and\ \citenamefont
  {Nikuni}(1998)}]{zhitomirsky98}%
  \BibitemOpen
  \bibfield  {author} {\bibinfo {author} {\bibfnamefont {M.~E.}\ \bibnamefont
  {Zhitomirsky}}\ and\ \bibinfo {author} {\bibfnamefont {T.}~\bibnamefont
  {Nikuni}},\ }\bibfield  {title} {\bibinfo {title} {{Magnetization curve of a
  square-lattice Heisenberg antiferromagnet}},\ }\href
  {https://doi.org/10.1103/PhysRevB.57.5013} {\bibfield  {journal} {\bibinfo
  {journal} {Phys. Rev. B}\ }\textbf {\bibinfo {volume} {57}},\ \bibinfo
  {pages} {5013} (\bibinfo {year} {1998})}\BibitemShut {NoStop}%
\bibitem [{\citenamefont {Coletta}\ \emph {et~al.}(2012)\citenamefont
  {Coletta}, \citenamefont {Laflorencie},\ and\ \citenamefont
  {Mila}}]{coletta12}%
  \BibitemOpen
  \bibfield  {author} {\bibinfo {author} {\bibfnamefont {T.}~\bibnamefont
  {Coletta}}, \bibinfo {author} {\bibfnamefont {N.}~\bibnamefont
  {Laflorencie}},\ and\ \bibinfo {author} {\bibfnamefont {F.}~\bibnamefont
  {Mila}},\ }\bibfield  {title} {\bibinfo {title} {{Semiclassical approach to
  ground-state properties of hard-core bosons in two dimensions}},\ }\href
  {https://doi.org/10.1103/PhysRevB.85.104421} {\bibfield  {journal} {\bibinfo
  {journal} {Phys. Rev. B}\ }\textbf {\bibinfo {volume} {85}},\ \bibinfo
  {pages} {104421} (\bibinfo {year} {2012})}\BibitemShut {NoStop}%
\bibitem [{\citenamefont {C\^onsoli}\ \emph {et~al.}(2020)\citenamefont
  {C\^onsoli}, \citenamefont {Janssen}, \citenamefont {Vojta},\ and\
  \citenamefont {Andrade}}]{consoli20}%
  \BibitemOpen
  \bibfield  {author} {\bibinfo {author} {\bibfnamefont {P.~M.}\ \bibnamefont
  {C\^onsoli}}, \bibinfo {author} {\bibfnamefont {L.}~\bibnamefont {Janssen}},
  \bibinfo {author} {\bibfnamefont {M.}~\bibnamefont {Vojta}},\ and\ \bibinfo
  {author} {\bibfnamefont {E.~C.}\ \bibnamefont {Andrade}},\ }\bibfield
  {title} {\bibinfo {title} {Heisenberg-Kitaev model in a magnetic field: $1/S$
  expansion},\ }\href {https://doi.org/10.1103/PhysRevB.102.155134} {\bibfield
  {journal} {\bibinfo  {journal} {Phys. Rev. B}\ }\textbf {\bibinfo {volume}
  {102}},\ \bibinfo {pages} {155134} (\bibinfo {year} {2020})}\BibitemShut
  {NoStop}%
\bibitem [{\citenamefont {Chernyshev}\ and\ \citenamefont
  {Zhitomirsky}(2009)}]{chernyshev09}%
  \BibitemOpen
  \bibfield  {author} {\bibinfo {author} {\bibfnamefont {A.~L.}\ \bibnamefont
  {Chernyshev}}\ and\ \bibinfo {author} {\bibfnamefont {M.~E.}\ \bibnamefont
  {Zhitomirsky}},\ }\bibfield  {title} {\bibinfo {title} {Spin waves in a
  triangular lattice antiferromagnet: Decays, spectrum renormalization, and
  singularities},\ }\href {https://doi.org/10.1103/PhysRevB.79.144416}
  {\bibfield  {journal} {\bibinfo  {journal} {Phys. Rev. B}\ }\textbf {\bibinfo
  {volume} {79}},\ \bibinfo {pages} {144416} (\bibinfo {year}
  {2009})}\BibitemShut {NoStop}%
\bibitem [{\citenamefont {Janssen}\ and\ \citenamefont
  {Vojta}(2019)}]{janssen19}%
  \BibitemOpen
  \bibfield  {author} {\bibinfo {author} {\bibfnamefont {L.}~\bibnamefont
  {Janssen}}\ and\ \bibinfo {author} {\bibfnamefont {M.}~\bibnamefont
  {Vojta}},\ }\bibfield  {title} {\bibinfo {title} {Heisenberg-Kitaev physics
  in magnetic fields},\ }\href {https://doi.org/10.1088/1361-648x/ab283e}
  {\bibfield  {journal} {\bibinfo  {journal} {J. Phys. Condens. Matter}\
  }\textbf {\bibinfo {volume} {31}},\ \bibinfo {pages} {423002} (\bibinfo
  {year} {2019})}\BibitemShut {NoStop}%
\bibitem [{\citenamefont {Price}\ and\ \citenamefont
  {Perkins}(2012)}]{price12}%
  \BibitemOpen
  \bibfield  {author} {\bibinfo {author} {\bibfnamefont {C.~C.}\ \bibnamefont
  {Price}}\ and\ \bibinfo {author} {\bibfnamefont {N.~B.}\ \bibnamefont
  {Perkins}},\ }\bibfield  {title} {\bibinfo {title} {Critical Properties of
  the Kitaev-Heisenberg Model},\ }\href
  {https://doi.org/10.1103/PhysRevLett.109.187201} {\bibfield  {journal}
  {\bibinfo  {journal} {Phys. Rev. Lett.}\ }\textbf {\bibinfo {volume} {109}},\
  \bibinfo {pages} {187201} (\bibinfo {year} {2012})}\BibitemShut {NoStop}%
\bibitem [{\citenamefont {Price}\ and\ \citenamefont
  {Perkins}(2013)}]{price13}%
  \BibitemOpen
  \bibfield  {author} {\bibinfo {author} {\bibfnamefont {C.}~\bibnamefont
  {Price}}\ and\ \bibinfo {author} {\bibfnamefont {N.~B.}\ \bibnamefont
  {Perkins}},\ }\bibfield  {title} {\bibinfo {title} {Finite-temperature phase
  diagram of the classical Kitaev-Heisenberg model},\ }\href
  {https://doi.org/10.1103/PhysRevB.88.024410} {\bibfield  {journal} {\bibinfo
  {journal} {Phys. Rev. B}\ }\textbf {\bibinfo {volume} {88}},\ \bibinfo
  {pages} {024410} (\bibinfo {year} {2013})}\BibitemShut {NoStop}%
\bibitem [{\citenamefont {Janssen}\ \emph {et~al.}(2016)\citenamefont
  {Janssen}, \citenamefont {Andrade},\ and\ \citenamefont {Vojta}}]{janssen16}%
  \BibitemOpen
  \bibfield  {author} {\bibinfo {author} {\bibfnamefont {L.}~\bibnamefont
  {Janssen}}, \bibinfo {author} {\bibfnamefont {E.~C.}\ \bibnamefont
  {Andrade}},\ and\ \bibinfo {author} {\bibfnamefont {M.}~\bibnamefont
  {Vojta}},\ }\bibfield  {title} {\bibinfo {title} {Honeycomb-Lattice
  Heisenberg-Kitaev Model in a Magnetic Field: Spin Canting, Metamagnetism, and
  Vortex Crystals},\ }\href {https://doi.org/10.1103/PhysRevLett.117.277202}
  {\bibfield  {journal} {\bibinfo  {journal} {Phys. Rev. Lett.}\ }\textbf
  {\bibinfo {volume} {117}},\ \bibinfo {pages} {277202} (\bibinfo {year}
  {2016})}\BibitemShut {NoStop}%
\bibitem [{\citenamefont {Janssen}\ \emph {et~al.}(2017)\citenamefont
  {Janssen}, \citenamefont {Andrade},\ and\ \citenamefont {Vojta}}]{janssen17}%
  \BibitemOpen
  \bibfield  {author} {\bibinfo {author} {\bibfnamefont {L.}~\bibnamefont
  {Janssen}}, \bibinfo {author} {\bibfnamefont {E.~C.}\ \bibnamefont
  {Andrade}},\ and\ \bibinfo {author} {\bibfnamefont {M.}~\bibnamefont
  {Vojta}},\ }\bibfield  {title} {\bibinfo {title} {Magnetization processes of
  zigzag states on the honeycomb lattice: Identifying spin models for
  $\alpha$-RuCl$_3$ and ${\mathrm{Na}}_{2}{\mathrm{IrO}}_{3}$},\ }\href
  {https://doi.org/10.1103/PhysRevB.96.064430} {\bibfield  {journal} {\bibinfo
  {journal} {Phys. Rev. B}\ }\textbf {\bibinfo {volume} {96}},\ \bibinfo
  {pages} {064430} (\bibinfo {year} {2017})}\BibitemShut {NoStop}%
\bibitem [{\citenamefont {Koga}\ \emph {et~al.}(2018)\citenamefont {Koga},
  \citenamefont {Tomishige},\ and\ \citenamefont {Nasu}}]{koga18}%
  \BibitemOpen
  \bibfield  {author} {\bibinfo {author} {\bibfnamefont {A.}~\bibnamefont
  {Koga}}, \bibinfo {author} {\bibfnamefont {H.}~\bibnamefont {Tomishige}},\
  and\ \bibinfo {author} {\bibfnamefont {J.}~\bibnamefont {Nasu}},\ }\bibfield
  {title} {\bibinfo {title} {Ground-state and Thermodynamic Properties of an S
  = 1 Kitaev Model},\ }\href {https://doi.org/10.7566/JPSJ.87.063703}
  {\bibfield  {journal} {\bibinfo  {journal} {Journal of the Physical Society
  of Japan}\ }\textbf {\bibinfo {volume} {87}},\ \bibinfo {pages} {063703}
  (\bibinfo {year} {2018})}\BibitemShut {NoStop}%
\bibitem [{\citenamefont {Stavropoulos}\ \emph {et~al.}(2019)\citenamefont
  {Stavropoulos}, \citenamefont {Pereira},\ and\ \citenamefont
  {Kee}}]{stavropoulos19}%
  \BibitemOpen
  \bibfield  {author} {\bibinfo {author} {\bibfnamefont {P.~P.}\ \bibnamefont
  {Stavropoulos}}, \bibinfo {author} {\bibfnamefont {D.}~\bibnamefont
  {Pereira}},\ and\ \bibinfo {author} {\bibfnamefont {H.-Y.}\ \bibnamefont
  {Kee}},\ }\bibfield  {title} {\bibinfo {title} {Microscopic Mechanism for a
  Higher-Spin Kitaev Model},\ }\href
  {https://doi.org/10.1103/PhysRevLett.123.037203} {\bibfield  {journal}
  {\bibinfo  {journal} {Phys. Rev. Lett.}\ }\textbf {\bibinfo {volume} {123}},\
  \bibinfo {pages} {037203} (\bibinfo {year} {2019})}\BibitemShut {NoStop}%
\bibitem [{\citenamefont {Dong}\ and\ \citenamefont {Sheng}(2020)}]{dong20}%
  \BibitemOpen
  \bibfield  {author} {\bibinfo {author} {\bibfnamefont {X.-Y.}\ \bibnamefont
  {Dong}}\ and\ \bibinfo {author} {\bibfnamefont {D.~N.}\ \bibnamefont
  {Sheng}},\ }\bibfield  {title} {\bibinfo {title} {Spin-1 Kitaev-Heisenberg
  model on a honeycomb lattice},\ }\href
  {https://doi.org/10.1103/PhysRevB.102.121102} {\bibfield  {journal} {\bibinfo
   {journal} {Phys. Rev. B}\ }\textbf {\bibinfo {volume} {102}},\ \bibinfo
  {pages} {121102} (\bibinfo {year} {2020})}\BibitemShut {NoStop}%
\bibitem [{\citenamefont {Jin}\ \emph {et~al.}(2022)\citenamefont {Jin},
  \citenamefont {Natori}, \citenamefont {Pollmann},\ and\ \citenamefont
  {Knolle}}]{jin22}%
  \BibitemOpen
  \bibfield  {author} {\bibinfo {author} {\bibfnamefont {H.-K.}\ \bibnamefont
  {Jin}}, \bibinfo {author} {\bibfnamefont {W.~M.~H.}\ \bibnamefont {Natori}},
  \bibinfo {author} {\bibfnamefont {F.}~\bibnamefont {Pollmann}},\ and\
  \bibinfo {author} {\bibfnamefont {J.}~\bibnamefont {Knolle}},\ }\bibfield
  {title} {\bibinfo {title} {{Unveiling the S=3/2 Kitaev honeycomb spin
  liquids}},\ }\href {https://doi.org/10.1038/s41467-022-31503-0} {\bibfield
  {journal} {\bibinfo  {journal} {Nature Communications}\ }\textbf {\bibinfo
  {volume} {13}},\ \bibinfo {pages} {3813} (\bibinfo {year}
  {2022})}\BibitemShut {NoStop}%
\bibitem [{\citenamefont {Chaloupka}\ and\ \citenamefont
  {Khaliullin}(2015)}]{chaloupka15}%
  \BibitemOpen
  \bibfield  {author} {\bibinfo {author} {\bibfnamefont {J.}~\bibnamefont
  {Chaloupka}}\ and\ \bibinfo {author} {\bibfnamefont {G.}~\bibnamefont
  {Khaliullin}},\ }\bibfield  {title} {\bibinfo {title} {Hidden symmetries of
  the extended Kitaev-Heisenberg model: Implications for the honeycomb-lattice
  iridates ${A}_{2}{\mathrm{IrO}}_{3}$},\ }\href
  {https://doi.org/10.1103/PhysRevB.92.024413} {\bibfield  {journal} {\bibinfo
  {journal} {Phys. Rev. B}\ }\textbf {\bibinfo {volume} {92}},\ \bibinfo
  {pages} {024413} (\bibinfo {year} {2015})}\BibitemShut {NoStop}%
\bibitem [{\citenamefont {Winter}\ \emph {et~al.}(2018)\citenamefont {Winter},
  \citenamefont {Riedl}, \citenamefont {Kaib}, \citenamefont {Coldea},\ and\
  \citenamefont {Valent\'{\i}}}]{winter18}%
  \BibitemOpen
  \bibfield  {author} {\bibinfo {author} {\bibfnamefont {S.~M.}\ \bibnamefont
  {Winter}}, \bibinfo {author} {\bibfnamefont {K.}~\bibnamefont {Riedl}},
  \bibinfo {author} {\bibfnamefont {D.}~\bibnamefont {Kaib}}, \bibinfo {author}
  {\bibfnamefont {R.}~\bibnamefont {Coldea}},\ and\ \bibinfo {author}
  {\bibfnamefont {R.}~\bibnamefont {Valent\'{\i}}},\ }\bibfield  {title}
  {\bibinfo {title} {Probing
  $\ensuremath{\alpha}\ensuremath{-}{\mathrm{RuCl}}_{3}$ Beyond Magnetic Order:
  Effects of Temperature and Magnetic Field},\ }\href
  {https://doi.org/10.1103/PhysRevLett.120.077203} {\bibfield  {journal}
  {\bibinfo  {journal} {Phys. Rev. Lett.}\ }\textbf {\bibinfo {volume} {120}},\
  \bibinfo {pages} {077203} (\bibinfo {year} {2018})}\BibitemShut {NoStop}%
\bibitem [{\citenamefont {Rau}\ \emph {et~al.}(2018)\citenamefont {Rau},
  \citenamefont {McClarty},\ and\ \citenamefont {Moessner}}]{rau18}%
  \BibitemOpen
  \bibfield  {author} {\bibinfo {author} {\bibfnamefont {J.~G.}\ \bibnamefont
  {Rau}}, \bibinfo {author} {\bibfnamefont {P.~A.}\ \bibnamefont {McClarty}},\
  and\ \bibinfo {author} {\bibfnamefont {R.}~\bibnamefont {Moessner}},\
  }\bibfield  {title} {\bibinfo {title} {Pseudo-Goldstone Gaps and
  Order-by-Quantum Disorder in Frustrated Magnets},\ }\href
  {https://doi.org/10.1103/PhysRevLett.121.237201} {\bibfield  {journal}
  {\bibinfo  {journal} {Phys. Rev. Lett.}\ }\textbf {\bibinfo {volume} {121}},\
  \bibinfo {pages} {237201} (\bibinfo {year} {2018})}\BibitemShut {NoStop}%
\bibitem [{\citenamefont {Khatua}\ \emph {et~al.}(2023)\citenamefont {Khatua},
  \citenamefont {Gingras},\ and\ \citenamefont {Rau}}]{khatua23}%
  \BibitemOpen
  \bibfield  {author} {\bibinfo {author} {\bibfnamefont {S.}~\bibnamefont
  {Khatua}}, \bibinfo {author} {\bibfnamefont {M.~J.~P.}\ \bibnamefont
  {Gingras}},\ and\ \bibinfo {author} {\bibfnamefont {J.~G.}\ \bibnamefont
  {Rau}},\ }\bibfield  {title} {\bibinfo {title} {Pseudo-Goldstone Modes and
  Dynamical Gap Generation from Order by Thermal Disorder},\ }\href
  {https://doi.org/10.1103/PhysRevLett.130.266702} {\bibfield  {journal}
  {\bibinfo  {journal} {Phys. Rev. Lett.}\ }\textbf {\bibinfo {volume} {130}},\
  \bibinfo {pages} {266702} (\bibinfo {year} {2023})}\BibitemShut {NoStop}%
\bibitem [{\citenamefont {Marshall}(1955)}]{marshall55}%
  \BibitemOpen
  \bibfield  {author} {\bibinfo {author} {\bibfnamefont {W.}~\bibnamefont
  {Marshall}},\ }\bibfield  {title} {\bibinfo {title} {{Antiferromagnetism}},\
  }\href {https://doi.org/10.1098/rspa.1955.0200} {\bibfield  {journal}
  {\bibinfo  {journal} {Proc. Roy. Soc.}\ }\textbf {\bibinfo {volume} {A232}},\
  \bibinfo {pages} {48} (\bibinfo {year} {1955})}\BibitemShut {NoStop}%
\bibitem [{\citenamefont {Lieb}\ \emph {et~al.}(1961)\citenamefont {Lieb},
  \citenamefont {Schultz},\ and\ \citenamefont {Mattis}}]{lieb61}%
  \BibitemOpen
  \bibfield  {author} {\bibinfo {author} {\bibfnamefont {E.}~\bibnamefont
  {Lieb}}, \bibinfo {author} {\bibfnamefont {T.}~\bibnamefont {Schultz}},\ and\
  \bibinfo {author} {\bibfnamefont {D.}~\bibnamefont {Mattis}},\ }\bibfield
  {title} {\bibinfo {title} {{Two soluble models of an antiferromagnetic
  chain}},\ }\href
  {https://doi.org/https://doi.org/10.1016/0003-4916(61)90115-4} {\bibfield
  {journal} {\bibinfo  {journal} {Ann. Phys.}\ }\textbf {\bibinfo {volume}
  {16}},\ \bibinfo {pages} {407} (\bibinfo {year} {1961})}\BibitemShut
  {NoStop}%
\bibitem [{\citenamefont {Lieb}\ and\ \citenamefont {Mattis}(1962)}]{lieb62}%
  \BibitemOpen
  \bibfield  {author} {\bibinfo {author} {\bibfnamefont {E.}~\bibnamefont
  {Lieb}}\ and\ \bibinfo {author} {\bibfnamefont {D.}~\bibnamefont {Mattis}},\
  }\bibfield  {title} {\bibinfo {title} {{Ordering Energy Levels of Interacting
  Spin Systems}},\ }\href {https://doi.org/10.1063/1.1724276} {\bibfield
  {journal} {\bibinfo  {journal} {J. Math. Phys.}\ }\textbf {\bibinfo {volume}
  {3}},\ \bibinfo {pages} {749} (\bibinfo {year} {1962})}\BibitemShut {NoStop}%
\bibitem [{\citenamefont {C\^onsoli}\ \emph {et~al.}(2021)\citenamefont
  {C\^onsoli}, \citenamefont {Fornoville},\ and\ \citenamefont
  {Vojta}}]{consoli21}%
  \BibitemOpen
  \bibfield  {author} {\bibinfo {author} {\bibfnamefont {P.~M.}\ \bibnamefont
  {C\^onsoli}}, \bibinfo {author} {\bibfnamefont {M.}~\bibnamefont
  {Fornoville}},\ and\ \bibinfo {author} {\bibfnamefont {M.}~\bibnamefont
  {Vojta}},\ }\bibfield  {title} {\bibinfo {title} {Fluctuation-induced
  ferrimagnetism in sublattice-imbalanced antiferromagnets with application to
  ${\mathrm{SrCu}}_{2}$(${\mathrm{BO}}_{3}{)}_{2}$ under pressure},\ }\href
  {https://doi.org/10.1103/PhysRevB.104.064422} {\bibfield  {journal} {\bibinfo
   {journal} {Phys. Rev. B}\ }\textbf {\bibinfo {volume} {104}},\ \bibinfo
  {pages} {064422} (\bibinfo {year} {2021})}\BibitemShut {NoStop}%
\bibitem [{\citenamefont {Winter}\ \emph {et~al.}(2016)\citenamefont {Winter},
  \citenamefont {Li}, \citenamefont {Jeschke},\ and\ \citenamefont
  {Valent\'{\i}}}]{winter16}%
  \BibitemOpen
  \bibfield  {author} {\bibinfo {author} {\bibfnamefont {S.~M.}\ \bibnamefont
  {Winter}}, \bibinfo {author} {\bibfnamefont {Y.}~\bibnamefont {Li}}, \bibinfo
  {author} {\bibfnamefont {H.~O.}\ \bibnamefont {Jeschke}},\ and\ \bibinfo
  {author} {\bibfnamefont {R.}~\bibnamefont {Valent\'{\i}}},\ }\bibfield
  {title} {\bibinfo {title} {Challenges in design of Kitaev materials: Magnetic
  interactions from competing energy scales},\ }\href
  {https://doi.org/10.1103/PhysRevB.93.214431} {\bibfield  {journal} {\bibinfo
  {journal} {Phys. Rev. B}\ }\textbf {\bibinfo {volume} {93}},\ \bibinfo
  {pages} {214431} (\bibinfo {year} {2016})}\BibitemShut {NoStop}%
\bibitem [{\citenamefont {Maksimov}\ and\ \citenamefont
  {Chernyshev}(2020)}]{maksimov20}%
  \BibitemOpen
  \bibfield  {author} {\bibinfo {author} {\bibfnamefont {P.~A.}\ \bibnamefont
  {Maksimov}}\ and\ \bibinfo {author} {\bibfnamefont {A.~L.}\ \bibnamefont
  {Chernyshev}},\ }\bibfield  {title} {\bibinfo {title} {Rethinking
  $\ensuremath{\alpha}\text{{-}}{\mathrm{RuCl}}_{3}$},\ }\href
  {https://doi.org/10.1103/PhysRevResearch.2.033011} {\bibfield  {journal}
  {\bibinfo  {journal} {Phys. Rev. Res.}\ }\textbf {\bibinfo {volume} {2}},\
  \bibinfo {pages} {033011} (\bibinfo {year} {2020})}\BibitemShut {NoStop}%
\bibitem [{\citenamefont {Möller}\ \emph {et~al.}(2025)\citenamefont
  {Möller}, \citenamefont {Maksimov}, \citenamefont {Jiang}, \citenamefont
  {White}, \citenamefont {Valenti},\ and\ \citenamefont
  {Chernyshev}}]{moeller25}%
  \BibitemOpen
  \bibfield  {author} {\bibinfo {author} {\bibfnamefont {M.}~\bibnamefont
  {Möller}}, \bibinfo {author} {\bibfnamefont {P.~A.}\ \bibnamefont
  {Maksimov}}, \bibinfo {author} {\bibfnamefont {S.}~\bibnamefont {Jiang}},
  \bibinfo {author} {\bibfnamefont {S.~R.}\ \bibnamefont {White}}, \bibinfo
  {author} {\bibfnamefont {R.}~\bibnamefont {Valenti}},\ and\ \bibinfo {author}
  {\bibfnamefont {A.~L.}\ \bibnamefont {Chernyshev}},\ }\bibinfo {title} {The
  Saga of $\alpha$-RuCl$_3$: Parameters, Models, and Phase Diagrams},\ \Eprint
  {https://arxiv.org/abs/2502.08698} {arXiv:2502.08698}\BibitemShut {NoStop}%
\bibitem [{\citenamefont {Rau}\ and\ \citenamefont {Kee}(2014)}]{rau14b}%
  \BibitemOpen
  \bibfield  {author} {\bibinfo {author} {\bibfnamefont {J.~G.}\ \bibnamefont
  {Rau}}\ and\ \bibinfo {author} {\bibfnamefont {H.-Y.}\ \bibnamefont {Kee}},\
  }\bibinfo {title} {Trigonal distortion in the honeycomb iridates: Proximity
  of zigzag and spiral phases in Na$_2$IrO$_3$},\ \Eprint
  {https://arxiv.org/abs/1408.4811} {arXiv:1408.4811}\BibitemShut {NoStop}%
\bibitem [{\citenamefont {Chen}\ \emph {et~al.}(2023)\citenamefont {Chen},
  \citenamefont {Luo}, \citenamefont {Zhou}, \citenamefont {He}, \citenamefont
  {Xi}, \citenamefont {Jia}, \citenamefont {Luo},\ and\ \citenamefont
  {Zhao}}]{chen23}%
  \BibitemOpen
  \bibfield  {author} {\bibinfo {author} {\bibfnamefont {K.}~\bibnamefont
  {Chen}}, \bibinfo {author} {\bibfnamefont {Q.}~\bibnamefont {Luo}}, \bibinfo
  {author} {\bibfnamefont {Z.}~\bibnamefont {Zhou}}, \bibinfo {author}
  {\bibfnamefont {S.}~\bibnamefont {He}}, \bibinfo {author} {\bibfnamefont
  {B.}~\bibnamefont {Xi}}, \bibinfo {author} {\bibfnamefont {C.}~\bibnamefont
  {Jia}}, \bibinfo {author} {\bibfnamefont {H.-G.}\ \bibnamefont {Luo}},\ and\
  \bibinfo {author} {\bibfnamefont {J.}~\bibnamefont {Zhao}},\ }\bibfield
  {title} {\bibinfo {title} {Triple-meron crystal in high-spin Kitaev
  magnets},\ }\href {https://doi.org/10.1088/1367-2630/acb5bb} {\bibfield
  {journal} {\bibinfo  {journal} {New J. Phys.}\ }\textbf {\bibinfo {volume}
  {25}},\ \bibinfo {pages} {023006} (\bibinfo {year} {2023})}\BibitemShut
  {NoStop}%
\bibitem [{\citenamefont {Stavropoulos}\ \emph {et~al.}(2024)\citenamefont
  {Stavropoulos}, \citenamefont {Yang}, \citenamefont {Rousochatzakis},\ and\
  \citenamefont {Perkins}}]{stavropoulos24}%
  \BibitemOpen
  \bibfield  {author} {\bibinfo {author} {\bibfnamefont {P.~P.}\ \bibnamefont
  {Stavropoulos}}, \bibinfo {author} {\bibfnamefont {Y.}~\bibnamefont {Yang}},
  \bibinfo {author} {\bibfnamefont {I.}~\bibnamefont {Rousochatzakis}},\ and\
  \bibinfo {author} {\bibfnamefont {N.~B.}\ \bibnamefont {Perkins}},\
  }\bibfield  {title} {\bibinfo {title} {Complex orders and chirality in the
  classical Kitaev-$\mathrm{\ensuremath{\Gamma}}$ model},\ }\href
  {https://doi.org/10.1103/PhysRevB.110.214406} {\bibfield  {journal} {\bibinfo
   {journal} {Phys. Rev. B}\ }\textbf {\bibinfo {volume} {110}},\ \bibinfo
  {pages} {214406} (\bibinfo {year} {2024})}\BibitemShut {NoStop}%
\bibitem [{\citenamefont {Yang}\ \emph {et~al.}(2012)\citenamefont {Yang},
  \citenamefont {Albuquerque}, \citenamefont {Capponi}, \citenamefont
  {L{\"a}uchli},\ and\ \citenamefont {Schmidt}}]{yang12}%
  \BibitemOpen
  \bibfield  {author} {\bibinfo {author} {\bibfnamefont {H.-Y.}\ \bibnamefont
  {Yang}}, \bibinfo {author} {\bibfnamefont {A.~F.}\ \bibnamefont
  {Albuquerque}}, \bibinfo {author} {\bibfnamefont {S.}~\bibnamefont
  {Capponi}}, \bibinfo {author} {\bibfnamefont {A.~M.}\ \bibnamefont
  {L{\"a}uchli}},\ and\ \bibinfo {author} {\bibfnamefont {K.~P.}\ \bibnamefont
  {Schmidt}},\ }\bibfield  {title} {\bibinfo {title} {Effective spin couplings
  in the Mott insulator of the honeycomb lattice Hubbard model},\ }\href
  {https://doi.org/10.1088/1367-2630/14/11/115027} {\bibfield  {journal}
  {\bibinfo  {journal} {New J. Phys.}\ }\textbf {\bibinfo {volume} {14}},\
  \bibinfo {pages} {115027} (\bibinfo {year} {2012})}\BibitemShut {NoStop}%
\bibitem [{\citenamefont {Wang}\ and\ \citenamefont {Liu}(2023)}]{wang23}%
  \BibitemOpen
  \bibfield  {author} {\bibinfo {author} {\bibfnamefont {J.}~\bibnamefont
  {Wang}}\ and\ \bibinfo {author} {\bibfnamefont {Z.-X.}\ \bibnamefont {Liu}},\
  }\bibfield  {title} {\bibinfo {title} {Effect of ring-exchange interactions
  in the extended Kitaev honeycomb model},\ }\href
  {https://doi.org/10.1103/PhysRevB.108.014437} {\bibfield  {journal} {\bibinfo
   {journal} {Phys. Rev. B}\ }\textbf {\bibinfo {volume} {108}},\ \bibinfo
  {pages} {014437} (\bibinfo {year} {2023})}\BibitemShut {NoStop}%
\bibitem [{\citenamefont {Francini}\ \emph {et~al.}()\citenamefont {Francini},
  \citenamefont {Cônsoli},\ and\ \citenamefont {Janssen}}]{data-availability}%
  \BibitemOpen
  \bibfield  {author} {\bibinfo {author} {\bibfnamefont {N.}~\bibnamefont
  {Francini}}, \bibinfo {author} {\bibfnamefont {P.~M.}\ \bibnamefont
  {Cônsoli}},\ and\ \bibinfo {author} {\bibfnamefont {L.}~\bibnamefont
  {Janssen}},\ }\bibinfo {title} {Data for ``Ferrimagnetism from quantum
  fluctuations in Kitaev materials''},\ \bibinfo {note}
  {\href{https://doi.org/10.25532/OPARA-895}{https://doi.org/10.25532/OPARA-895}}\BibitemShut
  {NoStop}%
\bibitem [{\citenamefont {Colpa}(1978)}]{colpa78}%
  \BibitemOpen
  \bibfield  {author} {\bibinfo {author} {\bibfnamefont {J.}~\bibnamefont
  {Colpa}},\ }\bibfield  {title} {\bibinfo {title} {Diagonalization of the
  quadratic boson hamiltonian},\ }\href
  {https://doi.org/https://doi.org/10.1016/0378-4371(78)90160-7} {\bibfield
  {journal} {\bibinfo  {journal} {Physica A: Statistical Mechanics and its
  Applications}\ }\textbf {\bibinfo {volume} {93}},\ \bibinfo {pages} {327}
  (\bibinfo {year} {1978})}\BibitemShut {NoStop}%
\end{thebibliography}%

\end{document}